\documentclass{aastex}
\usepackage{emulateapj5}
\pdfoutput=1

\shortauthors{Chiboucas,Karachentsev,Tully}
\shorttitle{Discovery of New Dwarf Galaxies in M81}

\begin{document}

\title{Discovery of New Dwarf Galaxies in the M81 Group}

\author{Kristin Chiboucas\altaffilmark{1}, Igor D. Karachentsev\altaffilmark{2}, and R. Brent Tully\altaffilmark{1}}
\email{kchib@ifa.hawaii.edu, ikar@luna.sao.ru, tully@ifa.hawaii.edu}

\altaffiltext{1}{Institute for Astronomy, University of Hawaii, 2680 Woodlawn Dr., Honolulu, HI 96821}
\altaffiltext{2}{Special Astrophysical Observatory (SAO)}

\begin{abstract}
An order of magnitude more dwarf galaxies are expected to inhabit the Local Group,
based on currently accepted galaxy formation models, than have been observed. This
discrepancy has been noted in environments ranging from the field to rich clusters.
However, no complete census of dwarf galaxies exists in any environment.  The
discovery of the smallest and faintest dwarfs is hampered by the limitations
in detecting such faint and low surface brightness galaxies.  An even greater
difficulty is establishing distances to or group/cluster membership for such faint 
galaxies.  The M81 Group provides an almost unique opportunity for establishing 
membership for galaxies in
a low density region complete to magnitudes as faint as $M_{r^{\prime}} = -10$.  With
a distance modulus of 27.8, the tip of the red giant branch just resolves in ground-based
surveys.   We have surveyed a 65 square degree region around M81 with CFHT/MegaCam.
From these images we have detected 22 new dwarf galaxy candidates.  Photometric, 
morphological, and structural properties are presented for the candidates.   
The group luminosity function has a faint end slope characterized by the parameter
$\alpha = -1.27\pm0.06$. We discuss implications of this
dwarf galaxy population for cosmological models.

\end{abstract}


\section{Introduction}\label{intro}

Standard $\Lambda$CDM, hierarchical structure formation models predict an order of magnitude 
more low mass halos in 
group environments and as satellites around Milky Way-like galaxies than there are 
observed dwarf galaxies \citep{klypin99,moore99}.   The steeper slope of the
predicted mass function over the observed luminosity function (LF) is
the source of the 'missing galaxy' problem.  There is growing evidence that there are
fewer dwarfs than predicted in all environments,
but the situation is even worse in field and group environments.  Simulations predict 
low density environments to have fractionally more dwarfs than the high density cluster
regions, whereas observationally the reverse has been measured \citep{rds04,tsb05,rds07,bls05,bark07}.

Several explanations to resolve this discrepancy 
between theory and observations have been proposed.  Cosmological solutions include
modifying the power spectrum on small scales \citep{kl00}, perhaps by
invoking warm dark matter \citep{colin00}.  Numerical simulations generally
predict subhalo counts as a function of total mass or maximum circular
velocity.  For Local Group satellites, these values are not well determined, making
direct comparison between these models and observations difficult unless other
parameters, such as the mass within a specific radius, are used \citep{sbkd07}. 
However, these more appropriate
comparisons still find predictions inconsistent with observations.

A large number of astrophysical solutions have been put forth to explain
the apparent underabundance of dwarf galaxies.
Number counts can be brought into agreement with models if it is assumed the observed
dwarf galaxies reside in relatively massive dark matter halos while lower mass halos 
remain invisible \citep{stoehr02}.  If star formation efficiency
is lower in lower density halos, or if large gas reservoirs are expelled from
small potential wells after an initial burst of star formation and subsequent
supernova activity, low mass galaxies will remain fairly dark and difficult to detect. 
Greater numbers of visible dwarfs might then be expected to inhabit the high density,
high gas pressure environment of rich clusters if 
the intracluster medium served to contain the dwarf galaxy outflows \citep{br92}.
For very small masses, cooling times may be greater than the age of the universe \citep{haiman96}.
Destructive processes which act in rich environments such as clusters may decrease
dwarf number counts or lower luminosities making them more difficult to detect.  Such
processes include harassment, strangulation, and ram pressure stripping 
\citep{mkldo96,st84,bcs01,mb00} and would be
expected to play a larger role in shaping galaxy populations in denser, more massive, and
hotter clusters.  Instead, high density regions have been observed to contain a 
larger fraction of dwarf to giant galaxies than low density regions. 

Another theory argues that reionization suppresses the collapse of gas in 
low mass halos thereby producing a shallower faint-end slope in the luminosity 
function \citep{tw96}.  This process will not affect gas accretion in halos that 
collapsed before the era of reionization - those that resided in denser regions, 
such as rich clusters, would preferentially collapse earlier when
intergalactic gas was still cold and halos could accumulate the gas.  
Reionization will therefore predominantly suppress baryon accumulation in low mass 
halos in lower density regions such as the poorer group and field 
environment, naturally explaining the trend of lower dwarf-to-giant ratios found in
lower density regions \citep{tstv02}.  In support of this theory, \citet{sbkd07} and 
\citet{simon07} find good agreement with 
theoretical expectations if the observed Local Group satellites correspond to the
earliest collapsed halos, or the most massive at the time of accretion.  

Distinguishing between all these possibilities has proven difficult, but before 
we can draw any conclusions as to the source of the discrepancy,
we must first constrain the number counts of dwarf galaxies and the faint-end slope of the
luminosity function.
No complete census of dwarf galaxies has been obtained in any environment, although
much recent progress has been made in the detection of faint Local Group satellites.
With SDSS survey releases, the number of known Milky Way and M31 satellites
has nearly doubled in just the past 4 years \citep{simon07}. This new information 
has mitigated the missing galaxy problem, but has not brought the number counts 
into alignment with $\Lambda$CDM predictions.  
Due to the sky coverage required for a complete survey and restrictions 
caused by the zone of obscuration, a complete census
of Local Group dwarf galaxies is not on the horizon.

The determination of the faint-end of the galaxy luminosity function
is hampered by the limitations in detecting the very faintest and lowest surface
brightness galaxies, and this is compounded by the the great difficulty in 
determining accurate distances to, or ascertaining group and cluster membership for,
such faint objects.  To measure the faint-end of the luminosity function
for nearby clusters of galaxies, researchers have often resorted to using
statistical means to estimate cluster memberships (see e.g. 
\citet{bntuw95, t97, KCMM1}).  Others use
colors or morphology in an attempt to directly assess membership probability 
\citep{shp97, mtt05}.
All these methods are fraught with large uncertainty.  Spectroscopy can
be used to establish membership, but to observe the large numbers of galaxies to
the necessary depth is prohibitively time consuming.  

The M81 Group provides
an alternative means to directly establish membership for galaxies to
magnitudes as faint as $M_{R} \sim -7$ in a low density environment.  
Residing at a distance of only 3.6 Mpc, the galaxies
in this group resolve into stars.  Supergiants and giants near the tip of
the red giant branch (TRGB) resolve over the diffuse
main sequence population light.  The degree of resolution is enough
to establish the group membership status with high confidence.  From
color-magnitude diagrams, one can determine TRGB distances to unambiguously 
assign membership.
Given this fortuitous circumstance, we can probe the faint-end of the
LF in a group environment down to $M_{R} \sim -9$ to high accuracy. 

\begin{figure}[t]
\begin{centering}
\includegraphics[angle=0, totalheight=8.0in]{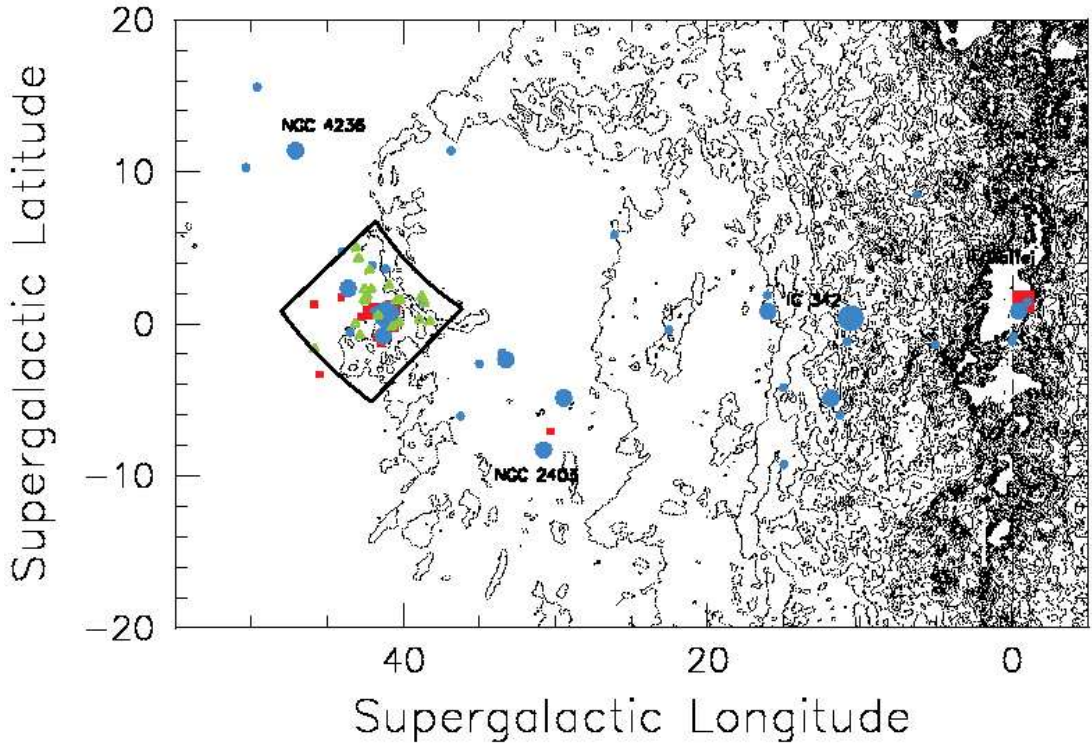}
\caption[The extended region around M81]{The extended region around M81.  
All galaxies with velocities in the Galactic standard of rest less than 
400 km s$^{-1}$ in an area of $60 \times 40$ degrees in supergalactic coordinates.  
Circles: spirals and irregulars.  Squares: early types.
Large, mid, and small symbols identify galaxies with $M_B<-20$, 
$-20 < M_B < -16$, and $M_B>-16$, respectively.  The box outlines the 
area of the CFHT MegaCam survey and the small open triangles identify the newly 
discovered dwarfs.  Contours illustrate the dust maps of \citet{sfd98}
at intervals of 0.2 $r^{\prime}$ magnitudes of extinction.
\label{fulreg}}
\end{centering}
\end{figure}

The M81 Group is embedded in a filament of galaxies that runs up from behind the 
Galactic plane, shown in Figure \ref{fulreg}.  The M81 Group has traditionally 
been considered to include the galaxies around NGC~2403 and those around
NGC~4236, each about 800~kpc removed in projection from M81.  
Maffei 1-2 and IC~342 are considered to be the dominant members of a 
separate group, 2~Mpc away in projection.   The nearest neighbors at higher 
supergalactic longitude are a group around M101 at SGL=64, SGB=+23 and the near 
Canes Venatici group at SGL=76, SGB=+9.  It has been shown by
\citet{K2002} that the subcomponents of 
the M81 Group around NGC~2403 and M81 itself are falling toward each 
other, consistent with a total mass within the zero velocity surface 
bounding the group of $1.6 \times 10^{12} M_{\odot}$.  The interest 
of the present study is the more dynamically evolved
subcomponent around M81 contained within the box in Fig. \ref{fulreg}.

We surveyed a  $8 \times 8$ degree area centered on the M81 Group 
at $\alpha=$ 10:05:00 $\delta=$ 68:15:00 J2000 to search for new
dwarf members of the group. Apart from the five brightest systems: M81,
M82, NGC 2976, NGC 3077, and IC 2574, the inspected area contains 13 previously known
dwarf galaxies which were discovered in the
1960-80s on POSS-I prints and the Tautenburg Schmidt telescope plates:  Holm I, 
BK3N, KDG 61, Holm IX, Anon 0952 (Arp's loop), Garland, BK5N, KDG 63, KDG 64,
HS 117, DDO 78, DDO 82, and BK6N 
\citep{vdB71, Ka68, BK82, BK85}.
Later, 4 low surface brightness dwarfs: F8D1, FM1, KK 77,
and IKN were found on POSS-II films by \citet{cads98},\citet{FM00},
\citet{KKp00}, and \citet{KKHM04}.

Here we report on the discovery of 22 new M81 Group galaxy candidates.
In Section \ref{obs} we present the observations and data reduction for our survey
along with a discussion of the dwarf galaxy detection techniques, survey limits, 
and follow-up observations.  We provide measurements of the properties of the
newly discovered dwarf candidates in Section \ref{propcands} and construct the total
group LF in Section \ref{lf}.  We discuss the dwarf population characteristics in 
Section \ref{popchar}, the distribution of M81 satellites in Section \ref{distrib},
and implications of our survey results on the faint-end of the galaxy luminosity
function and for cosmological models in Sections \ref{lfconc} and \ref{cosmo}.  
Conclusions are presented in 
Section \ref{conc}.  Throughout this work we assume a distance modulus to M81
of 27.80 (3.6 Mpc).

\section{Observations and Data Reduction}\label{obs}

\subsection{Initial survey}\label{obssuvey}
                                                                                
The M81 Group was chosen in large part because it 
is near enough at 3.6 Mpc that stars near the tip of the red giant branch 
just resolve in ground-based images, enabling derivation of the group luminosity 
function complete down to very faint magnitudes.  Stars in Local Group
galaxies resolve down to fainter magnitudes but a complete survey
would require all-sky coverage. For a comparable survey of dwarf satellites around
M31 alone, imaging of over 800 square degrees would be required,
including observations within the zone of obscuration.  The scale
on which dwarfs cluster around parent galaxies is found to 
depend on the mass of the parent halo according to the relation
$r_{2t} = 193(M_{12})^{1/3}$ kpc \citep{ttm06} where $M_{12}$ is the 
mass of the halo in units of $10^{12} M_{\odot}$ and $r_{2t}$ is the 
second turnaround radius of spherical collapse, similar to the 
overdensity radius $r_{200}$.  At a distance of 3.6 Mpc and halo 
mass of $1.6 \times 10^{12} M_{\odot}$ \citep{K2002}, M81 has a second
turnaround radius of 230 kpc, or only 3.6 degrees.  A survey of
this group with complete areal coverage is therefore made possible with 
large format CCD detectors such as the square degree MegaCam detector
mounted on the 3.6m  Canada-France-Hawaii Telescope.
In addition, the M81 Group is interesting in its own right as a 
rich group with a large
late-type dwarf galaxy population, unlike our own well studied 
dSph-rich Local Group.

\begin{figure}[t]
\begin{centering}
\includegraphics[angle=0, totalheight=8.0in]{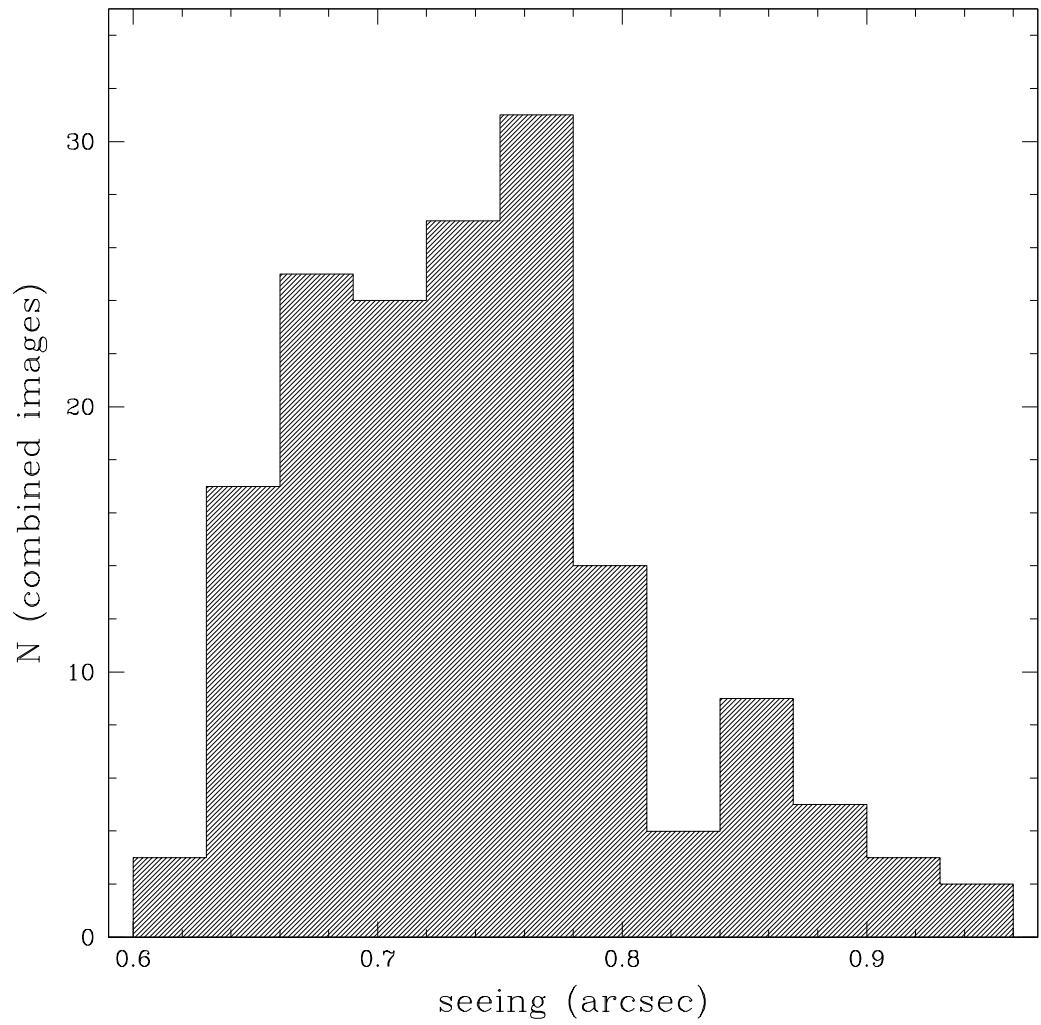}
\caption[Seeing spread]{Histogram displaying the seeing range for all 
combined images in our survey.
\label{seeplot}}
\end{centering}
\end{figure}

MegaCam is a wide field camera composed of a mosaic of 36 CCDs that
cover a total area of $0.96 \times 0.94$ deg with chip gaps 13
and 80 arcsec wide.  Each CCD is composed of 
$2048 \times 4612$ pixels of $0.187^{\prime\prime}$. 
We imaged a 65 square degree region centered around M81 in a raster
pattern that resulted in each location being observed twice for a total 
integration time of 1096s.  
The survey was conducted entirely in the $r'$-band because
we were primarily interested in the resolved giant population.
Typical seeing was 0.75 arcsec, but ranged
between 0.6 and 0.9 arcsec (Figure \ref{seeplot}).
Most of the data were observed on photometric nights, although a few
fields were imaged through thin cirrus.
Observations were carried out by CFHT staff in queue mode
over 3 semesters in the period 2005B-2006.

\subsection{MegaCam data processing methods}\label{methods}

The raw images were initially processed by TERAPIX, an astronomical data 
reduction center located at the Institut d'Astrophysique in Paris, 
which provides pipeline data reduction for MegaCam.  The TERAPIX pipeline
first evaluates data for image quality, identifies chip defects and 
cosmic rays, and generates weight and flag-map images. 
Each image is warped onto a common
astronomical projection (TAN) and coordinate system (J2000) using 
SWarp software and an astrometric reference catalog. 
SExtractor is run on all images to perform the object detection and photometry.
The astrometric and photometric calibrations are determined using SCAMP.
Astrometry is performed by downloading reference astrometic 
catalogs e.g. USNO-B1 \citep{Monet03}. These are  
cross correlated with each exposure and a pattern matching
algorithm is run.  Final astrometric uncertainties are dominated by errors in
the reference catalogs. On large scales, the astrometric accuracy is
expected to be better than $\sim 0.3$ arcsec.
Photometric calibration uses standard stars observed on photometric nights.
The weighted, quadratic sum of SExtractor parameter MAG\_AUTO differences
from overlapping detections is used to put all fields onto a common
zero-point scale.  Residuals for bright objects are typically less than
0.05 magnitudes rms with even smaller zero-point uncertainties.
These raw and processed MegaCam survey data are available online through the 
Canadian Astronomy Data Centre.

Our fields were mosaicked in such a way that every region was imaged
twice with some additional overlap along the sides of each field.  
These pairs of 548 sec images had still to be stacked.  Because 
MegaCam images are huge, we first split each image and its
associated mask image into quadrants.  The mask image contained
flagged chip defects such as bad columns and pixels as well as
cosmic rays.  Images were combined using IRAF tasks for image
registration and image combination.  Pixels that were flagged in
each image were ignored during the average operation.  Two to four
individual images were used to produce final stacked quadrant
images of size $30 \times 30$ arcmin.   

SExtractor was subsequently
re-run on these combined images.  Weight images were created during
the stacking step with the weight of each pixel set to the square
of the number of image pixels averaged together for the final
pixel value.  This weight image was used during the SExtractor detection
of objects to put a significantly lower weight on those regions
such as chip gaps where data came from only a single image as opposed to at least
two images as is the case over the majority of the field.  This prevented SExtractor
from detecting excess noise in these regions while still reaching 
to $\sim0.75$ magnitudes brighter than full exposure limits.  
SExtractor parameters were tuned to go deep enough to 
recover most of the mottles in the known resolved M81 Group galaxies 
by setting detection and deblending
levels to quite low values. Object detection was set to 
1$\sigma$ with a minimum area of 3 contiguous pixels while the 
object deblending parameter 
NTHRESH was set to $1\sigma$ and DEBLEND\_MINCONT to 5E-7.  
These low detection settings have the side
effect of inducing SExtractor to recover excess noise, primarily in the 
halos of bright stars and
galaxies, and occasionally in the numerous galactic cirrus patches obscuring
this part of the sky.  We err on the side of picking up spurious candidates
rather than losing real candidates.

\begin{figure}[t]
\begin{centering}
\includegraphics[angle=0, totalheight=8.0in]{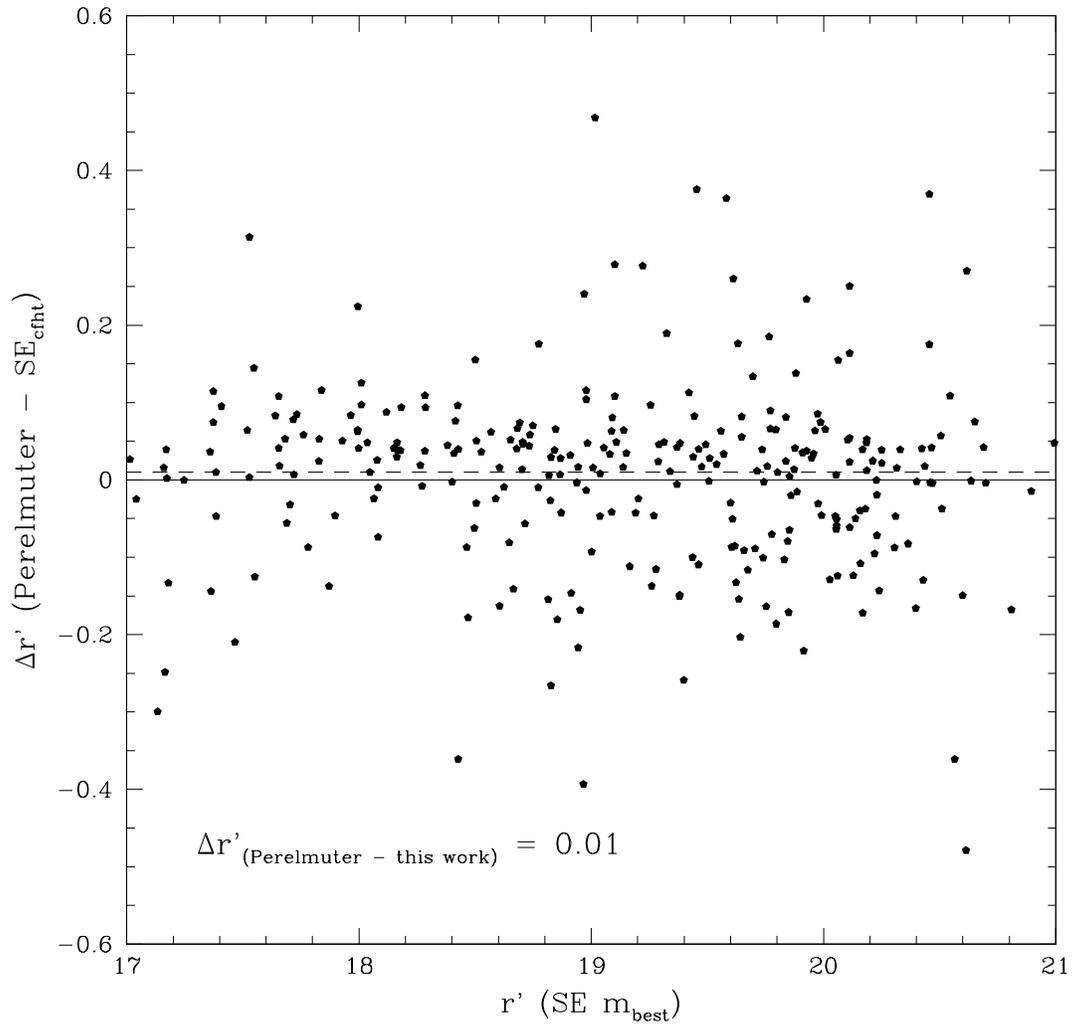}
\caption[Testing photometry]{Tests of photometric accuracy.  We compare
our r' total magnitudes to transformed R magnitudes of \citet{perel}.
\label{pdif}}
\end{centering}
\end{figure}

We checked the reliability of the photometry by comparison with
stars of known magnitudes. \citet{perel} provide a catalog of $B,V,R$ 
magnitudes for all objects in the vicinity of M81.  We transform the
$R$ magnitudes to the SDSS $r^{\prime}$ system using the rough transformation
$r^{\prime}$ = V - 0.84(V-R) + 0.13 provided in \citet{fuku}.  
The average difference in $r^{\prime}$ (Perelmuter - this work)
for stars with $17 < r^{\prime} < 19.5$ is 0.01 with
$\sigma = 0.11$ (Figure \ref{pdif}).  Since this is only an approximate color transformation,
we take this as strong support for the reliability of the photometry.

\subsection{Detection of 22 new candidate M81 companions}\label{cands}

The tip of the red giant branch lies at $M_{I} = -4.05$, with $(V-I)_{tip}\sim1.5$.
For $(V-R)_{tip} \sim 0.8$ and using color transformations for metal poor
population II stars from \citet{jordicol06}, we find $M_{r^{\prime}tip} \sim -3.1\pm0.1$.
With our 18 minute combined exposures, we reach limiting magnitudes for
resolved stars of $r^{\prime} \sim 25.0-25.5$.
From \citet{sfd98} maps, galactic reddening $A_{r^{\prime}}$ in this region is
typically 0.1-0.2 but as high as 0.7 magnitudes.  Because the M81 Group is so nearby
at a distance of only 3.6 Mpc ($\langle m - M\rangle_{\circ} = 27.8$),
we can resolve stars down to limiting magnitudes of $M_{r^{\prime}} \sim -2.9\pm0.5$.
The radial distance within the group will also affect the limiting
absolute magnitude by $\pm0.15$.
Thus, stars at the tip of the red giant branch are just
at the edge of detection in our
MegaCam images while young main sequence and red supergiant stars, and
intermediate age AGB stars are well resolved.  

Two different methods were used to search the MegaCam images for new M81 Group
dwarf galaxy candidates.  
The first method used visual inspection of the images.
Final images were examined
by eye by two of the authors (IK and KC) for new candidates by
identifying objects that had resolved and semi-resolved faint stars.
Good candidates also had an associated diffuse low surface brightness (LSB)
component, although this was not observed in every case.  17 new
candidate M81 dwarf members were detected in this way.

The second method used to discover new candidate dwarfs
implemented a 2-point correlation auto-detection routine.
The auto-detection code calculates the 2-point correlation using the
natural estimator for objects having magnitudes
$22 < r^{\prime} < 27$ on scales of 1.1-6.4 arcsec 
(0.02 - 0.11 kpc at the distance of M81) and within boxes of size $\sim40$ arcsec
(0.7 kpc).
The code is set up to use any box or scale size but we were
focusing on smaller galaxies with a concentration of resolved
stars.  Presumably large diffuse fields of stars could be missed,
although the code did pick up a number of these as well.
As the box marches across the image, the correlation values within the box are
summed and any region with an excess value is flagged.  
This value was empirically chosen to be 
80 based on recovery of previously known and newly visually detected
galaxies, and based on tests of the recovery of artificial galaxies (see below).
The value chosen maximized the ratio of the rate of recovery of artificially 
added galaxies to noise detections, but does then necessarily impose
minimum  magnitude and concentration limits on our survey.
All regions that were flagged were checked
by eye. Typically $40-200$ such regions were found in each square degree MegaCam
image.  The great majority turned out to be noise detections along
still remaining bad columns, or an excess of noise spikes around
halos of bright stars and galaxies or within cirrus.  
Several detections in each image proved to be distant galaxy clusters.
The remainder of the detections include the M81 candidates and previously 
known group members.
                                                                            
The code recovered all previously known dwarfs and most of the
ones found by eye in this paper, the 3 exceptions being
d0957+70, d0944+69, and d1016+69. For these 3, too few resolved
points were detected by SExtractor.  It did pick
out 2 dwarfs, d1009+70 and d0946+68, which have more of a diffuse 
LSB component than a resolved stellar population and are thus more 
likely to be galaxies at somewhat larger distances.
We retain all 5 questionable objects in our full candidate list but 
although we still believe 
d1016+69 to be a good candidate, we consider d0957+70 to be a possible artifact
and the other 3 to be potentially more distant galaxies.
                                                                             
The code also found a number of candidates not picked out by eye.  A couple
of these appear to have both stellar and diffuse
LSB components: d1006+67 is the best of these candidates with a 
resolved stellar component contained within a diffuse low surface brightness region.
d0934+70 is hidden behind cirrus but appears to have both a faint resolved
stellar component and excess diffuse emission visible over the cirrus.
d1048+70 appears to have a diffuse component but only a semi-resolved stellar
component.  If real, this may perhaps be a more distant dwarf.
A couple more have an excess of faint stellar-like detections but
in these cases the objects do not appear to be concentrated nor
have an LSB component and may prove to be distant galaxy clusters or
chance concentrations of stars.
Coordinates and basic parameters for a total of 22 new candidate
dwarf galaxy group members are listed in Table \ref{dwarftab1}.

\begin{figure}[t!]
\begin{centering}
\resizebox{\columnwidth}{!}{\includegraphics{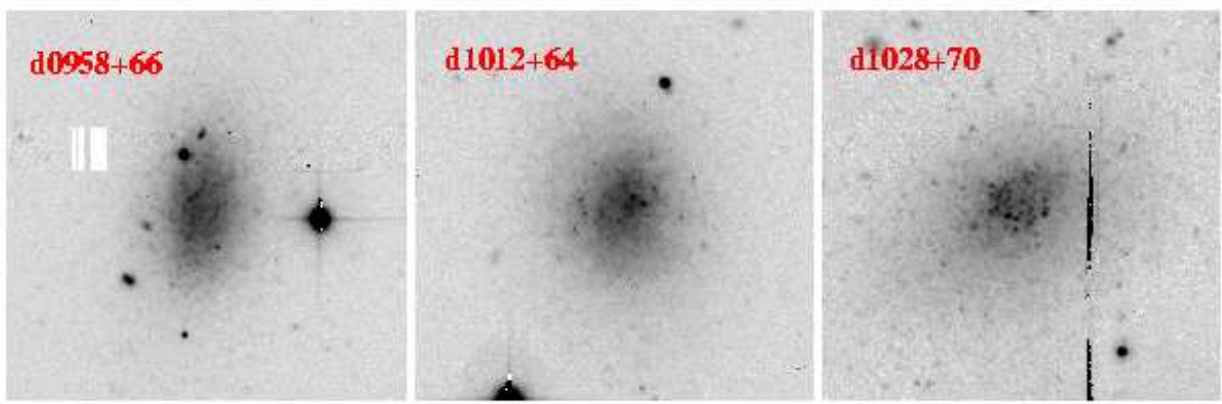}}
\caption[Images of 3 new M81 BCD candidates]{Thumbnails 
of the 3 new M81 candidate BCDs.  Images are 1.5 arcmin on a side.
\label{imbcd}}
\end{centering}
\end{figure}

\begin{figure}[b!]
\begin{centering}
\includegraphics[angle=0, totalheight=7.0in]{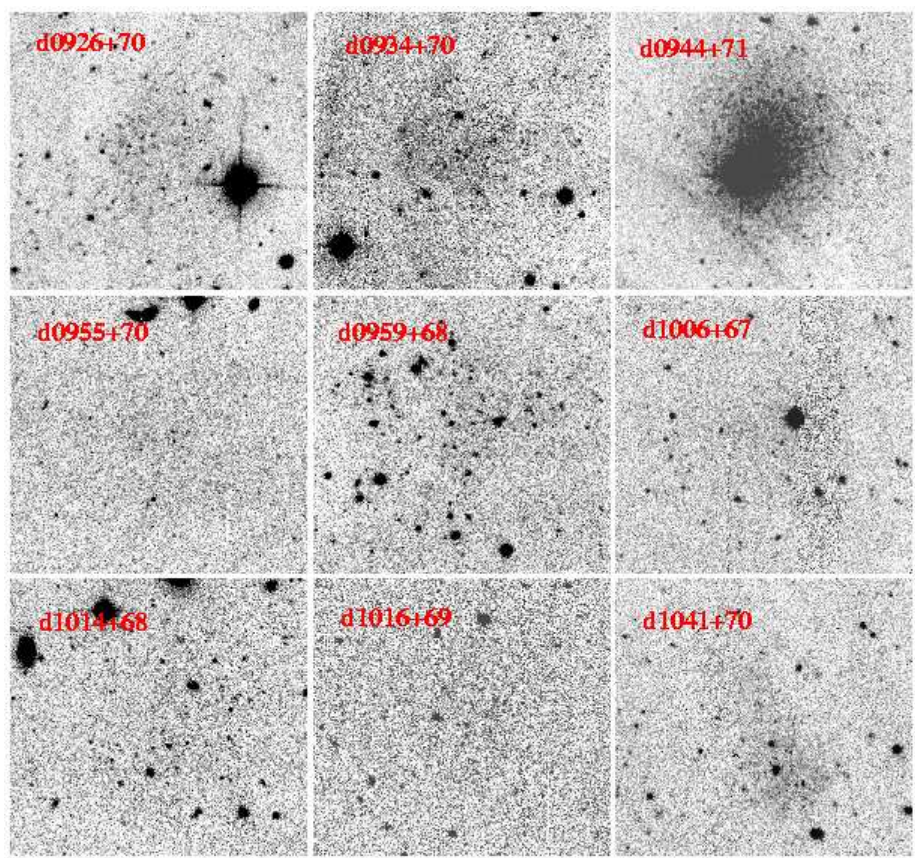}
\caption[Images of 9 new M81 candidates]{We display a mosaic of images
of 9 new candidates that were detected by eye and/or 2-point
correlation of resolved points.  Images are 1.5 arcmin on a side.
\label{imbest}}
\end{centering}
\end{figure}

\begin{figure}
\begin{centering}
\includegraphics[angle=0, totalheight=7.0in]{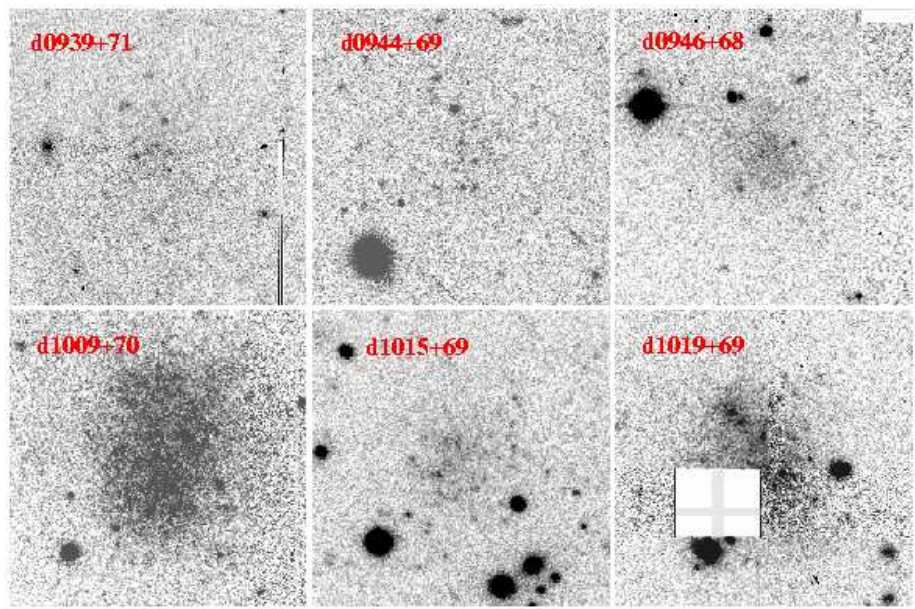}
\caption[Images of 6 possible background dwarfs]{While these six candidates 
have a slight degree of resolution, many of these may prove to lie 
in the background of the M81 Group.
\label{imback}}
\end{centering}
\end{figure}

\begin{figure}
\begin{centering}
\resizebox{\columnwidth}{!}{\includegraphics{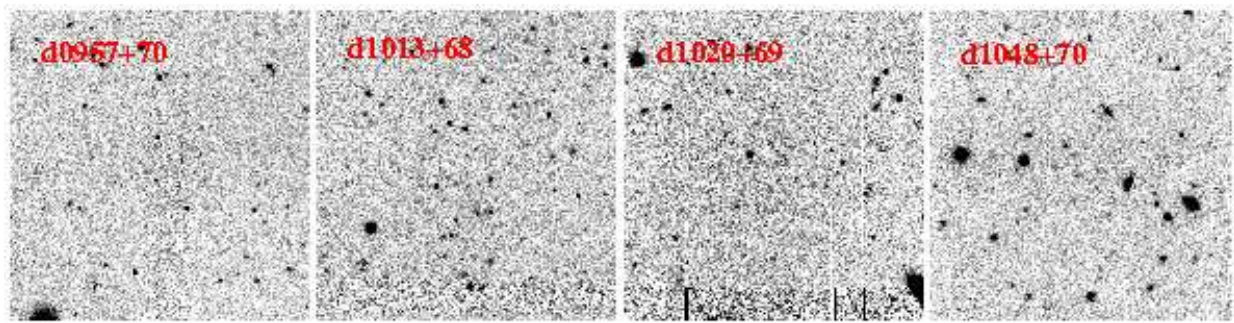}}
\caption[Images of 4 final candidates]{Four candidates which may turn out
to be artifacts (excess noise or foreground concentrations of stars) 
or distant galaxy clusters.
\label{imart}}
\end{centering}
\end{figure}

\begin{figure}[t]
\begin{centering}
\includegraphics[angle=0, totalheight=8.5in]{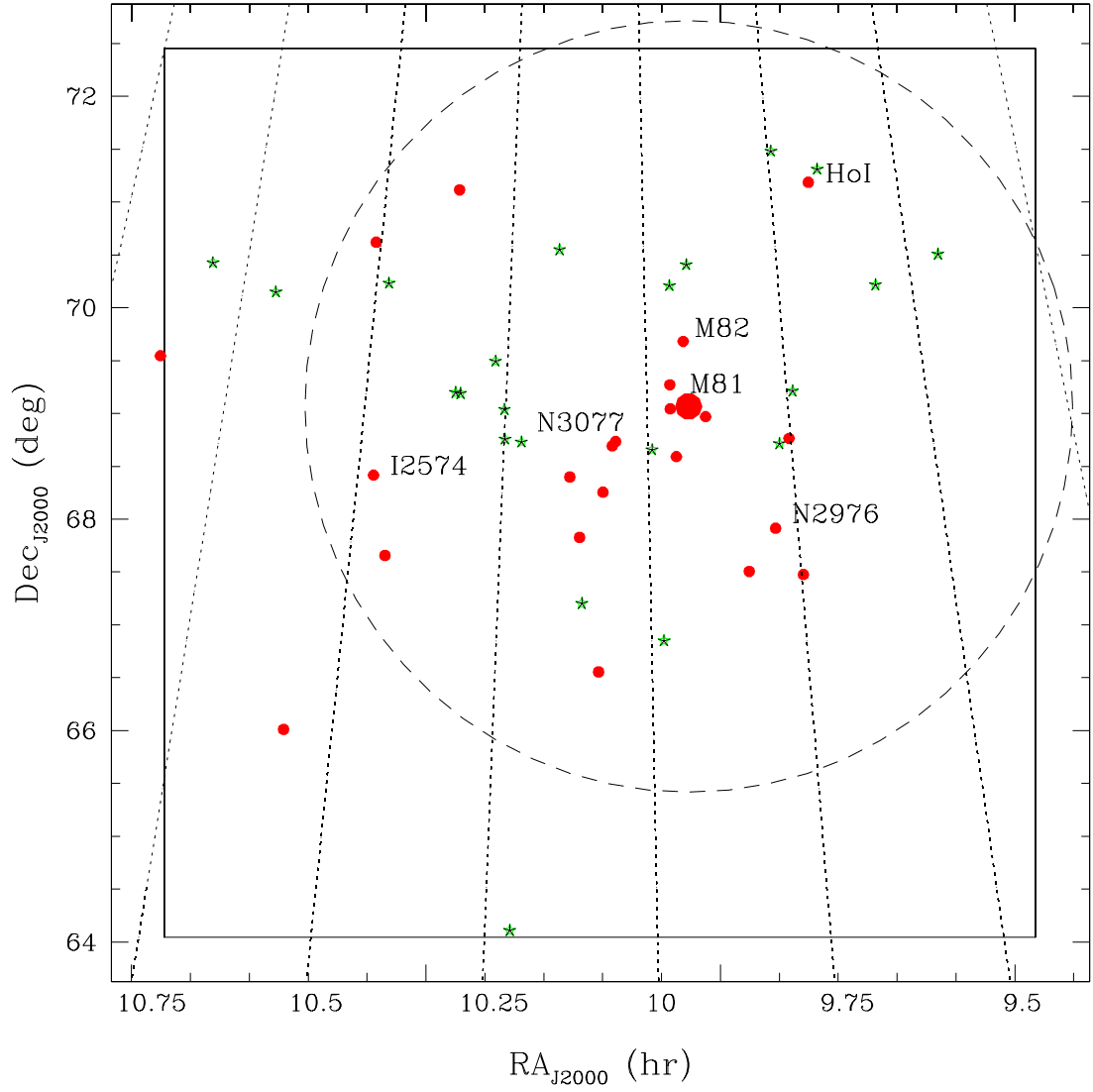}
\caption[Map of our M81 survey region]{Map of our
M81 survey region (large box almost filling the figure).  
Circles indicate previously known
group members while stars denote the location of the
new group candidates.  The large dashed circle is the projection
of the putative surface of second turnaround at 230 kpc from M81.
\label{smap}}
\end{centering}
\end{figure}

\begin{figure}[t]
\begin{centering}
\includegraphics[angle=0, totalheight=8.5in]{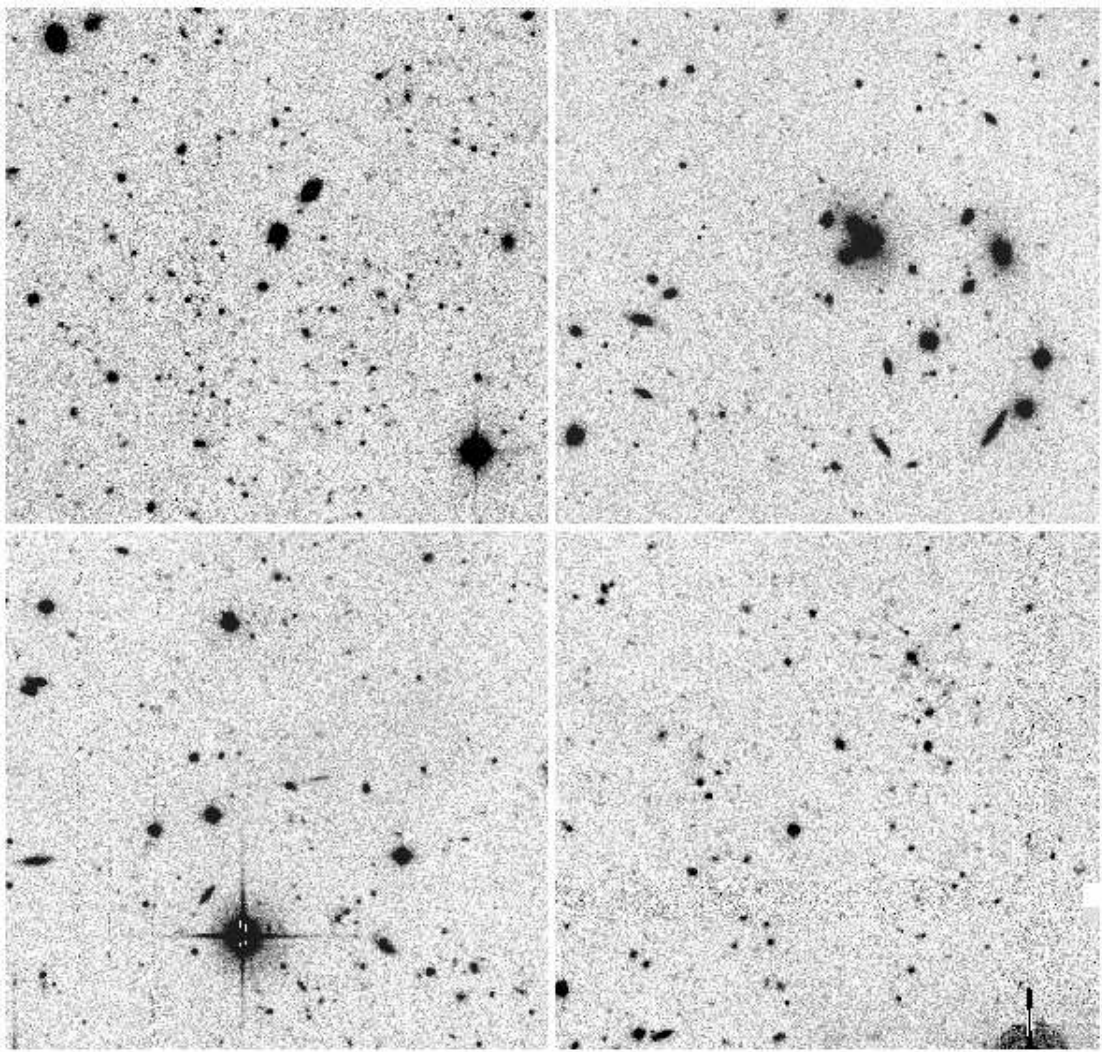}
\caption[Images of 4 final background galaxy clusters]{Four examples of
object concentrations detected with our 2-point correlation routine that 
we believe to be distant galaxy clusters.  Image sections are 2.2 arcmin across.
\label{bgals}}
\end{centering}
\end{figure}

\vskip 0.1cm 
We display thumbnails of 3 candidate M81 Group blue compact dwarfs (BCDs) 
in Figure \ref{imbcd} along
with a mosaic of another 9 good dwarf member candidates in Figure \ref{imbest}.  Each
galaxy image is 1.5 arcmin on a side. 
In Figures \ref{imback} - \ref{imart} we display candidates which may turn
out to lie slightly in the background of the M81 Group or simply be artifacts
in the data, foreground clumps of stars, or distant galaxy clusters.
Figure \ref{smap} displays a map of our 65 square degree survey region
and the distribution of the new candidate sources along with previously known 
group member galaxies.   We show a few examples of galaxy clusters detected
by our 2-point correlation routine in Figure \ref{bgals}.

\begin{figure}[t]
\begin{centering}
\includegraphics[angle=0, totalheight=9.0in]{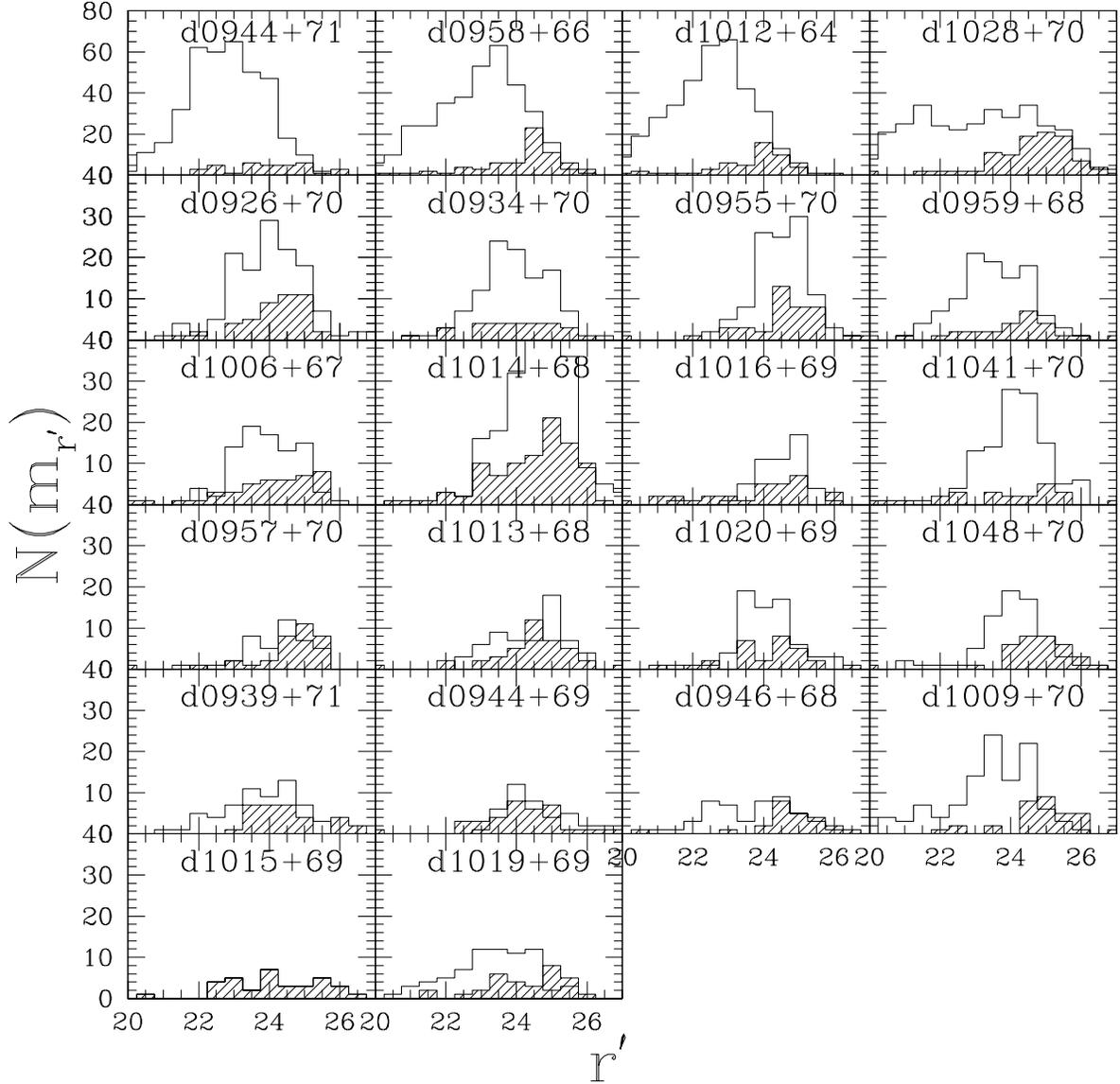}
\caption[Histograms of new M81 candidates]{We display histograms
of the resolved objects in the new M81 candidates. Number counts of
objects recovered by SExtractor within a 30-45 arcsec radius (see text) centered
on the new candidates are binned as a function of magnitude. 
The top row includes 4 objects, including 3 BCD candidates, with the 
largest resolved stellar populations.
The next 2 rows include a further 8 good candidates. 
The 4th row contains 4 objects that may prove to be artifacts, while the
bottom 2 rows include possible more distant galaxies.
The shaded
histograms represent number counts in an identical sized region 2.5 arcmin
west of each galaxy.
\label{histocand}}
\end{centering}
\end{figure}

\vskip 0.1cm 
In Figure \ref{histocand}, we display histograms of the number counts,
N($m_R$), of the SExtractor detections
within a radius of 30 arcsec centered on each of the new M81 candidates.  For the
3 BCDs and d0944+71, another bright candidate, we construct histograms
for detections out to a radius of 45 arcsec.  A rough estimate of the 
fore/background object counts is obtained by measuring the number counts in
a same sized region 2.5 arcmin west of each galaxy. These are displayed as
shaded histograms overlaying the candidate population histograms.

\subsection{Limits of our Detection Methods}\label{false}
                                                                                 
We tested our automated method of detecting dwarf candidates through
the 2-point correlation of resolved stellar populations using
false stars.  We wish to determine to what limiting stellar densities and
magnitudes our galaxy detection routines are sensitive.
To simulate artificial dwarf galaxies, large numbers of stars are added to 
real images with the IRAF package ARTDATA and convolved with the seeing of 
the field.  Stars were distributed according to an exponential spatial density 
profile with an effective radius, r$_{e}$, ranging from 4-90 arcsec 
(0.07-1.6 kpc).  The stellar luminosity function has the form
of a power law with power $\sim0.3$ for the RGB \citep{mmrt06}.
This value was allowed to vary randomly between 0.25 - 0.35 in the simulations.
We add between 100 stars and 120,000 stars to each galaxy realization ranging in
magnitude from $r^{\prime} \sim 30.0$ at the faint end to
24.7 - 25.3 at the bright end. 
We take the tip of the RGB magnitude as our bright limit (24.7) but include
up to 0.6 magnitudes of extinction.
We also include a number of simulations with a bright limit up to 22.  This allows
us to simulate galaxies with a range of stellar populations, from
those having entirely old stellar populations to those also containing a young,
bright component.
                                                                                 
\begin{figure}[t]
\begin{centering}
\plotone{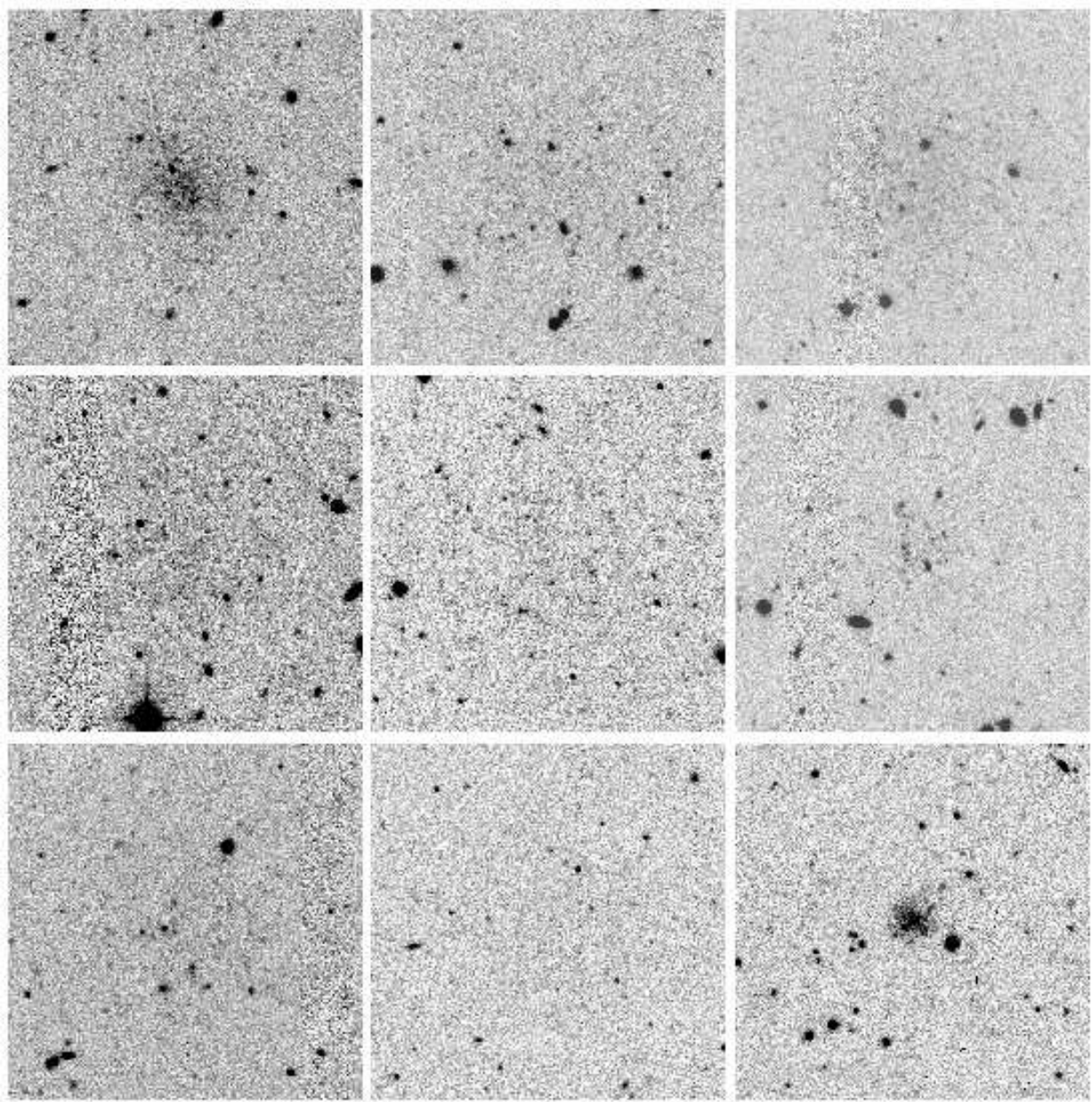}
\caption[Images of several artificial galaxies]{Examples of 
artificial galaxies that were added to real images with which we test
the recovery efficiency of our
detection techniques. Images are 1.5 arcmin on a side.
\label{falsestar}}
\end{centering}
\end{figure}

The STARLIST task was used to generate a file
suitable for input to MKOBJECTS.  Simulated stars were first convolved with the
image seeing and Poisson noise was included before objects were added to the real images.
Sets of 40 artificial galaxies were added
at a time to single MegaCam quadrant images and run through our detection procedure. 
A total of 1900 galaxies were generated in this manner.
In Figure \ref{falsestar}, we display a few examples of our simulated galaxies.
SExtractor was re-run on these images and the 2-point correlation
automated method was then used to search for these artificial dwarf galaxies.
                                                                                 
\begin{figure}[t]
\begin{centering}
\includegraphics[angle=0,totalheight=8.0in]{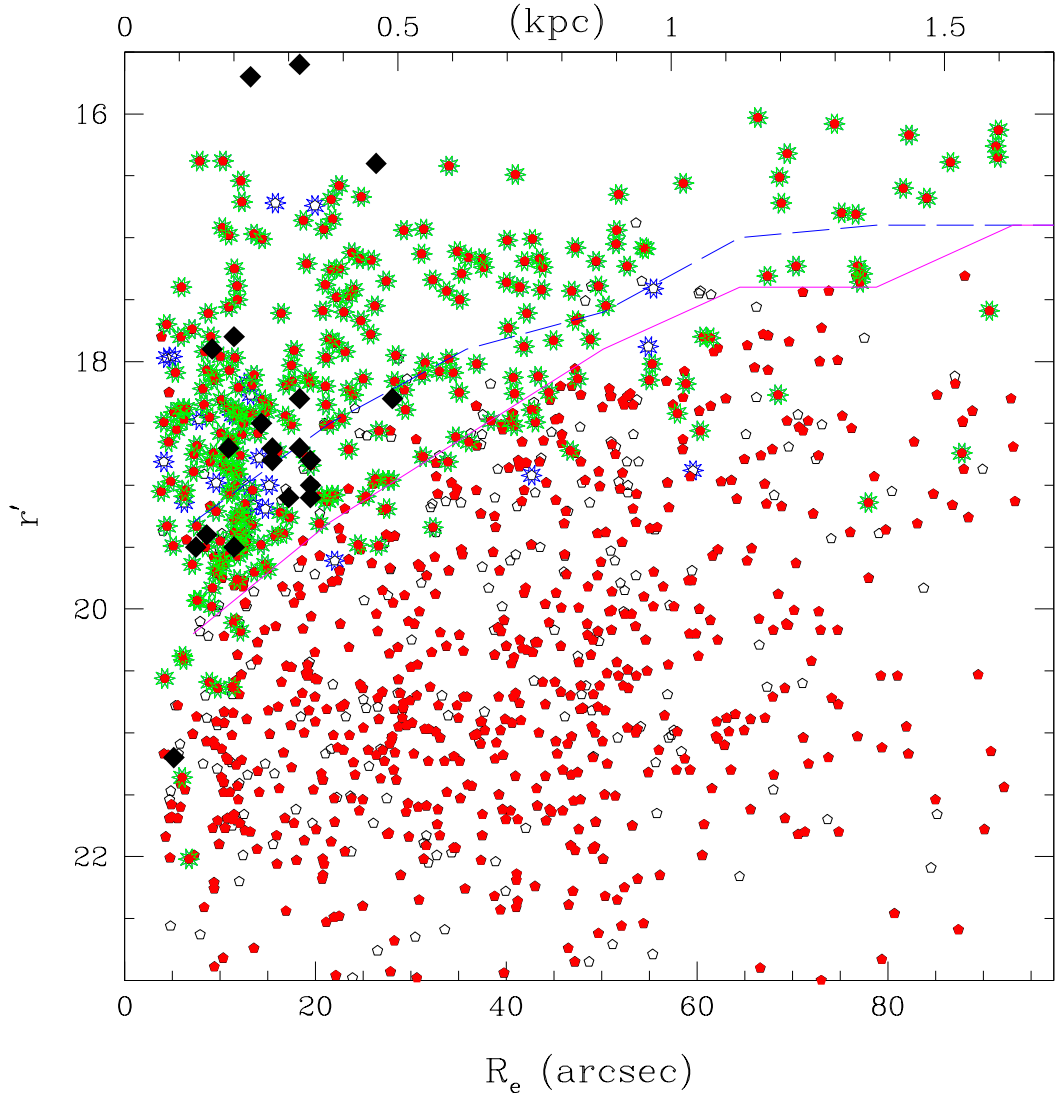}
\caption[Results of artificial galaxy recovery tests]{Results
of 1200 artificial galaxy recovery tests.  Small pentagons represent
all artificial galaxies. Open pentagons are those galaxies that landed
at least partially in chip gap regions of the MegaCam field or
are largely obscured by bright saturated stars or other objects and
are less likely to be recovered.
Stars around these points denote that the simulated
galaxy was recovered.  Overlaid as black diamonds are the candidate real galaxies 
in this magnitude - effective radius plane.  The solid line denotes the
magnitude at which the recovery drops to 50\% (in 2-magnitude bins), while
the dashed line marks the boundary of 90\% completeness.
\label{falsermres}}
\end{centering}
\end{figure}

Although we expect the detectability of a galaxy to depend only on
whether the dwarf galaxy has a significant resolved stellar population,
we find that the detection rate and limiting depth of our survey depends on
a number of factors.  The primary factors are the total number of resolved
stars and the concentration of these stars.  These correspond to the typical
measured quantities of total integrated magnitude and surface brightness of
the galaxy. Results of 1200 artificial galaxy realizations generated with
typical 0.64 - 0.75 arcsec seeing and with extinction values $< 0.2$
are displayed in Figure \ref{falsermres}.  We plot the integrated total
magnitude and half-light radius for each simulated galaxy.  Open points
are those objects which landed on single exposure chip gap regions of the
MegaCam field, or fell on bright stars or galaxies and are less likely
to be recovered.  Large black diamonds represent real candidate dwarf
galaxies discovered in this survey.  The dashed line displays the magnitude
at which recovery drops to 90\% within 2-magnitude bins, while the
solid line represents the 50\% completeness limit.  We find
that we are able to detect compact galaxies with R$_{e}$ $< 11$ arcsec
down to $r^{\prime} = 20.5$, or about $M_{r^{\prime}} = -7.3$.
For galaxies with R$_{e}$ $< 30$ arcsec, we
are able to reach $r^{\prime} = 19.0$ (-8.8) with at least 50\% completeness.
For galaxies larger than R$_{e}$ $= 30$ arcsec ($\sim 0.5$ kpc) we are
at least 50\% complete to $r^{\prime} = 18$ (-9.8).  Assuming real dwarf galaxy
sizes of R$_{e}$ $< 1$ kpc, we expect our survey is 90\% complete down to
about $r^{\prime} = 18.0$, or about -9.8 at the distance of M81.

In Figure \ref{falsemmu}, we display the survey detection limits in
the $mag-\mu$ plane.  Symbols are as before.  Lines of constant
effective radius are labeled.   A curved line is drawn that best bounds
the region within which simulated galaxies are recovered.  This is the
traditional curve that describes isophotal limits for exponential profile
galaxies \citep{as79}.  The completeness of this survey depends on other
factors, but the recovery boundary is adequately described by isophotal
limits.  At a given magnitude, as one goes to fainter surface brightnesses,
recovery fails because of increasingly low concentrations of stars and a failure
of the eye or 2-point correlation methods to distinguish these from the
background. At a given surface brightness, as one goes to fainter magnitudes,
objects become smaller with decreasing  
stellar population sizes.  Therefore, while not necessarily isophotal limits, the detection
limits imposed on this survey can be translated to limiting angular sizes and 
surface brightness.
                                                                                 
\begin{figure}[t]
\begin{centering}
\includegraphics[angle=0,totalheight=8.0in]{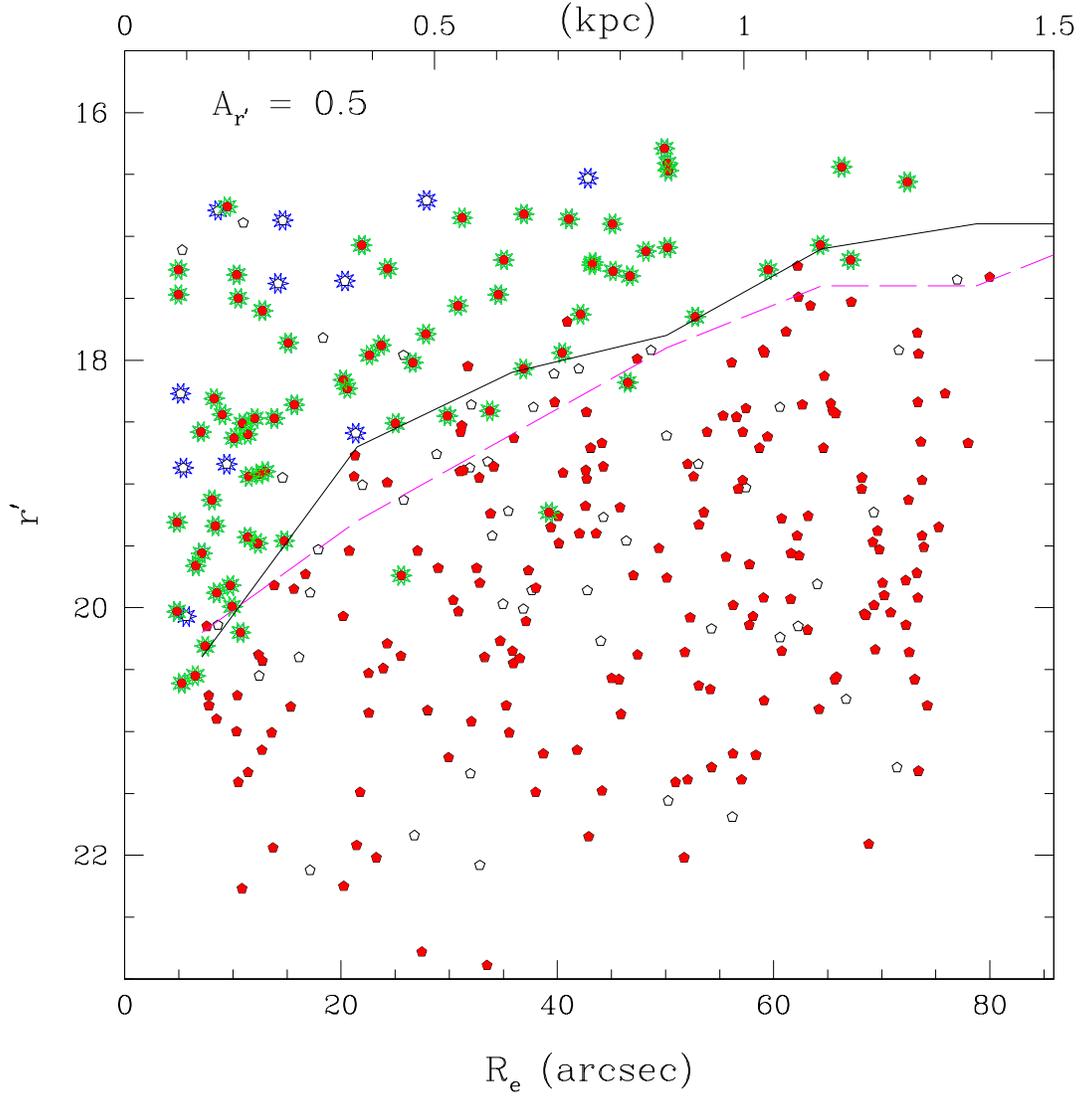}
\caption[Tests of artificial galaxy recovery with extinction]{Recovery results
for objects simulated with 0.5 magnitudes of extinction.  Symbols as in previous figures.
The solid line represents the magnitude at which
the recovery drops to 50\% for these galaxies, while the
dashed line denotes the magnitude at which the recovery for
artificial galaxies with 0.-0.2 magnitudes of extinction drops to 50\%.
\label{falsermext}}
\end{centering}
\end{figure}

However, the detection efficiency will also vary by location in
these fields.  Because our survey just reaches the tip of the RGB,
any galaxy with a pure old population is detected only at brighter magnitudes
and surface brightnesses in regions of high extinction and fields with worse
seeing.  We test this with 240 artificial galaxies generated with 0.5 magnitudes 
of extinction.  See Figure \ref{falsermext}.
On average, the 50\% limit in the
recovery is 0.3 magnitudes brighter than for the sample added with $0 - 0.2$ magnitudes
of extinction.  Although the number statistics in this sample is small, this
suggests that the completeness limit in magnitude depends directly on the level of
extinction and that our completeness limits may vary by as much as 0.6 magnitudes 
throughout our survey region.  Similarly, the depth of the group may affect 
completeness limits.  Variations in the distance with respect to M81
of $\pm 300$ kpc affect apparent magnitudes by $\pm 0.18$ mag.
This is a very small
effect but implies that the dwarfs on the near side of the group with pure old
populations are more likely to exhibit detectable resolved stellar populations
than those on the far side.
                                                                                 
\begin{figure}[t]
\begin{centering}
\includegraphics[angle=0,totalheight=8.0in]{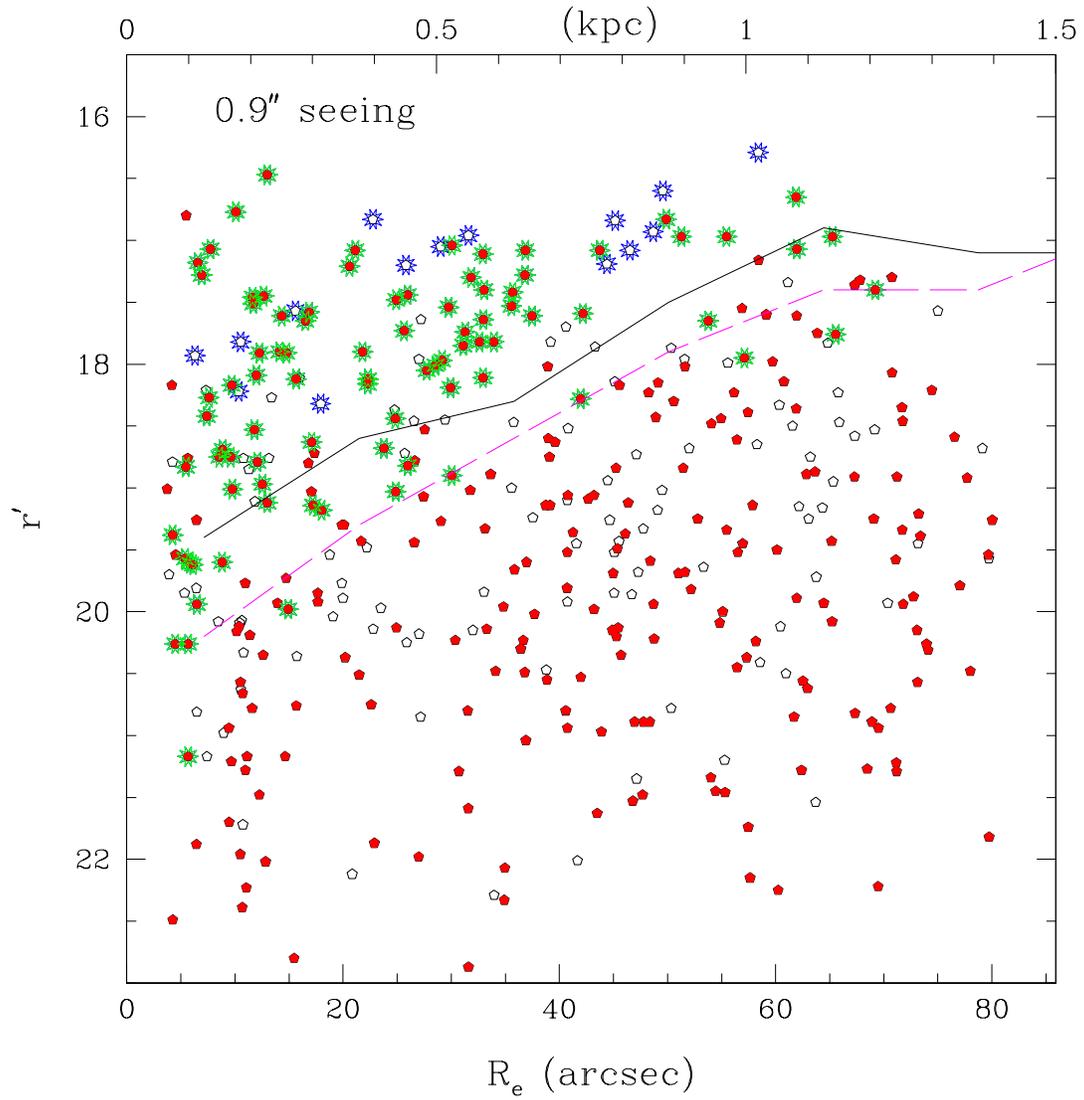}
\caption[Seeing effects of artificial galaxy recovery]{Recovery of
400 galaxies added with 0.9 arcsec seeing.  Symbols as in previous
figures.  We compare results to
that of the original 1200 simulations.   The dashed line represents the
magnitude at which the completeness drops to 50\% for artificial galaxies
added with seeing less that 0.75 arcsec.  The solid line displays the
50\% completeness for artificial galaxies added with 0.9 arcsec seeing.
\label{falsesee}}
\end{centering}
\end{figure}

We test the effect of poorer seeing with 400 galaxies that were simulated with 
0.9 arcsec seeing.  Results
are presented in Figure \ref{falsesee}.  The solid line denotes the location
where completeness drops to 50\% with 0.9 arcsec seeing, while the dashed line 
represents the 50\% completeness for seeing no worse than 0.75 arcsec.  We
find evidence that seeing has an effect on the recovery of resolved dwarfs,
and this is most significant for compact galaxies (R$_{e} < 20$ arcsec) where the   
completeness shifts 0.8 magnitudes brightward.  This is to be expected.
In these cases worse seeing blends the light of these small objects to
produce a smooth, less resolved low surface brightness object.  With the small number
statistics, it is not clear that at larger R$_{e}$ the 0.2 arcsec worse seeing
will affect recovery by the measured 0.4 magnitudes, but the results do 
indicate that seeing has an affect on the detection rates of these objects.  
Fortunately, only $\sim19$ combined images,
or 11\% of our survey area, will be affected by such poor seeing.
                                                                                 
To a lesser extent, one might expect the bright cirrus patches present
throughout much of this region to affect detection limits through
increased sky noise.  From artificial galaxy tests, however, we do not find this to
significantly affect our recovery rates.
Regions of high galactic stellar density can also limit detectability
for the more diffuse dwarfs.  At galactic latitudes ranging between $39 - 44$ deg
this is not too much of a problem, but a few fields exhibited non-uniformly
distributed excess numbers of stars throughout the images, increasing the
confusion level.  This will primarily affect fainter, lower surface brightness
dwarfs with fewer total numbers of resolved stars spread over a larger area.
                                                                                 
Finally, two areal coverage issues will affect recovery.  Any small dwarf
lying behind bright saturated stars or galaxies will likely be missed.  From
false galaxy tests, where the galaxies are randomly placed within the images,
we find less than 4\% of recoverable galaxies remain undetected because they landed on
bright objects.  This primarily affects only the smallest galaxies which can hide entirely
behind brighter foreground objects, although a few more extended simulations
were obscured by bright stars or galaxies.   Also, due
to the survey strategy of observing each field twice by offsetting between
images by half a field, all chip gap regions in the MegaCam field are observed
for a total of only 10 minutes.  With half the exposure time as the rest
of the field, we expect the magnitude completeness limits in these regions
to be about 0.75 magnitudes brighter than the above described survey limits.
Because the detected dwarf galaxies in M81 have angular sizes generally greater than
16 arcsec, the vertical chips gaps of 13 arcsec should not affect recovery of any      
but the tiniest galaxies if by misfortune they should fall in these regions.  There
are however two large horizontal chip
gaps which have single exposure widths of 80 arcsec each across the length of each MegaCam
field, and this gap size is essentially doubled in our mosaic of two images per field.
Since all our newly discovered dwarf galaxies have angular extents smaller than 160 arcsec,
it is therefore entirely possible for dwarf galaxies to be lost in these
regions.  The total area of these single exposure regions, including the area
at the top and bottom of our full mosaic, is 9.7 square degrees,
or $\sim 15$\% of the total survey area.
                                                                                 
Based on these results, we find that we are nearly 100\% complete to $r^{\prime} = 17$
and 90\% to $r^{\prime} = 18 (M_{r^{\prime}} = -9.8)$.
At fainter magnitudes, a large number of factors act to diminish our ability to detect
group galaxies, although the primary factors are simply the limiting magnitude and
surface brightness to which we are sensitive.  Factors such as worse seeing, higher
extinction, areal coverage by brighter objects, and single exposure regions
serve to push these limits to slightly brighter magnitudes.

\subsection{Follow-up Spectroscopy}\label{specres} 

Follow-up spectroscopic observations were carried out with FOCAS 
on the 8m Subaru Telescope and SCORPIO on the Special Astrophysical Observatory 
6m BTA telescope. 
Spectra were obtained for our three BCD
candidates along with several globular clusters in a couple of previously
known M81 dwarf galaxies.

Observations with Subaru were made 22-24 Nov 2006 using the R300/mm grating in
second order and VPH 600-450 grating in first order.
A two arcsec slit was used with the VPH grating, and images were
binned $4\times2$, providing a spectral resolution of 10\AA \ at 4000\AA.
For the R300 grating, a 1 arcsec slit and $4\times1$ binning
was used to obtain a spectral resolution of 10\AA.
Spectral coverage ranged from $3800-5250$\AA \ for the VPH
and $3650-5950$\AA \ for the R300 grating.  Exposure
times varied from object to object.
                                                                                
\begin{figure}[t]
\begin{centering}
\includegraphics[angle=0,totalheight=8.0in]{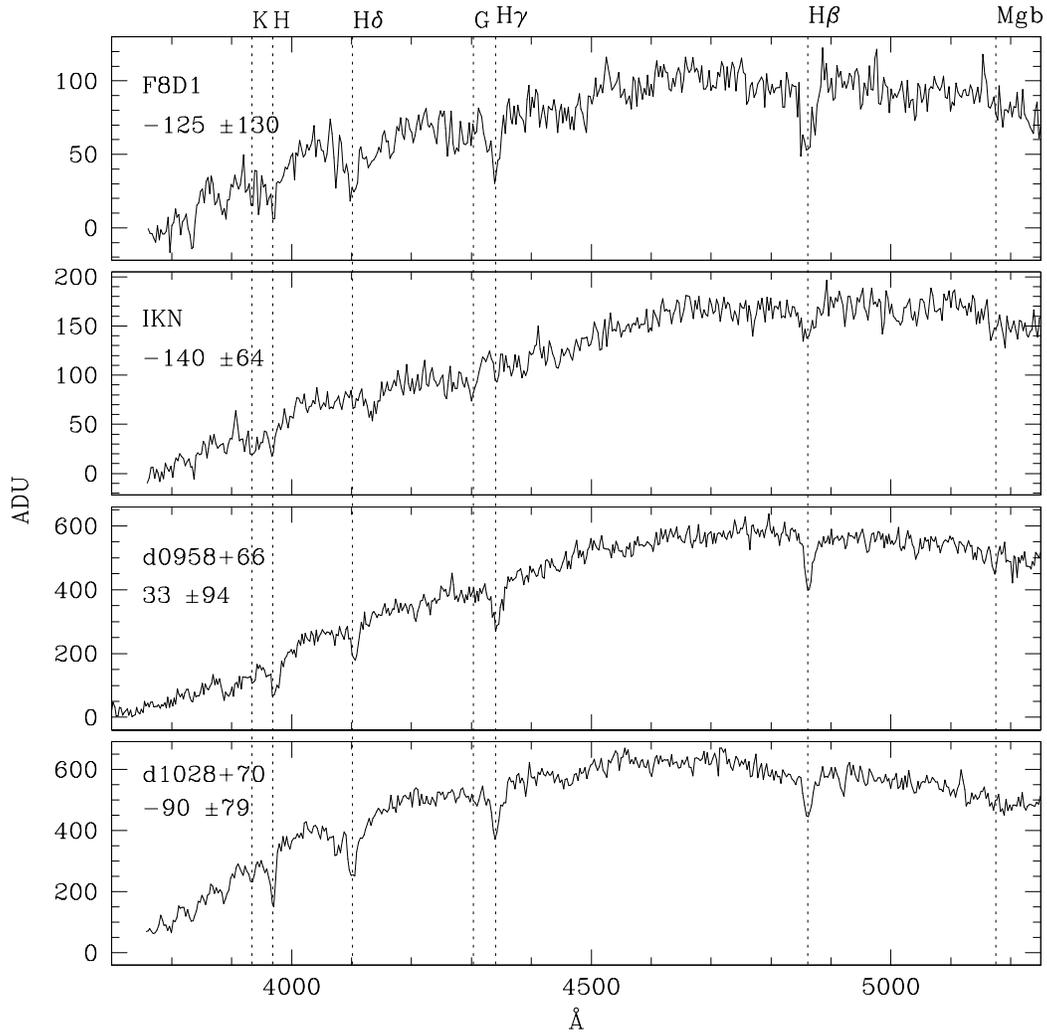}
\caption[Spectra of 4 M81 Group dwarfs]{Spectra of 4 M81 
group members obtained with Subaru/FOCAS. All spectra are shifted 
to the rest frame 
given the cross-correlation measured velocities as labeled.
\label{specplot}}
\end{centering}
\end{figure}

Longslit spectra were obtained with FOCAS of globular clusters in the previously 
known faint dwarf
galaxies F8D1 and IKN.  The globular cluster in F8D1 was originally noted
by \citet{cads98}. 
For two bright candidate galaxies (both BCDs) d1028+70 and d0958+66, we obtained 
spectra by simply placing the slit across the long axis of the galaxy. 
These data were processed using standard IRAF routines.  Images were
first cosmic ray corrected using CRMEDIAN and then bias subtracted
using 10 median combined bias frames.  Arcs were observed before or
after each observation and used to wavelength calibrate the spectra.
Lines in the arcs were identified using IDENTIFY and REIDENTIFY,
and FITCOORD was used to obtain a dispersion correction solution
which was applied using TRANSFORM.  The two-dimensional spectrum
for each object in the slit was extracted using APALL.  Finally,
RVSAO/XCSAO \citep{KM98} was used to obtain a cross correlation solution 
from template spectra for the
redshift of each object and hence the galaxy itself.
Spectra are displayed in Figure \ref{specplot}.  Radial velocities are
listed in Table \ref{tabspec}.

\begin{figure}[t]
\begin{centering}
\includegraphics[angle=270,totalheight=5.0in]{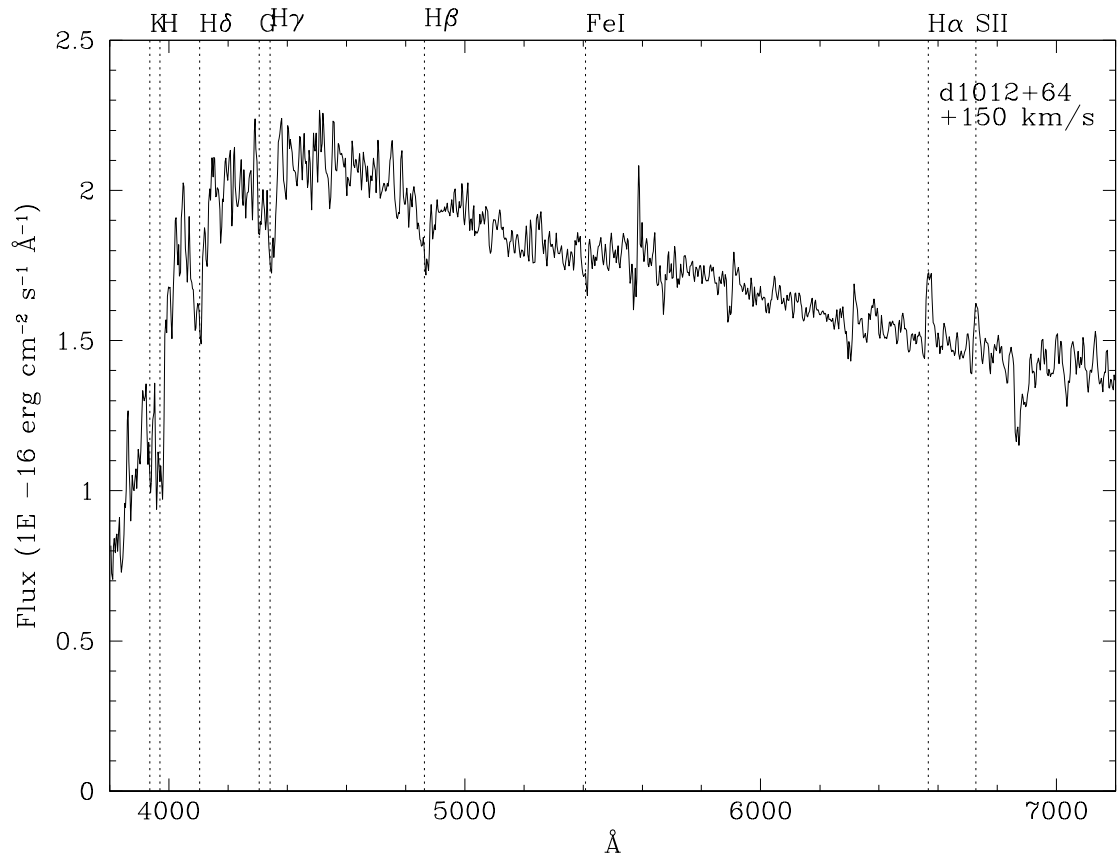}
\caption[Spectrum of d1012]{Spectrum of d1012+64 from the 6m BTA
telescope.  H$\alpha$ emission is detected and used to
measure a heliocentric radial velocity.
\label{bta0}}
\end{centering}
\end{figure}

\begin{figure}[t]
\begin{centering}
\includegraphics[angle=0,totalheight=8.0in]{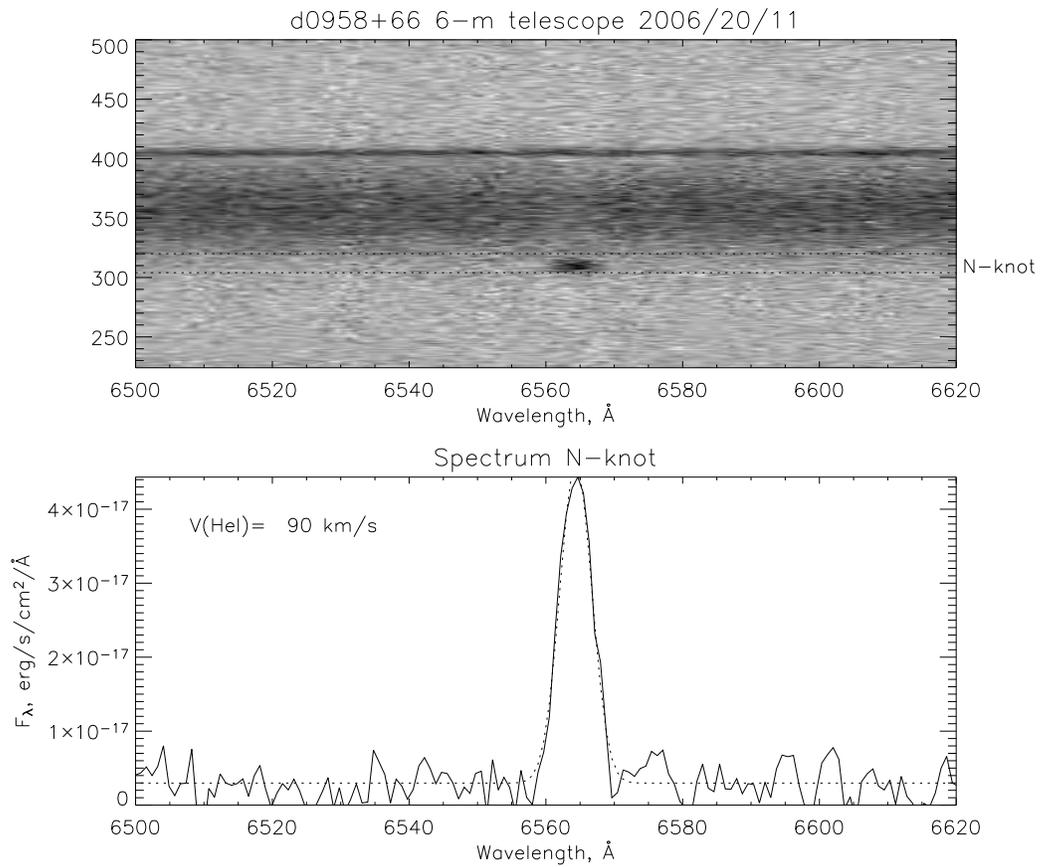}
\caption[Spectrum of d0958]{Spectrum of d0958+66 from the 6m BTA
telescope.  Strong H$\alpha$ emission is detected and used to
measure a heliocentric radial velocity.
\label{bta1}}
\end{centering}
\end{figure}

\begin{figure}[t]
\begin{centering}
\includegraphics[angle=0,totalheight=8.0in]{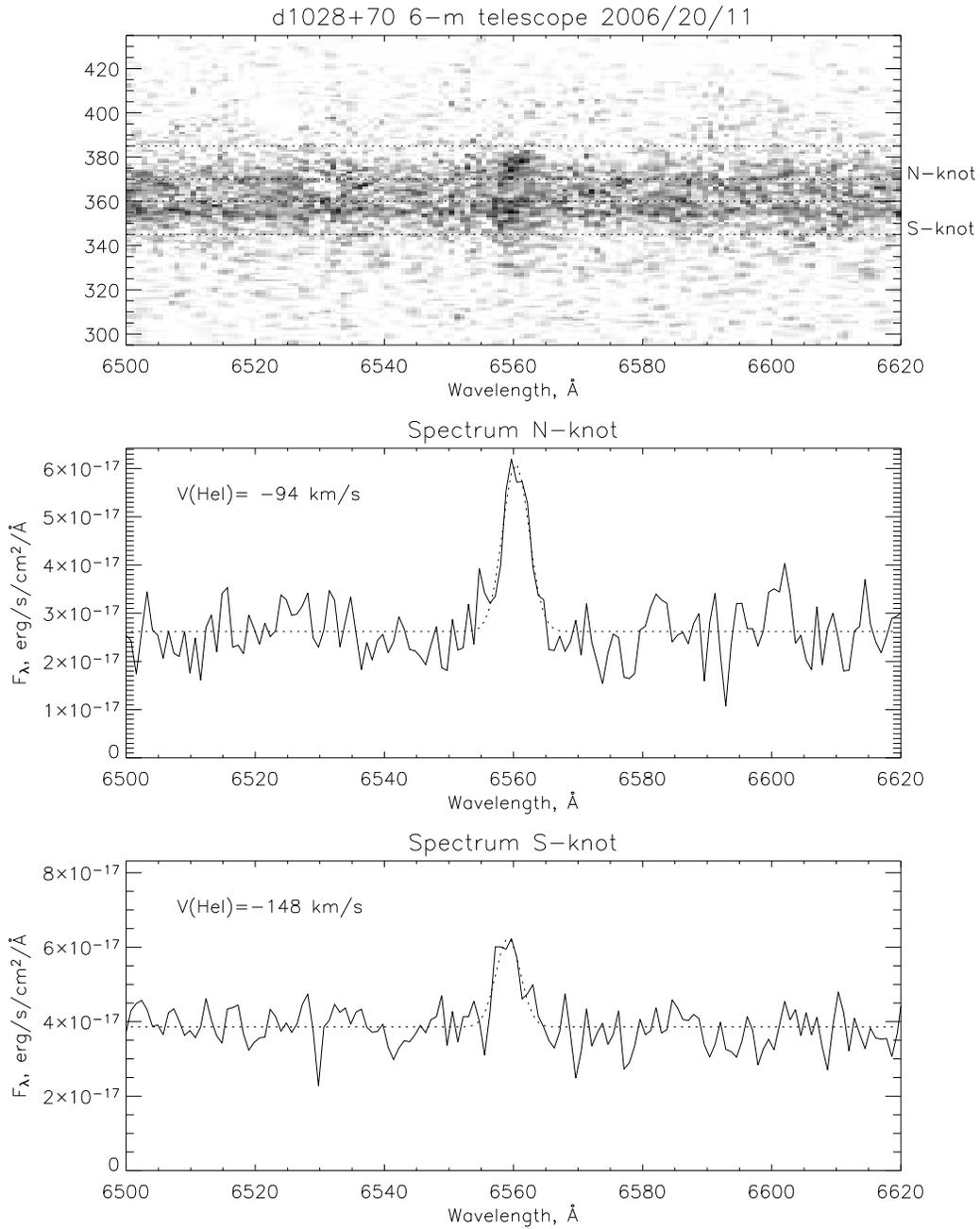}
\caption[Spectrum of d1028]{Spectrum of d1028+70 from the 6m BTA
telescope.  H$\alpha$ emission is detected in two separate knots which
we label as N and S-knots.
\label{bta2}}
\end{centering}
\end{figure}

Further observations were performed for the 3 BCD candidates with 
SCORPIO in long-slit mode installed at the BTA 6-meter telescope prime focus.
The grism VPHG400 was used with a 1 arcsec slit 
yielding the spectral range $\sim3500 - 7500$\AA \ chosen to include H$\alpha$ 
with a resolution of 3 \AA/pixel and FWHM $\sim20$ \AA/pixel. 
The objects d1028+70 and d0958+66 were observed during the night 
11 November 2006 with the exposure time of 600 sec under 
$2^{\prime\prime}$ seeing. The galaxy UGC 5497 (d1012+64) was observed on 20 October 2007 
with an exposure time of 900 sec under $2.5^{\prime\prime}$ seeing. 
Spectrophotometric standard stars were also observed for calibration.
The standard pipeline with the use of MIDAS, as described in \citet{AM05},
was applied for the reduction of these long-slit spectra.
All three BCD candidates show knotty H$\alpha$ emission.
We display the spectra in Figures \ref{bta0} - \ref{bta2}.  Due to the
short exposure times and low S/N it is difficult to measure line ratios.
However, we note that the [NII] $\lambda6584$ line is quite weak in comparison
to H$\alpha$, especially for d0958+66.  We estimate an upper limit for the line
ratio of 0.05 for this galaxy.  From the relations of \citet{kd02} and \citet{pp04}, 
this leads to an upper limit for the metallicity of 0.3 Z$_{\odot}$.
Mean heliocentric radial velocities of the objects are included in Table \ref{tabspec}.

\subsection{Multi-band Imaging Follow-up}\label{obsmult}

Using the UH 2.2m with Tek 2K CCD on 19-24 Dec 2006, we imaged 9 of the 
dwarf galaxy candidates in $V$ and $I-$bands.  Seeing ranged from $0.7 - 1.4$ arcsec
and 5 nights were photometric.  Images were $4 \times 900$s in the $V-$band and 
$5 \times 720$s in the $I-$band.  Standard stars were observed on 
each photometric night.  Images were processed using standard 
IRAF routines.  Due to the slow speed of the mechanical shutter,
a shutter mapping correction was also implemented.  This was necessary for the
short exposure images (5-20s) of the standard stars.

\begin{figure}[t]
\begin{centering}
\includegraphics[angle=0,totalheight=9.0in]{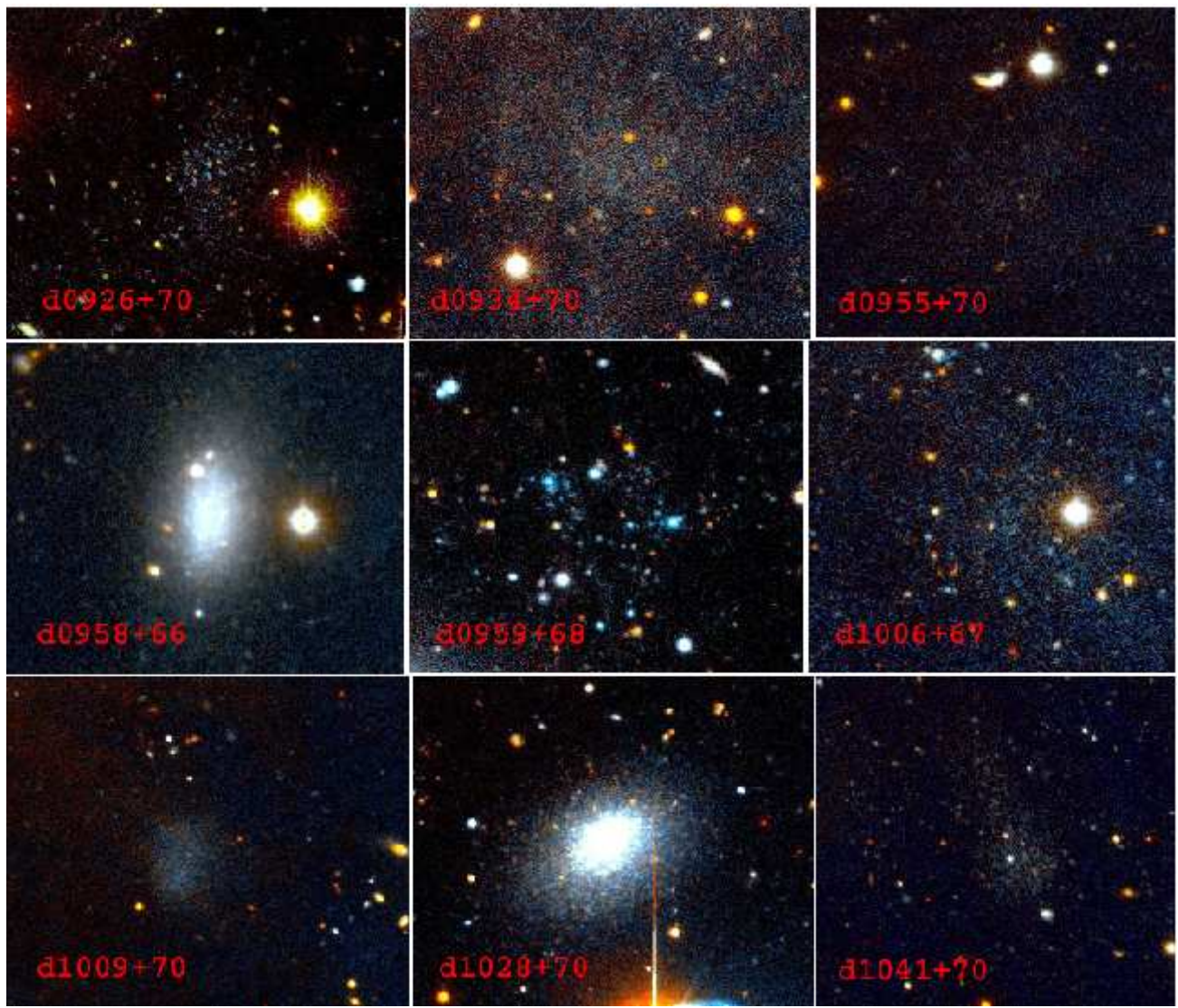}
\caption[Color images of 9 candidate dwarfs]{Color images based
on 2 color photometry of 9 dwarf candidates.
\label{colim}}
\end{centering}
\end{figure}

In most cases, the images did not go deep enough to produce useful
color-magnitude diagrams (CMD).  For 8 galaxies observed on photometric nights,
we measure integrated $V-I$ colors within circular apertures out to 
the effective radius as measured in the $r^{\prime}$ band (Section \ref{propcands}).
We correct for extinction using the maps 
of \citet{sfd98}.  $V-I$ colors are listed in Table \ref{tabX}. We also generate
color images of the 9 candidates.  To create images for the blue channel we take
$2*V_{image} - I_{image}$. The $V$ and $I$-band images are taken to be 
green and red respectively.  Color images are displayed in Figure \ref{colim}.  

Large young, blue stellar populations
are evident in d0926+70, d0958+66, d0959+68, and d1028+70.  These are
objects classified by us as BCDs, dIs, and tidal dwarfs.  These multi-band data
may not be going deep enough to reach the tip of the red giant branch.  While
resolved stars are seen in the MegaCam survey images, 
the lack of visible resolved stars in the UH 2.2m images of d0955+70 suggests it 
may consist predominantly
of an older stellar population.  Similarly, if members of the M81 Group,
d0934+70 and d1006+67 also appear to contain predominantly older stellar populations.

We have recently obtained HST/WFPC2 multiband imaging for 14 of the candidates
and have been awarded time to complete observations of the sample with HST/ACS.
These data will be deep enough to measure distances from the CMD tip of the red giant 
branch.  Results will be presented in a subsequent paper.

\section{Results}\label{results}

\subsection{Properties of the M81 companion candidates}\label{propcands}

To determine centroids of the new objects, we used the intensity 
weighted means of pixel values, $\frac{\sum{\vec{x} I}}{\sum{I}}$, 
around each galaxy.  Ellipticity
and position angle around these centroids were then derived from 
the intensity weighted second moments \citep{sh97}.  The measured ellipticities,
($1-b/a$), for the faintest objects are highly uncertain due to low S/N of the
diffuse light component and small visible stellar populations.  Measuring
total magnitudes for these very faint and diffuse extended resolved galaxies
is challenging because of expected contamination from foreground
stars, and in several cases, from excess foreground cirrus reflected flux, not
to mention extinction.  We therefore estimated total magnitudes using
several methods.  

We first mask all obvious bright foreground stars and background
galaxies around each of our candidates.
Simple aperture magnitudes were obtained for
each galaxy with radii 50, 100, 150, or 200 pixels (9.35, 18.7, 28.05,
and 37.4 arcsec) depending on the visible extent of the galaxy.  
The sky was determined in an annulus 4 pixels wide with a radius
of 400 pixels.  Alternatively, we measure total flux in an identical
sized aperture in 4-5 regions around each galaxy.  Taking
the average of these as an estimate for the contribution from
both sky and foreground/background object flux we subtract this
from our object flux for a second estimate of total magnitude.

To obtain structural parameter measurements for each galaxy,
we extracted fluxes in successive elliptical apertures around
each galaxy out to a radius of 40 arcsec and truncated the data if it exhibited
a $>$ 30\% increase in slope attributed to contamination by an unmasked 
neighboring object, or, more often,
to features like bright cirrus bands.  The central 4 arcsec were also 
discarded. We then fit both the surface brightness profile with a Sersic
function and the curve of growth with a cumulative
Sersic function.  We extracted parameters from the best of these two fits.
This provides a third estimate of the total 
magnitude along with central surface brightness, 
effective surface brightness, and half-light radius.
The generalized Sersic function is defined as 

\begin{equation}
 I(r) = I_{\circ} \ e^{-(r/r_{\circ})^{1/n}}
\end{equation}

\noindent
where $n$ is the Sersic parameter.  For $n = 1$, this reduces to 
an exponential profile which usually provides a good fit for
disk and dwarf galaxies, and for $n = 4.0$ becomes a de Vaucouleurs profile.
The cumulative Sersic function is then 

\begin{equation}
I(r) = 2\pi\sigma_{\circ}r_{\circ}^{2}n \cdot \gamma[2n,(r/r_{\circ})^{1/n}]
\end{equation}
                                                                                
\noindent
where $\gamma$[a,x] is the Incomplete Gamma function,
\begin{equation}
\int_{0}^{x}exp(-t)t^{a-1}dt.
\end{equation}
This can be integrated to obtain the total object flux,
\begin{equation}
I_{t} = 2 \pi \sigma_{o} r_{o}^{2}n \Gamma[2n]
\end{equation}

\noindent
with total magnitude,
                                                                                
\begin{equation}
m_{t} = \mu_{o} - 5\log(r_{o}) - 2.5\log(n\Gamma[2n]) - 2.0
\end{equation}
                                                                                
\noindent
\citep{bj98} where $\Gamma[a]$ is the Gamma function.

\begin{figure}[t]
\begin{centering}
\includegraphics[angle=270,totalheight=4.5in]{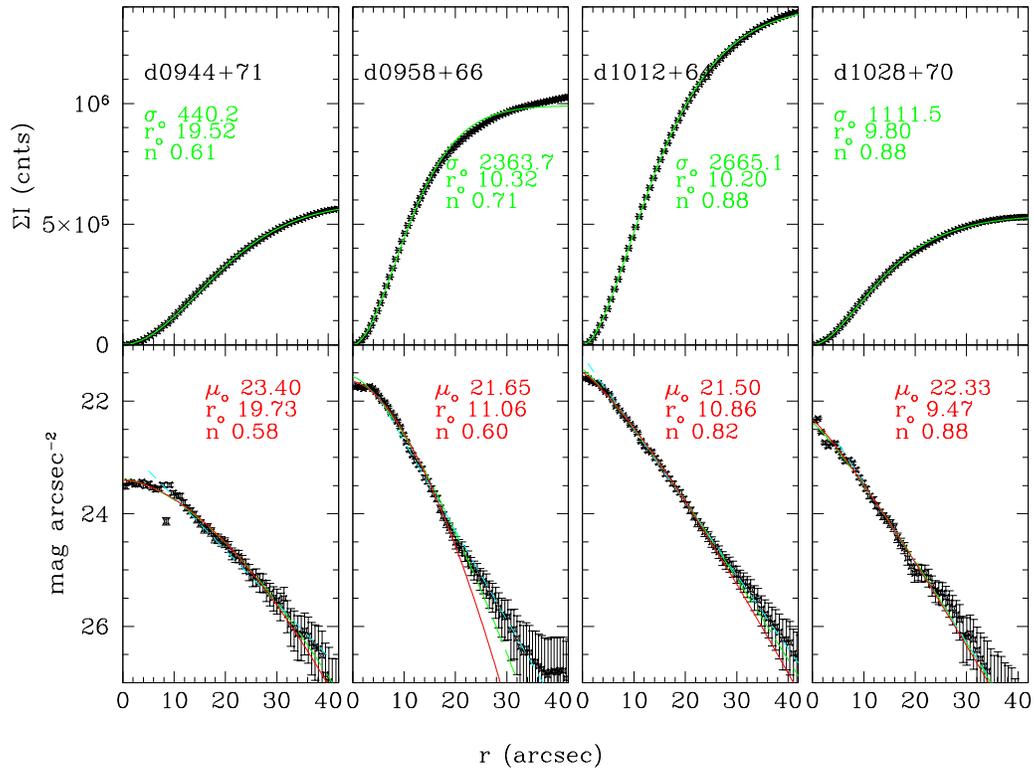}
\caption[Curve of growth and surface brightness profile fits]{
Curve of growth (top) and surface brightness profiles (bottom) for our
4 brightest candidates with 
best cumulative (top) and surface brightness profile (bottom) Sersic function 
fits overlaid.
In the bottom plots, we also overlay the best cumulative Sersic function 
fit (green long dash) and best exponential
surface brightness profile fit to outer radii only (cyan short dash).
\label{cogsbfitsA}}
\end{centering}
\end{figure}

\begin{figure}[t]
\begin{centering}
\includegraphics[angle=270,totalheight=4.5in]{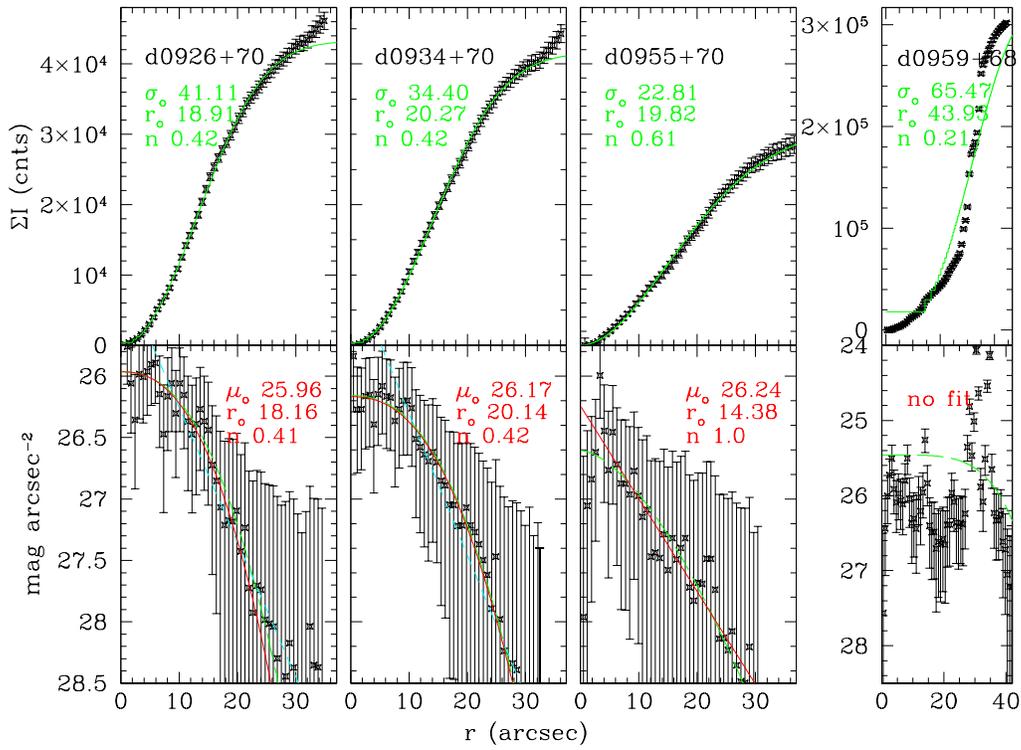}
\caption[Curve of growth and surface brightness profile fits]{Fits
for 4 of our candidate galaxies, as in Figure \ref{cogsbfitsA}.  
Curve of growth profile data points which exhibit a
sharp increase in slope beyond 10 arcsec were truncated during profile fitting.
For d0926+70 and d0934+70 this rise was determined to be due to  
nearby bright cirrus.
For d0959+68 we
were unable to obtain a surface brightness profile fit due to the fact
that this object consists almost entirely of bright stars without a significant low
surface brightness component.
\label{cogsbfitsB}}
\end{centering}
\end{figure}

\begin{figure}[t]
\begin{centering}
\includegraphics[angle=270,totalheight=4.5in]{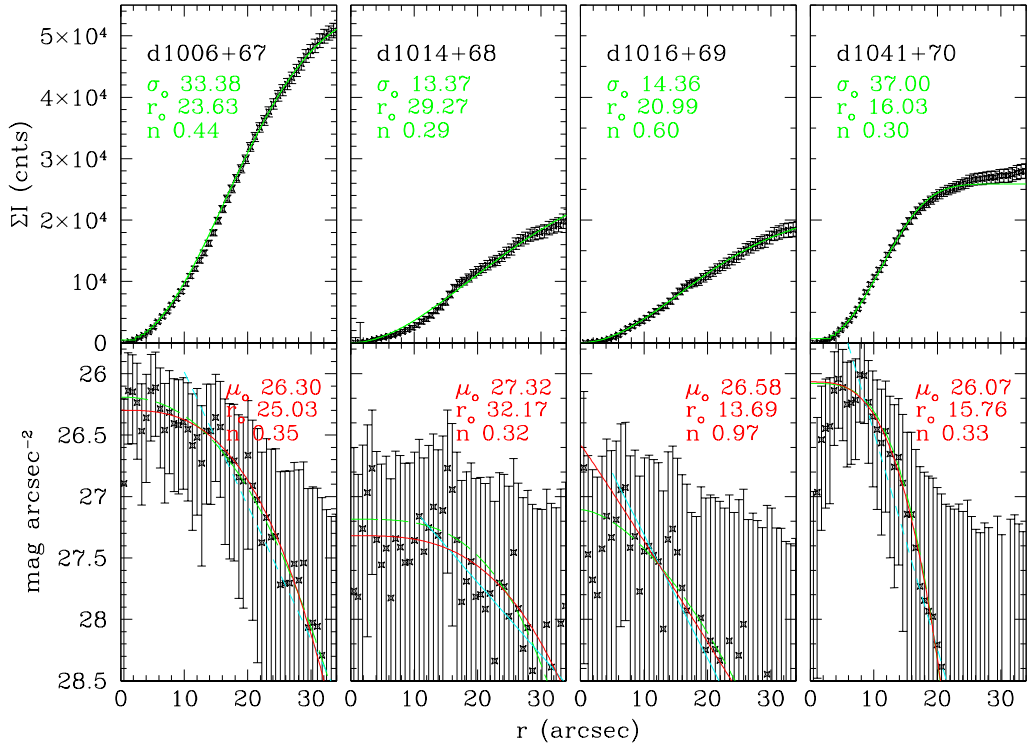}
\caption[Curve of growth and surface brightness profile fits]{Fits
for 4 of our candidate galaxies, as in Figure \ref{cogsbfitsA}.  
\label{cogsbfitsC}}
\end{centering}
\end{figure}

\begin{figure}[t]
\begin{centering}
\includegraphics[angle=270,totalheight=4.5in]{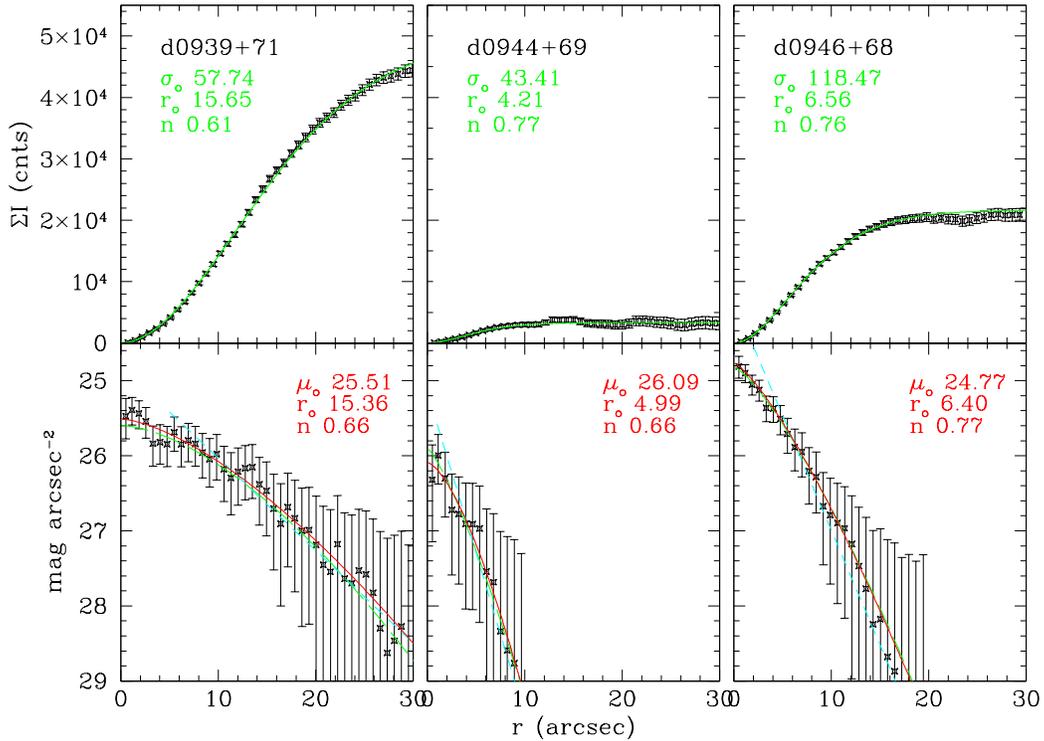}
\caption[Curve of growth and surface brightness profile fits]{Fits
for 3 potential background galaxies, as in Figure \ref{cogsbfitsA}.
\label{cogsbfitsD}}
\end{centering}
\end{figure}

\begin{figure}[t]
\begin{centering}
\includegraphics[angle=270,totalheight=4.5in]{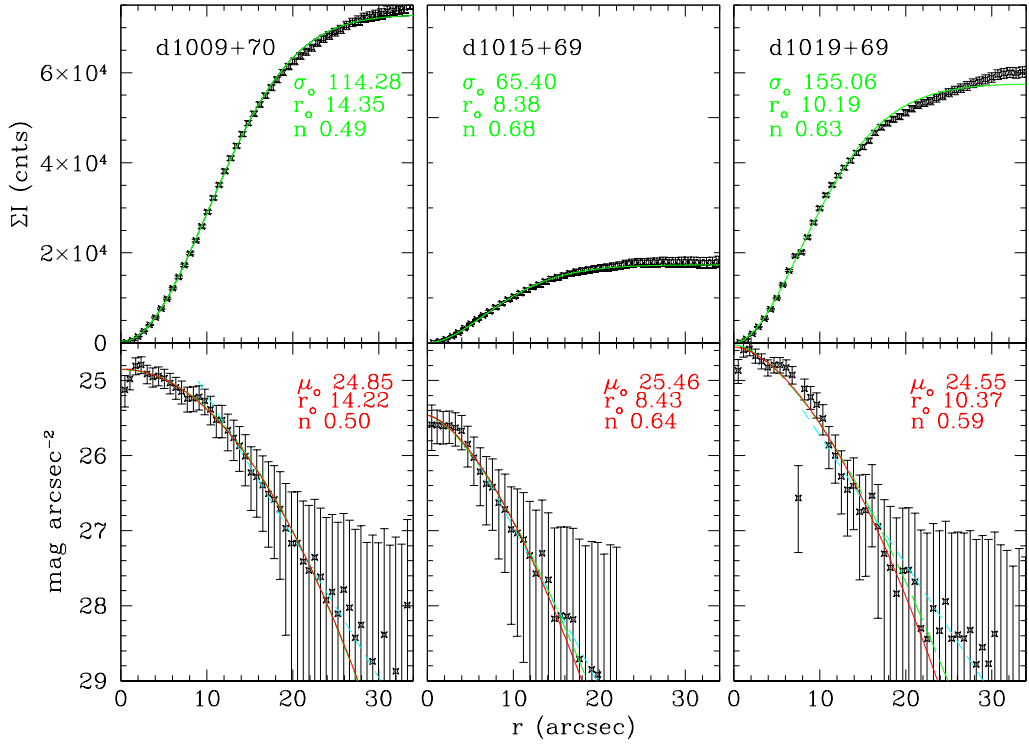}
\caption[Curve of growth and surface brightness profile fits]{Fits
for a further 3 likely background objects, as in Figure \ref{cogsbfitsA}.
\label{cogsbfitsE}}
\end{centering}
\end{figure}

\begin{figure}[t]
\begin{centering}
\includegraphics[angle=270,totalheight=4.5in]{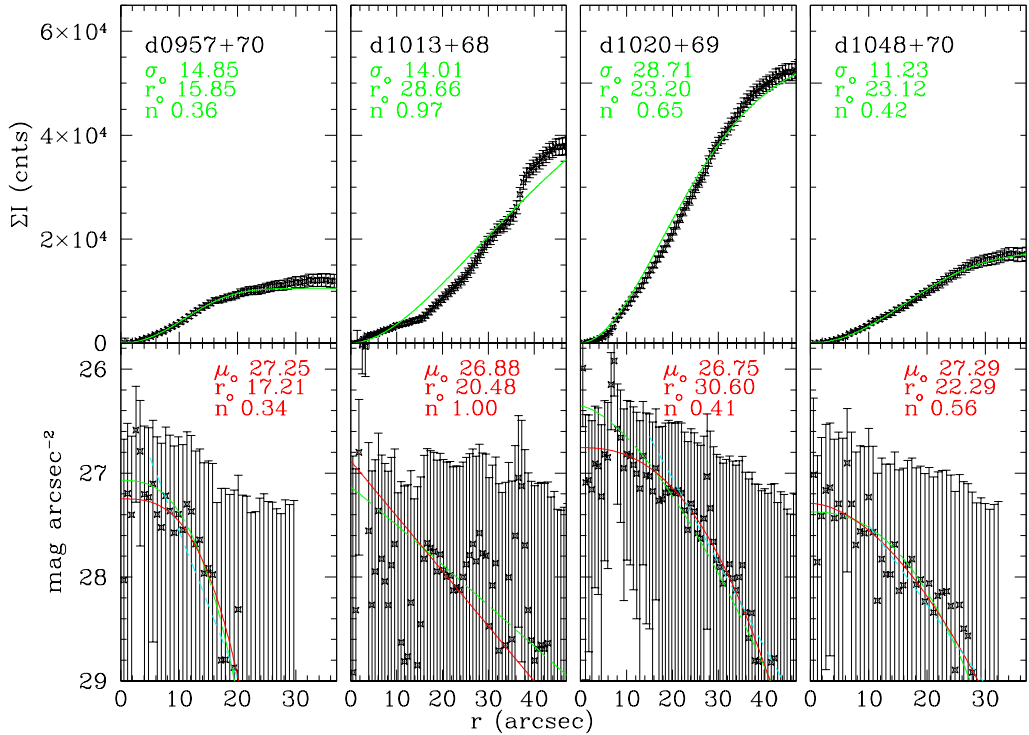}
\caption[Curve of growth and surface brightness profile fits]{Fits
for 4 of our poorer candidates, as in Figure \ref{cogsbfitsA}.  
\label{cogsbfitsF}}
\end{centering}
\end{figure}

The fitting of this nonlinear function to the data was done using a 
Levenberg-Marquard algorithm \citep{press92} which performs a 
$\chi^{2}$ minimization that implements an
inverse-Hessian method far from the minimum and switches to
a steepest decent method as the minimum is approached.
In Figures \ref{cogsbfitsA} - \ref{cogsbfitsF}, we
display both curve of growth and surface brightness profile fits 
for our candidate galaxies.
From these fits we directly obtained the central surface brightness,
scale length, and profile type, where scale length in the 
generalized Sersic function is directly
related to the half-light or effective radius through
                                                                                
\begin{equation}
R_{\circ}^{1/n} = R_{e}^{1/n} / (2.3026 b_{n})
\end{equation}

\noindent
with b$_{n}$ = 0.868 \ n - 0.142
\citep{ccd93}.

The majority of the galaxies are best fit with Sersic index
$n < 1.0$.  From inspection of the plots it is clear that these
low n fits are primarily driven by large cores, typical
of dwarf galaxies.  In the outer radii, the surface brightness
profile tends to drop exponentially.  We overlay best fit exponential
profiles to the outer radii in the surface brightness profiles
of Figures \ref{cogsbfitsA} - \ref{cogsbfitsF}.  A number of cases exhibit a
sharper drop-off, but are still consistent with an exponential
fall-off within the large uncertainties of the low surface
brightnesses at these radii.

We estimate typical magnitude measurement errors of 0.5.  Sources of uncertainty 
come from sky determination (0.4, estimated from curve of growth fitting when using
3 different estimates of sky values), foreground/background
contamination (typically 0.2, from magnitudes measured with and without masking of likely
foreground sources),
aperture magnitude measurement error from qphot (0.04), and the zeropoint uncertainty
(0.015, estimated from comparison with \citet{perel} catalog stars).
Cirrus may also effect measurements
of some of the galaxies. Sky estimates in these regions have an even larger
uncertainty, and faint old stellar populations may be lost in the excess sky
noise.  In particular, d0934+70, if real, appears largely obscured
by cirrus.   d0944+71, d0959+68,
and d1048+70 are strongly effected by fore or background objects.  We
estimate total magnitude uncertainties of  
0.8 for d0958+68 and 0.6 for d1048+70.  The magnitude for d0944+71 
is corrected by assuming a symmetric shape.  Flux from the star superposed
quadrant is replaced by the flux in the symmetrically opposite quadrant.
We assume a magnitude uncertainty of 0.8.  Candidate d1019+69 is affected by missing 
data in a chip gap; the measured magnitude is corrected in a similar way
as d0944+71.   

We estimate measurement errors in R$_{e}$ by comparing results from the surface
brightness profile and curve of growth fits and comparing differences from
fits using slightly different estimates for the mean sky value.  We find
fractional uncertainties in R$_{e}$ range from 1\% for the bright BCD d1028+70 to 
40\% for low surface brightness object d0955+70,
with a median uncertainty of 10\%.  Similarly, we find an uncertainty in the
central surface brightness ranging from 0.05 to 0.5 magnitudes arcsec$^{-2}$ 
with a median uncertainty of 0.2.  

Because of the shallow nature of our survey, we can assume that our
magnitude and size measurements are lower limits.  
To obtain estimates of total magnitudes, all measured magnitudes had to
be corrected for light lost in the sky noise.
Since many of the profile fits were influenced by the 
bright O/B stars, it is possible that measured profiles do not accurately 
reflect that of the underlying old stellar populations.  
We therefore calculate the missing fraction of light assuming both the
fitted profile shape and an exponential surface brightness profile.
Dwarf galaxies may be better fit with King profiles than exponential
due to the presence of large cores and possible truncation in the wings of the 
profile.  This is supported by our generalized Sersic function profile fits 
which find n $< 1.0$ for most of our galaxies, implying a steeper fall off 
of the galaxy light in the wings of the profile.
Magnitude corrections calculated by assuming exponential profiles may 
therefore be overestimates for these galaxies, 
and thus provide upper limits on the total galaxy luminosities.

From an isophotal limiting surface brightness (ranging from $27.0$
to 28.5 depending on the sky level and presence of cirrus), along with 
the measured central 
surface brightness and total magnitudes,
we extrapolate to obtain the intrinsic total magnitude.  Over $x$ scale lengths,
$x = r/r_{\circ}$, the surface brightness of an exponential profile drops by
\begin{equation}
-2.5 \log \ e^{-x^{1/n}} = 1.086x
\end{equation}
for $n = 1$ or
\begin{equation}
= 1.086x^{1/n}
\end{equation}
for a generalized Sersic profile.
The number of scale lengths observed between $\mu_o$ and $\mu_{lim}$ is
\begin{equation}
\Delta x = (\mu_{lim} - \mu_{o})/1.086
\end{equation}
or 
\begin{equation}
\Delta x = ((\mu_{lim} - \mu_{o})/1.086)^{n}.
\end{equation}
For $m_{T} - m_{lim} = m_{extrap}$ with
\begin{equation}
m_{i} \propto 2.5 \log [(1 + x)e^{-x}] \ \ (exponential)
\end{equation}
and
\begin{equation}
m_{i} \propto 2.5 \log [\int_{0}^{r_{i}} e^{-x^{1/n}} r dr ] \ \  (Sersic),
\end{equation}
we find the extrapolation beyond the measured magnitudes is 
\begin{equation}
\Delta m = 2.5 \log[1-(1 + \Delta x)e^{-\Delta x}]
\end{equation}
and
\begin{equation}
\Delta m = 2.5 \log[\gamma[2n,\Delta x^{1/n}] / \Gamma[2n]]
\end{equation}
for exponential and generalized Sersic profiles respectively. See also
\citet{TullyUM}.  Exponential profile corrections from the isophotal magnitude are 
less than $-0.1$ mag for our brightest
new candidates, but as high as $-1.2$ mag for the faintest
objects with a mean value of $-0.4$.  Corrections for the best fit Sersic profiles
range from $0.$ to $-0.8$ magnitudes with a mean of $-0.2$.  
Besides assuming a profile shape, these corrections further 
assume an old stellar population exists which may be too faint to detect, 
even as a diffuse low surface brightness component, in some cases.
For potential tidal dwarfs like d0959+68,
it is not clear that this assumption is valid.  There may
be evidence that d0959+68 is dominated by a very young population formed
less than 70 Myr ago \citep{durrell}.  We therefore do not apply a correction for
this object.

We provide the detailed structural and photometric properties of these new 
objects in Table \ref{tabX}.
If we assume all candidates are group members at the distance of M81,
we find M$_{r^{\prime}}$ magnitudes range from -6.7 for d0944+69 to -13.3
for d1012+64.
Half-light radii range from 90 to 490 pc for objects well fit with Sersic functions.
Because we do not yet have distances to these galaxies, we estimate 
that intrinsic magnitudes
and radii can vary by as much as 0.14 mag and 6\% in size (assuming a
distance range of $\pm$ r$_{2t}$).
We provide $r^{\prime}$ total magnitudes of previously known M81 Group
members that lie within our survey region in Table \ref{knownpar}.  For the
largest galaxies, we expect larger magnitude measurement errors, up to 1 mag, due to 
large errors in the local sky determination and subsequent negative features 
in the images. 
For measuring the local sky backgrounds, the TERAPIX reduction pipeline implements
box sizes much smaller than the size of our largest galaxies.

\subsection{M81 Group Luminosity Function}\label{lf}

\begin{figure}[t]
\begin{centering}
\includegraphics[angle=0,totalheight=8.0in]{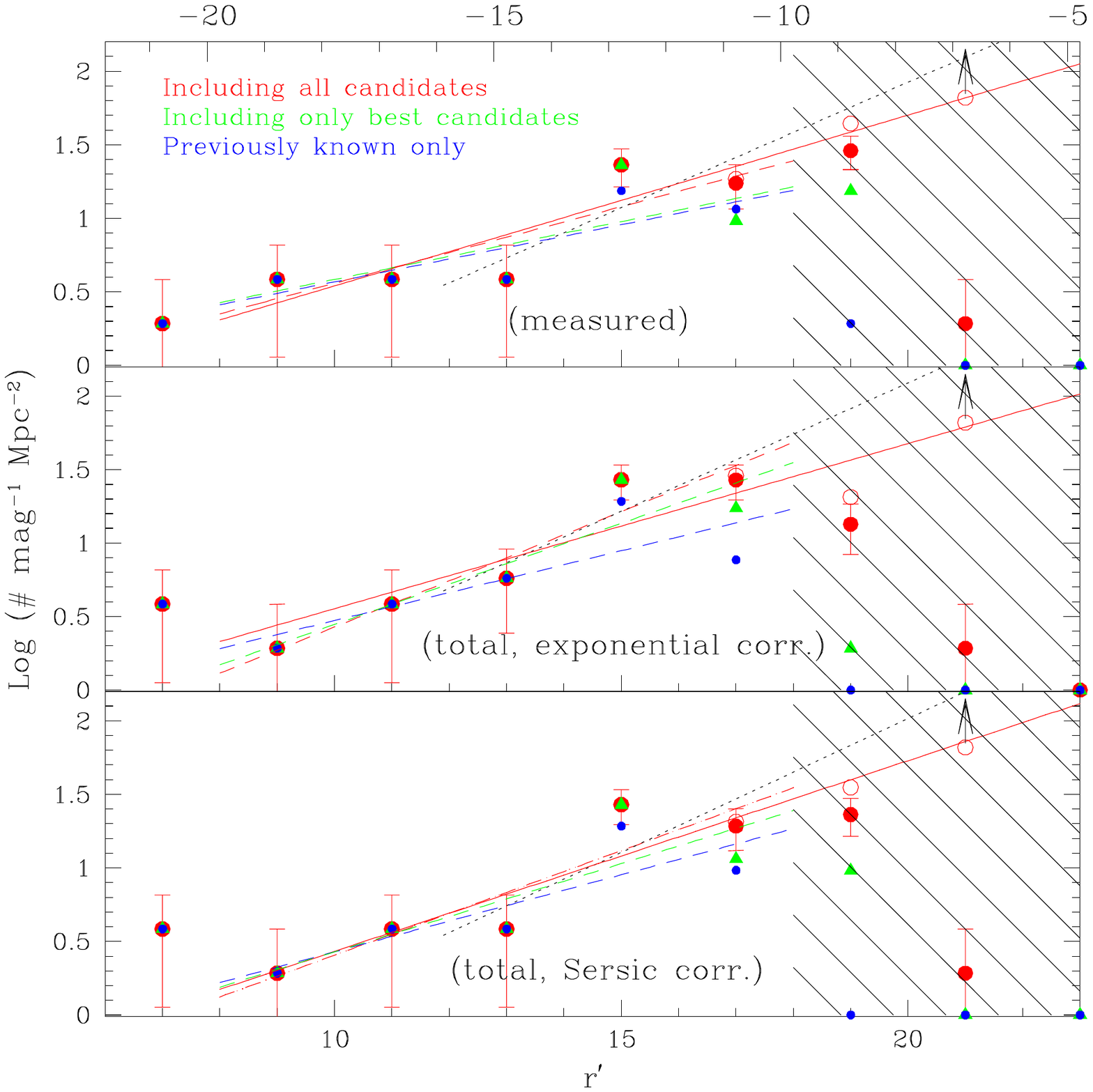}
\caption[Differential luminosity function for M81 Group]{Differential luminosity function
for the M81 Group, using measured magnitudes (top), total magnitudes corrected
assuming exponential profiles (middle), and total magnitudes assuming Sersic 
profiles (bottom).  Counts are normalized to the area of the survey
coverage.  Solid red circles include previously known group members and all candidates
from this survey.  Open circles are corrected for incompleteness.  Green
triangles include only the best candidates from this work.  Small blue circles
represent only previously known group members.  The hatched region 
denotes where this survey is incomplete; brightward of this we expect nearly
100\% completeness in the detection of member dwarfs.  The best power law fits to
the data brightward of our completeness limit are shown by the dashed lines,
while the solid line is the best fit for corrected counts. Dotted lines
are faint-end only fits to all counts within the magnitude range $-15 \leq r^{\prime} < -10$.  
The arrow in the final data point signifies that a minimum completeness correction based on 
good seeing/low extinction simulations was applied.
\label{LFdiff}}
\end{centering}
\end{figure}

To construct the M81 Group differential luminosity function (LF), we bin number
counts as a function of magnitude in 2-magnitude bins.  Such large
bin sizes are necessary because of the small number counts we are
working with and because of large magnitude measurement errors ($> 0.5$ mag).  
Counts are normalized by the size of our survey region.
We construct luminosity functions for all previously known members,
all previously known plus all new candidates from this work, and for 
previously known plus best candidates from this work.  In the case of 'best',
we include objects in Figures \ref{imbcd}-\ref{imbest}.  We have excluded
counts for objects we believe may be background or which
could potentially prove to be artifacts. 
Of the previously known objects, we also exclude from this 'best' sample 
Arp's loop and Garland 
since we do not have reliable magnitude estimates for these likely tidal objects.
The luminosity functions for these different samples are displayed in Figure \ref{LFdiff}.
We have constructed the LFs using measured magnitudes (top), exponential profile 
corrected total magnitudes (middle), and Sersic profile corrected total magnitudes (bottom).

\begin{figure}[t]
\begin{centering}
\includegraphics[angle=270,totalheight=5.5in]{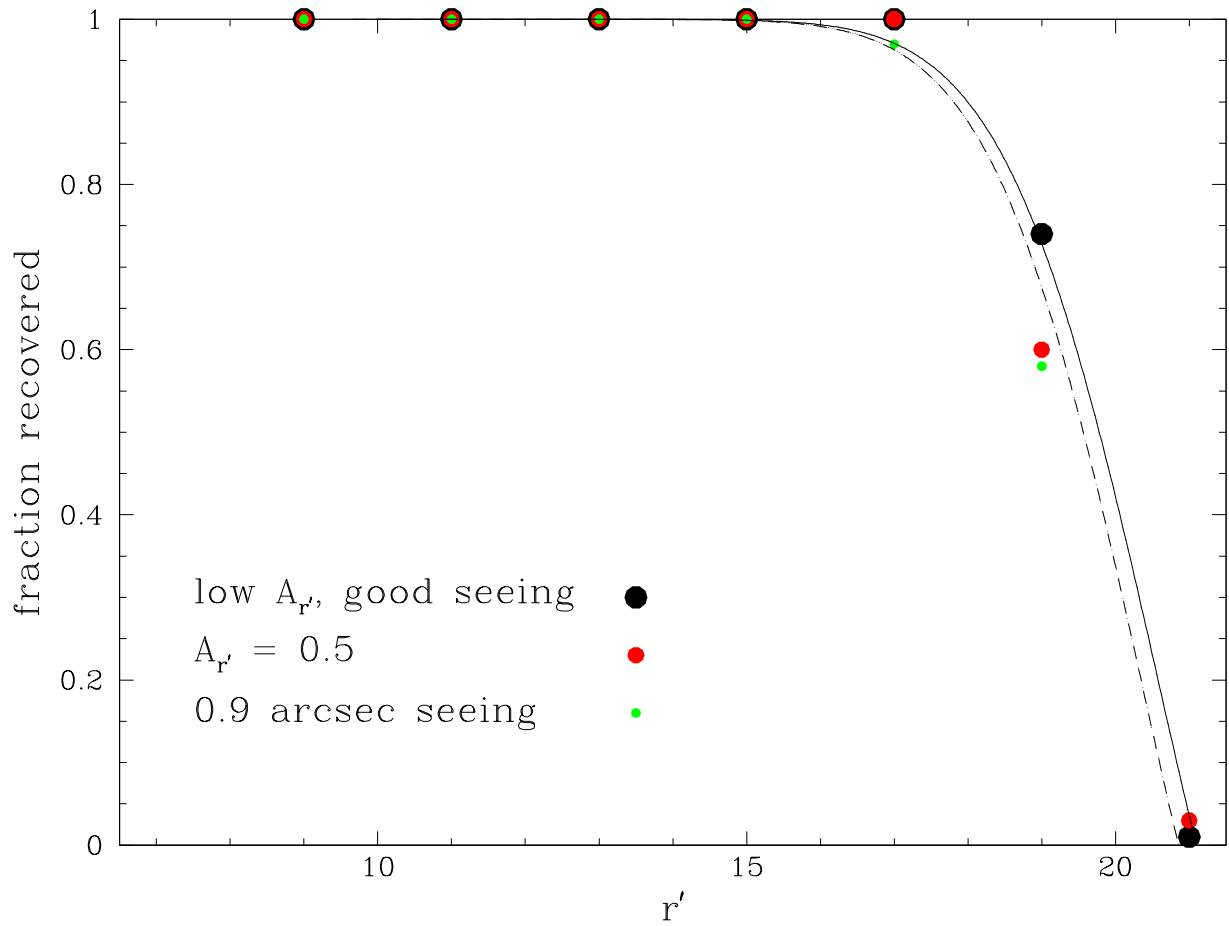}
\caption[Detection efficiency as a function of magnitude]{We plot detection
efficiency as a function of magnitude.  The solid line is the best fit
to the good seeing, low extinction case, while the dashed line 
incorporates the survey average effects of seeing, extinction,
single exposure regions, and loss of survey area due to coverage by
brighter objects.
\label{complim}}
\end{centering}
\end{figure}

Based on our artificial galaxy tests, 
we shade the region where incompleteness sets in.  Brighter than
$r^{\prime} = 17$, we expect we are 100\% complete, and to $r^{\prime} = 18$,
to be at least 90\% complete. We may
still be missing a small number of dwarfs which are lurking 
behind bright stars or other galaxies, but
find this should be less than a 4\% effect, primarily affecting
only the smallest objects. 
We use our simulations to determine estimates for our completeness at each magnitude
bin in the LF.
The lower limits in the detection efficiency imply that, at least
in regions with good seeing and low extinction, we should recover all artificial galaxies
down to $r^{\prime} = 17$.  At magnitudes brighter than
this, we find no candidate dwarf galaxies with effective radii greater than 30 arcsec.
Fainter galaxies will presumably be smaller.  As a conservative estimate, we assume
a maximum R$_{e}$ of 35 arcsec and uniform distribution in R$_{e}$ and calculate the 
fraction of galaxies we would expect to recover in each 2-magnitude bin of our 
luminosity function.  In reality, we expect galaxies to be weighted towards
smaller sizes at fainter magnitudes.  The assumption of a uniform distribution 
should therefore provide us with an
upper limit for the fraction of the real population we would expect to have missed, 
and an upper limit for corrected counts.
Results are plotted in Figure \ref{complim}.  The solid line is a fit to the recovery
results of simulations with good seeing, low extinction, and falling on full integration 
regions.  The dashed line incorporates the 15\% of the survey with single exposures
and takes into account any lower completeness due to worse seeing and extinction in
each field.  We apply these fractions as completeness corrections
to the counts in our luminosity function (open circles in Figure \ref{LFdiff}).  Applied 
corrections also include a 3\% correction for areal coverage by brighter objects for 
$r^{\prime} \geq 17.0$.
Although these are only crude corrections since we do not know the true distribution
of M81 Group galaxy sizes faintward of our 100\% completeness limits, we do not
find any evidence suggesting the faint-end slope steepens beyond what we
measure from our observed galaxy counts.

The luminosity distribution of galaxies is traditionally modeled by a \citet{s76}
function or with a simple power law for the faint end only.
To measure the faint-end slope of the magnitude distribution, we perform a power law 
fit to the 
complete set of galaxy counts between $8 < r^{\prime} < 18$ and find a 
faint-end slope, $\alpha$, ranging from $-1.26^{+0.09}_{-0.08}$ for the case 
of measured magnitudes to a maximum of $-1.39^{+0.05}_{-0.05}$ for 
exponential profile corrected total magnitudes.   The slope for the full, 
completeness corrected sample ranges from $-1.28^{+0.04}_{-0.05}$ 
to $-1.32^{+0.04}_{-0.05}$.  We measure a slightly shallower faint-end 
slope for only previously known group members of $-1.19^{+0.07}_{-0.07}$ 
to $-1.26^{+0.08}_{-0.08}$. 
However, it is clear from the figures that the slope does not increase much with the
addition of these new candidate dwarf group members; only the completeness
has shifted to fainter limiting magnitudes.
Completeness corrected counts as a function of both measured
and Sersic corrected magnitudes are well described by $\alpha = -1.3$ slopes.
It is possible that our luminosity functions display an upturn at the faint-end,
at magnitudes $r^{\prime} > -15$.  We therefore also fit all counts at the faint-end
in the range
$-15 \leq r^{\prime} < -10$ where we expect to be nearly complete.  This provides
us with the steepest slopes consistent with our data, with $\alpha \sim -1.45$.  
However, we caution that 
this measurement is based on few counts and low number statistics.
Best fit values for the faint-end slope, $\alpha$, for all samples are 
listed in Table \ref{slopediff}.

\begin{figure}[t]
\begin{centering}
\includegraphics[angle=0,totalheight=8.0in]{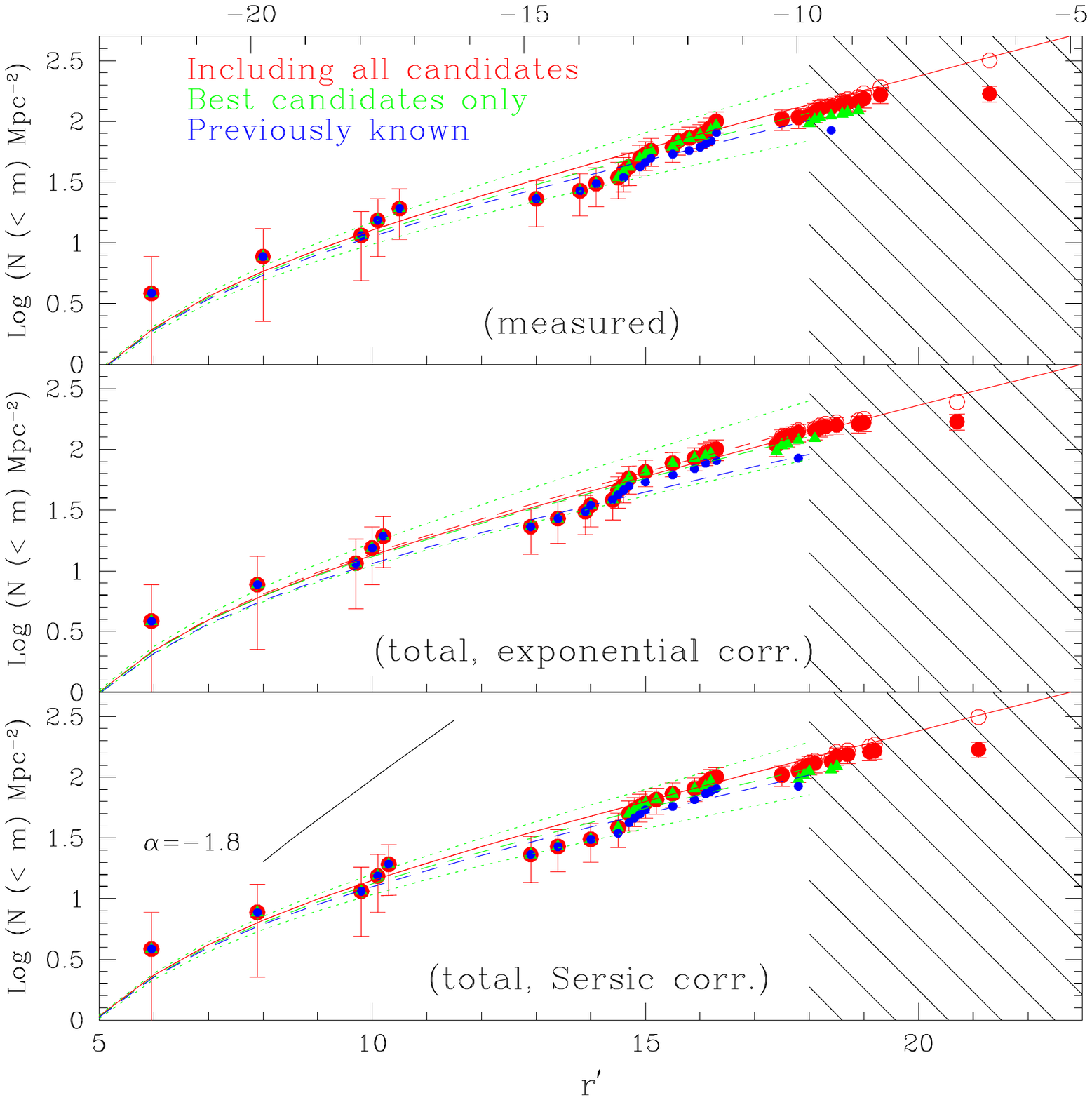}
\caption[Cumulative luminosity function for M81 Group]{Cumulative luminosity function
for the M81 Group, using measured magnitudes (top), magnitudes corrected assuming
exponential profiles (middle), and corrected magnitudes assuming Sersic profiles (bottom).
Counts are normalized to the area of the survey
coverage.  Large solid circles include previously known group members and all candidates
from this survey.  Open circles are corrected for incompleteness.  
The hatched region denotes where this survey is incomplete; brightward of this we 
expect nearly 100\% completeness in the detection of member dwarfs.  The best 
cumulative Schechter function fits to
the data brightward of our completeness limit are shown by the dashed lines.
The solid red line is the best fit to completeness corrected counts.  
Dotted lines are $\pm1\sigma$ fits for the sample of 'best' candidates. 
\label{LFint}}
\end{centering}
\end{figure}

\begin{figure}[t]
\begin{centering}
\includegraphics[angle=0,totalheight=8.0in]{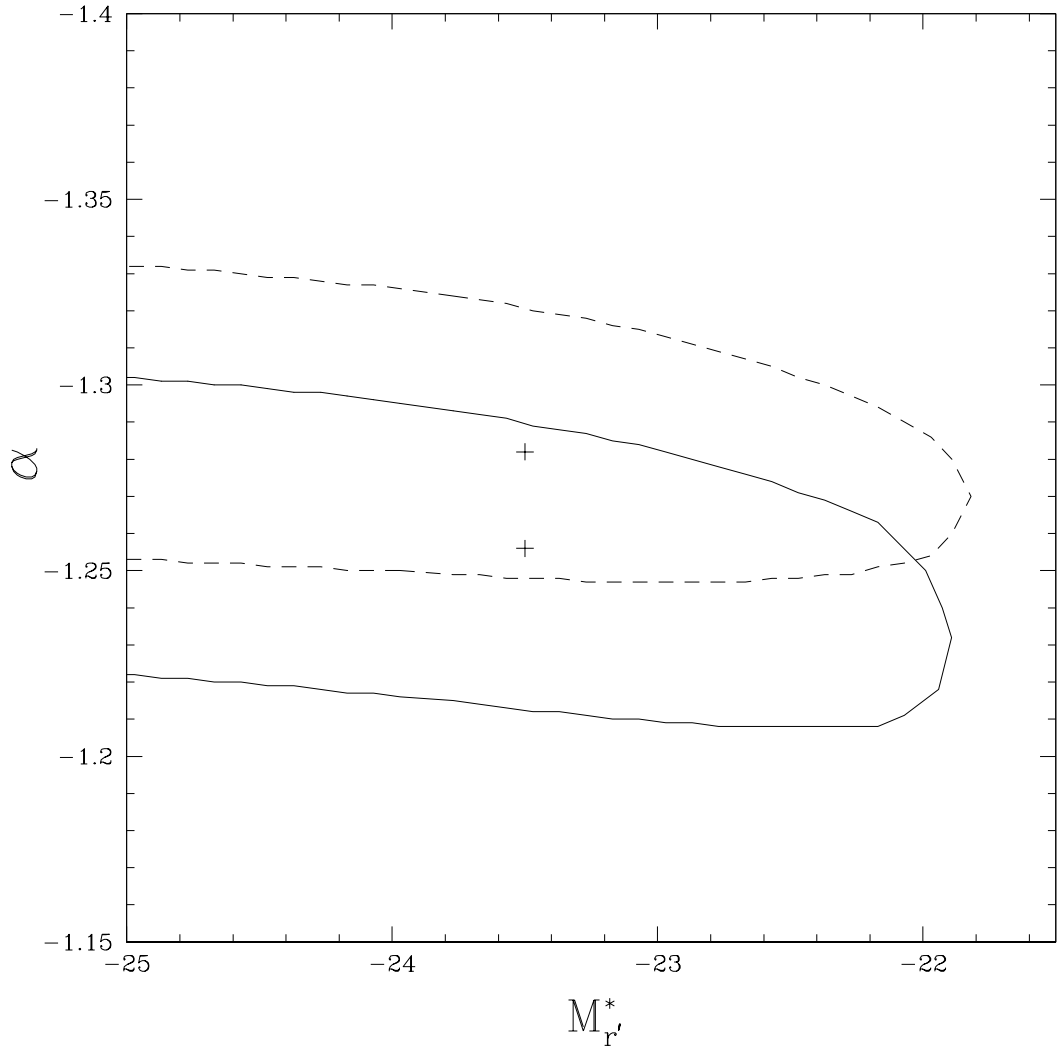}
\caption[Error ellipses]{We display 1 $\sigma$ error ellipses for
the luminosity function parameters $\alpha$ and $M_*$ for our
best (solid) and all (dashed) candidate samples using Sersic 
corrected magnitudes.
The values for $M_*$ are unconstrained at the bright end.
\label{eefig}}
\end{centering}
\end{figure}

Due to the overall small number counts of group galaxies, uncertainties in each 
bin of our differential luminosity function are large. This has the effect of
producing a number count distribution that does not monotonically rise as a function
of magnitude and, with so few data points, is not well fit with a power law.
We therefore also produce a plot of the unbinned cumulative distribution of galaxy counts as a function
of measured and corrected total magnitude for all subsamples in Figure \ref{LFint}.
We fit a cumulative Schechter function to the cumulative counts,
\begin{equation}
\int \Phi(M)dM = 0.4 \ln 10 \phi_{*} \int^{}_{} 10^{0.4(\alpha + 1)(M_{*} - M)} e^{-10^{0.4(M_{*} - M)}}dM
\end{equation}
such that
\begin{equation}
N(< M) = \phi_{*} \gamma[\alpha + 1,10^{0.4(M_{*} - M)}]
\end{equation}
using maximum likelihood techniques with a Poisson estimator.
$M_{*}$ is not only not well constrained due to the very small number of 
objects at the bright end of the LF, but is also affected by large errors 
in measured magnitude for the the brightest galaxies.  
For M81, we know our
measured total magnitude is strongly affected by improper sky subtraction.
For the luminosity function we therefore assume a total magnitude for this galaxy,
$M_{r^{\prime}} = -21.8$, from previously obtained $R$ band data \citep{tp00}.  For
fitting the cumulative Schechter function to our data, we hold $M_{*}$ 
and $\phi_{*}$ constant while solving for the best fit slope, $\alpha$.
Results for the different samples are presented in Table \ref{slopeint}.
To test the effect of $M_{*}$ on the fit, we refit the data allowing all
parameters to vary.  We find the slope changes typically by about
0.015, and no more than 0.04, smaller than the random errors in 
our fitted slope values.  Results are similar when we
use our original total magnitude measurement for M81.  
Therefore, the value of $M_{*}$, although 
coupled to $\alpha$, does not significantly affect the determination of 
the faint-end slope.  In Figure \ref{eefig}, we display the 
$1 \sigma$ $\alpha - M_*$ error ellipses for the luminosity functions of 
'best' and 'all' candidate samples with Sersic corrected magnitudes.  It
can be seen that $M_*$ is unconstrained at the bright end, but the choice
of $M_*$ will not strongly effect the value of $\alpha$.

Overall, we obtain very similar values for the faint-end slope as compared with the 
power law fit to the faint end of the differential LF.  For all samples, we find
a faint-end slope $\alpha$ $\sim -1.28$.  
Slopes for only originally known
and 'best' candidates are insignificantly shallower than for all candidates.  
We furthermore find the faint-end slope 
to be robust against corrections from measured to total magnitude.

Fits to completeness corrected counts do not produce steeper faint-end slopes, 
but do extend our fits to all counts down to $r^{\prime} \sim -7$.
Since we expect these completeness corrections
are already an upper limit given the expected true size distribution of real galaxies, 
we do not believe the slope steepens appreciably at magnitudes fainter than our completeness limit.
The steepest measured slopes are for the full sample of objects
which assume all candidate objects we have detected are real 
galaxies within the M81 Group.   Certainly some subset of these candidates
are not true M81 Group galaxies, so we can take these slopes as an upper limit.
We find the best fit slope for the subsample of most likely M81 members
using Sersic profile corrected magnitudes to be $-1.30\pm0.10$ and $-1.26\pm0.05$ for 
the differential and integrated LFs respectively. 

\section{Discussion}\label{disc}

\subsection{Dwarf Population Characteristics}\label{popchar}

\begin{figure}[t]
\begin{center}
\resizebox{\columnwidth}{!}{\includegraphics{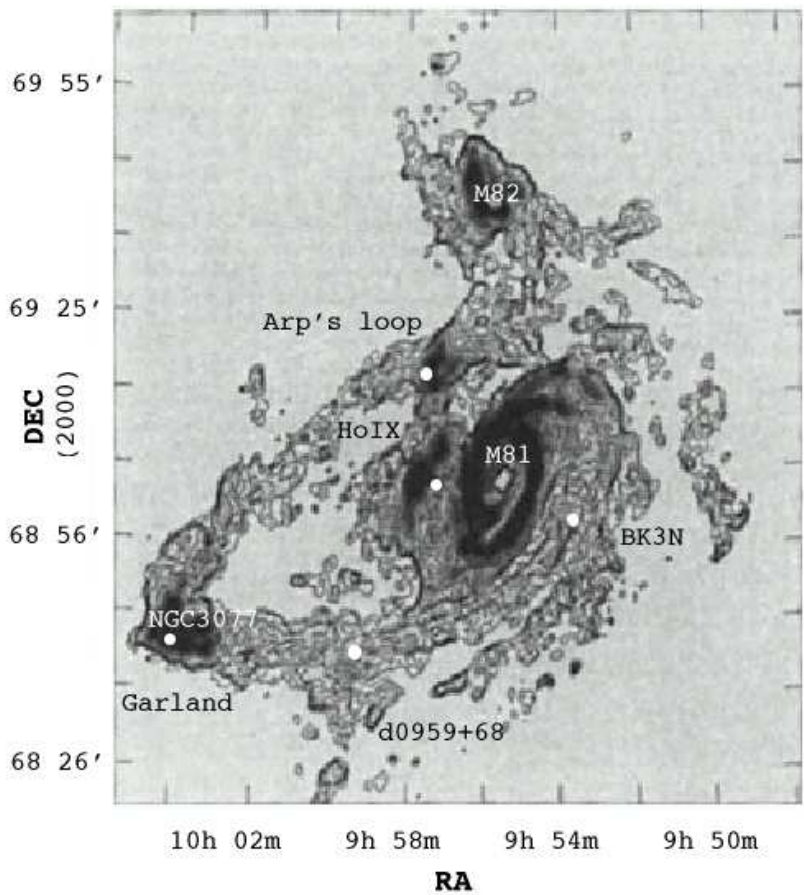}}
\end{center}
\caption[M81 HI map of \citet{yun94}]{We
display the M81 HI map of \citet{yun94}.  Overlaid (white dots) are the positions of
the five suspected tidal dwarfs.  d0959+68 is found to lie in the tidal 
bridge between M81 and NGC3077.  Adapted by permission of Macmillan Publishers Ltd:
Nature, copyright 1994.
\label{tdl}}
\end{figure}

Of the new candidate galaxies, roughly 45\% are believed to belong to the dwarf
spheroidal class with dominant old stellar populations, while the rest exhibit 
dwarf irregular or BCD morphology 
with younger stellar populations and ongoing/recent star formation.
At least one new object may be a tidal dwarf, formed in 
the tidal streams connecting the three closely interacting galaxies of
M81, M82, and NGC 3077.  This area contains a number
of knots of star formation and tidal debris.  Previously known or suspected
tidal objects include Arp's loop knots, Garland knots, Ho IX, and BK3N \citep{mak02}.
These objects are found to be dominated by young stellar populations and have
high gas contents \citep{dem08,sbsdm08}.  \citet{dav08} recently found an arc of
stars in the M82 halo which may have formed around the same time as these other
tidal features.
One of our candidate objects, d0959+68, was first noted by \citet{durrell} 
in a search for intragroup stars in the field of the M81-NGC 3077 HI
tidal tail.  
It has recently undergone a burst of star formation as evidenced by
the large population of bright blue stars seen in Figure \ref{colim}. 
\citet{durrell} match isochrones to the CMD of this object and find
evidence that star formation occurred 30-70 Myr ago, after the formation
of the HI tidal arm.
In a search for HI clouds in the M81 Group, \citet{bws07} find
a number of free floating clouds without optical counterparts which
they suggest may be tidal debris from the 3-body interaction 
involving M81, M82, and NGC 3007.  One of these clouds is just offset
from d0959+68, about $1^{\prime}.5$ arcmin to the SW, several times
the measured optical half-light radius. This HI cloudlet has a radial 
velocity distinct from both the Milky Way and M81 tidal arm. 
If it turns out the optical 
component has a similar radial velocity as the HI cloud, it would be
likely that the two were associated.  It may be possible that, like
the Local Group Phoenix dwarf \citep{yswd07}, blow out from supernovae
energetics from this young stellar population has displaced the 
neutral hydrogen gas.  
We display this new object along with previously suspected tidal dwarf galaxies 
superimposed on the HI map of \citet{yun94} in Figure \ref{tdl}.

Three new M81 candidate member galaxies (d1028+70, d0958+66, d1012+64) are of the 
poorly understood
blue compact dwarf class.  Spectra of these galaxies were obtained
with Subaru/FOCAS and BTA/SCORPIO.  Radial velocities of $\sim +60$, $-100$, 
and $+150$ km s$^{-1}$
support group membership for d0958+66, d1028+70 and d1012+64 respectively.  
All spectra 
exhibit strong Balmer absorption lines indicative of recent star formation. 
The BTA 6m spectra also find H$\alpha$ emission in all three cases indicating
active star formation, in agreement with the blue $(V-I) \sim 0.8$ 
colors found for these galaxies.  These three candidates bring the total number
of M81 Group BCDs to at least four, including UGC 6456, long
considered an isolated member of the M81 Group \citep{lynd98,tul81}. DDO 82 may
have a BCD component as well.  It has an optical structure and color that 
corresponds to that of BCDs but has an HI flux contaminated by local HI.  UGC 6456 lies
far outside our survey region, at SGL $=37^{\circ}$, SGB $=+11^{\circ}$ in 
Figure \ref{fulreg},
while DDO 82 lies at the projected distance of the second turnaround radius for
the M81 Group.
BCDs tend to be found more frequently in field environments than clusters
and are one of the least clustered galaxy types known \citep{vz01}.  The
three BCDs discovered here all lie in low density regions at a minimum 
projected distance of 
140 kpc from M81, with d1012+64 lying outside of the second 
turnaround radius at a projected distance of 315 kpc.  
The excess of this type of galaxy
observed in this group may indicate that the termination of star formation
is due to processes such as ram pressure stripping which occur preferentially in 
more massive and more dynamically evolved groups/clusters.
However, given the projected distances of the M81 BCD population, it is possible
that these galaxies are on first infall orbits and processes such as strangulation 
that serve to shut down star formation even in poor groups \citep{km08} 
have yet to act on these galaxies.

\begin{figure}[t]
\begin{centering}
\includegraphics[angle=0,totalheight=8.0in]{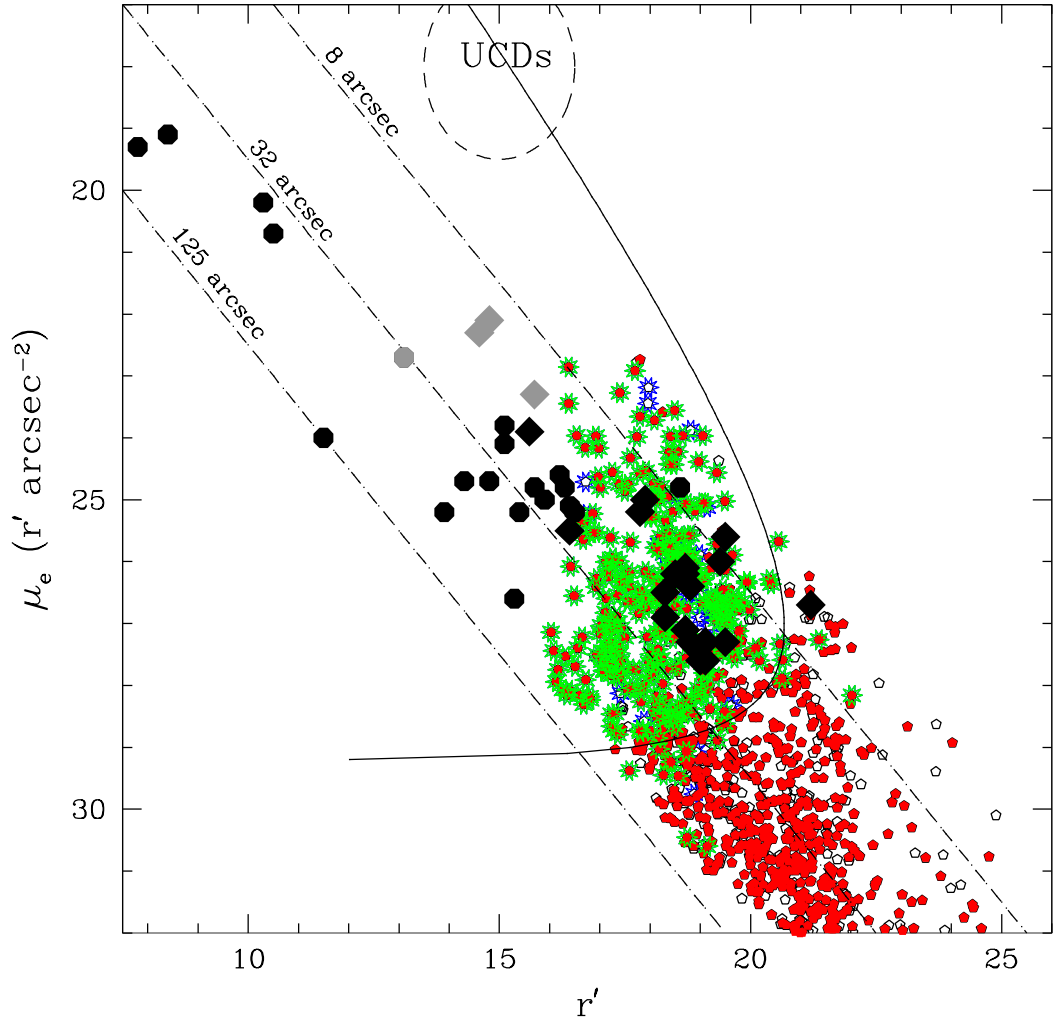}
\caption[Artificial galaxy recovery in mag-$\mu$]{We plot the recovery
of 1200 artificial galaxies in the magnitude-surface brightness plane.
As before, pentagons represent all artificial galaxies.
Stars around these points indicate that the simulated
galaxy was recovered.  Overlaid as black (gray) diamonds are the candidate 
real galaxies (BCDs)
in this magnitude - effective surface brightness plane.   Circles represent
previously known M81 Group galaxies.  DDO 82, shaded in gray, is a potential
BCD.  Lines of constant effective
radius for exponential profiles are labeled.  We draw a solid curved line 
which bounds the
region of recovered artificial galaxies.
\label{falsemmu}}
\end{centering}
\end{figure}

We find no ultra-compact dwarfs (UCDs) such as those found in the richer
Fornax and Virgo clusters \citep{djgp00, drink2}.  
Neither have UCDs
been discovered in the Local Group.  This may support the hypothesis
that UCD formation is driven by a mechanism that occurs 
preferentially in denser environments, such as tidal stripping
of nucleated dwarf ellipticals \citep{bcd01} or late-type spirals 
\citep{mlk98}. Alternatively, \citet{fk02} suggest the possible formation
of UCDs as objects evolved from young massive star clusters which are
formed within tidal tails of massive galaxy-galaxy mergers.  In this
scenario, one might expect the formation of such objects to be ongoing 
in the tidal tails of the galaxy interactions occurring in the M81 Group
although we find no evidence for this.
It is possible that we would be unable to detect the more compact UCDs in 
this survey using our current search techniques.
We show in Figure \ref{falsemmu} the location in $r^{\prime}-\mu_{e}$ space the
UCDs would occupy.  Our detection limit would enclose only the brighter
and larger UCDs. Due to the extremely compact nature of UCDs, it is probable
that the smaller ones would not appear in our images as resolved concentrations of 
stars. Rather, the light on these small scales would be blended making detection
based on a resolved stellar population difficult and in some cases impossible. 
However, we would expect to be able to detect the larger and brighter 
subpopulation.  As we do not detect a single UCD, this suggests that
UCDs may be absent in this poor, unevolved environment.

No new bright ($r^{\prime} < -12$) ultra-diffuse dwarfs like F8D1 and 
IKN have been found in this survey.  This could partially
be due to selection effects since neither our
eye nor the two-point correlation code would be particularly sensitive
to extremely diffuse objects, although these two objects were easily picked
out both by eye and recovered with the two-point correlation technique.
We would therefore expect to detect similarly low density objects if they existed.
We do find a number
of very diffuse, faint, and small candidates including d1013+68, d1014+68, and d1016+69 
which may or may not prove to be real objects.

In comparison with the Local Group, we find a very different morphological
mix of dwarf satellites.
In the M81 Group, we have found 3 new BCD candidates while no such objects 
have been discovered around M31 or the Milky Way.  
The fraction of late type, dwarf irregulars is also much higher in 
the M81 Group.  Including the recent detections of ultra-faint dwarfs in the Local Group, 
the fraction of late types is $\sim 25$\% while in the
M81 Group, including our new best candidates, the fraction of late types
could be as high as 55\%.  This different morphological mix had been
noted earlier for giant galaxies and brighter dwarfs, but recently 
discovered faint dwarfs in both groups
maintain the distinctly different population fractions.  It is inferred
that the M81 Group is dynamically less evolved than the Local Group.

We do not find any objects as faint as the recently discovered ultra-faint
Local Group galaxies. This is due to incompleteness at these
low magnitudes.  We do find one candidate, d0944+69, with an absolute r$^{\prime}$
magnitude of $-6.7 \pm 0.5$.  This would make it the faintest known
M81 Group dwarf.  It also would have one of the smallest sizes for known
dwarfs of R$_{e}$ = 90 pc, similar in size to some of the recent Local Group satellite 
discoveries with R$_{e}$ ranging
from 23 - 125 pc \citep{will1,zucker06,martin06,belo07,bootes07}. 
We do not yet know if this object is either a
real galaxy or a member of the M81 Group.  If this object does turn
out to be a group member, its lower limit half-light radius would 
impinge on the $40 < R_e < 100$ pc 'size gap' between dwarfs and globular 
clusters \citep{belo07}.
However, as most of the galaxy light likely falls below the level of the
sky, we are certainly underestimating its physical size and perhaps,
even with the magnitude correction, total luminosity.  

\begin{figure}[t]
\begin{centering}
\includegraphics[angle=0,totalheight=8.0in]{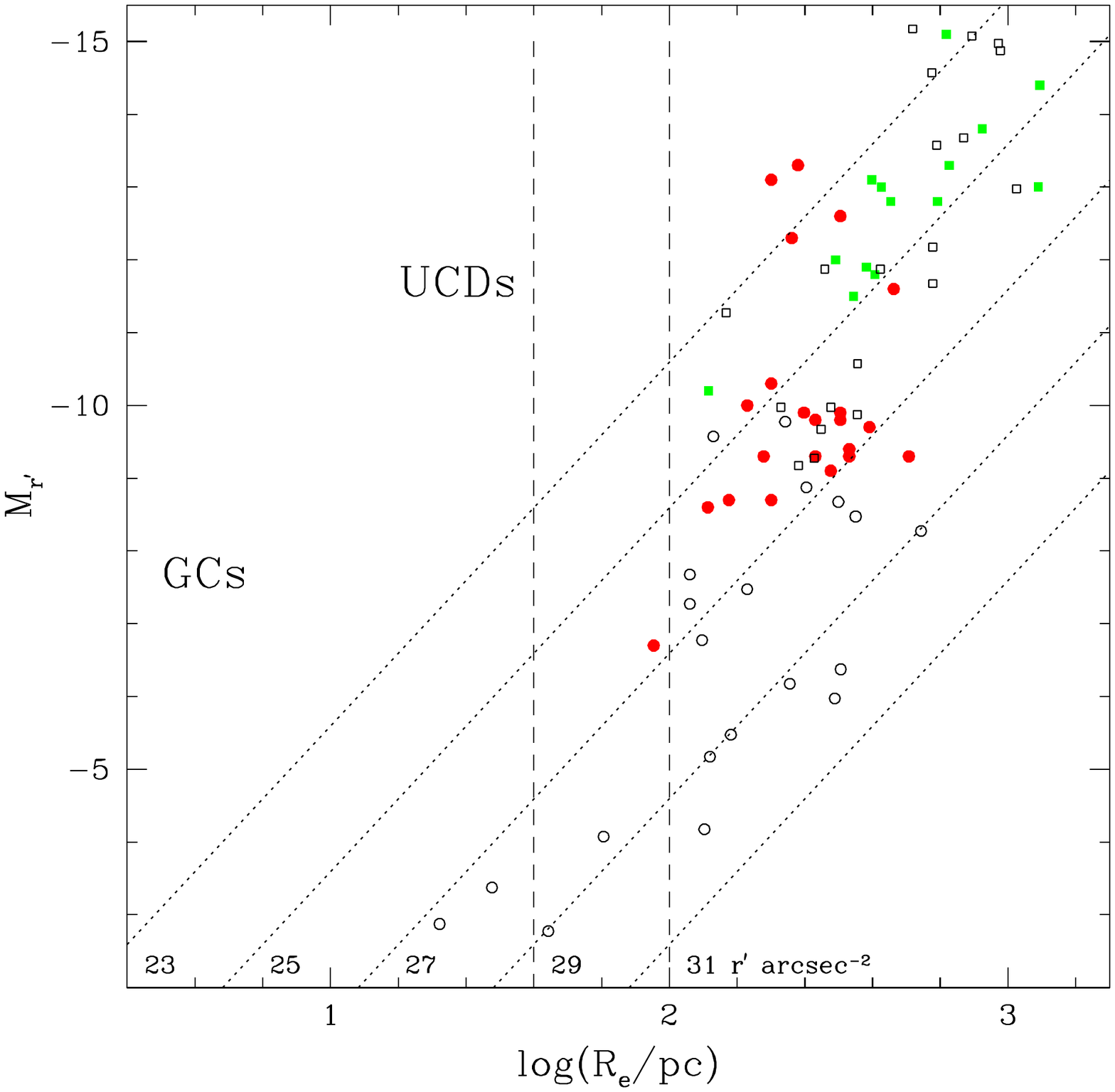}
\caption[Magnitude - half-light radius distribution]{Locations
of our M81 candidates on a total magnitude vs half-light radius plane
(filled circles), along with previously known M81 galaxies (filled 
squares), Local Group
previously known galaxies (open squares) \citep{IH95, mm98, mci06} and 
recent discoveries (open circles) \citep{simon07,belo07,will1,bootes07,zucker06,
martin06,imi07,ifh08}. To transform $V$ magnitudes to $r^{\prime}$, we assume
$r^{\prime} = V - 0.84(V-R) + 0.13$ \citep{figdss96}.
The rough locations of globular clusters and Fornax UCDs on this plane are 
shown.  Globular Clusters have sizes that range up to $\sim 40$ pc, while
Local Group galaxies have sizes generally larger than $\sim 100$ pc.  The
size gap region falls between the two dashed lines.  Lines of constant
effective surface brightness are also shown.
\label{mrplane}}
\end{centering}
\end{figure}

The other candidates all have properties consistent with normal faint dwarfs.
We plot $M_{r^{\prime}}$ vs log (half-light radius) in Figure \ref{mrplane}.
New candidates are shown as filled circles (assuming a distance modulus of 27.8)
while previously known group members are displayed as filled squares.  The
three candidate BCDs are distinguished by their high surface brightnesses.
For comparison, we include Local Group dwarfs in this plot.  Recent discoveries
are denoted by open circles while previously known galaxies are shown as
open squares.  To transform from $V$ to $r^{\prime}$ magnitudes, we use
$r^{\prime} = V - 0.84(V-R) + 0.13$ \citep{figdss96} assuming an
approximate $(V-R)$ color of 0.6 for dwarf galaxies.  Lines of constant surface 
brightness are shown.   New Local Group detections only became
possible with the SDSS survey reaching effective surface brightnesses below
27.0 mag$_{V}$ arcsec$^{-2}$ \citep{belo06}, corresponding to a stellar surface
density limit.  Incompleteness in our survey may be starting to set in
by 26.0 mag$_{r^{\prime}}$ arcsec$^{-2}$ and most
of our detections are brighter than 27.0 mag$_{r^{\prime}}$ arcsec$^{-2}$,
so we may still be missing a large population of ultra-faint dwarfs.
According to Figures \ref{falsemmu} and \ref{mrplane}, we would expect
these missing galaxies to have total magnitudes fainter than $M_{r^{\prime}} \sim -8$.

\subsection{Distribution of M81 Galaxies}\label{distrib}

\begin{figure}[t]
\begin{centering}
\includegraphics[angle=0,totalheight=8.0in]{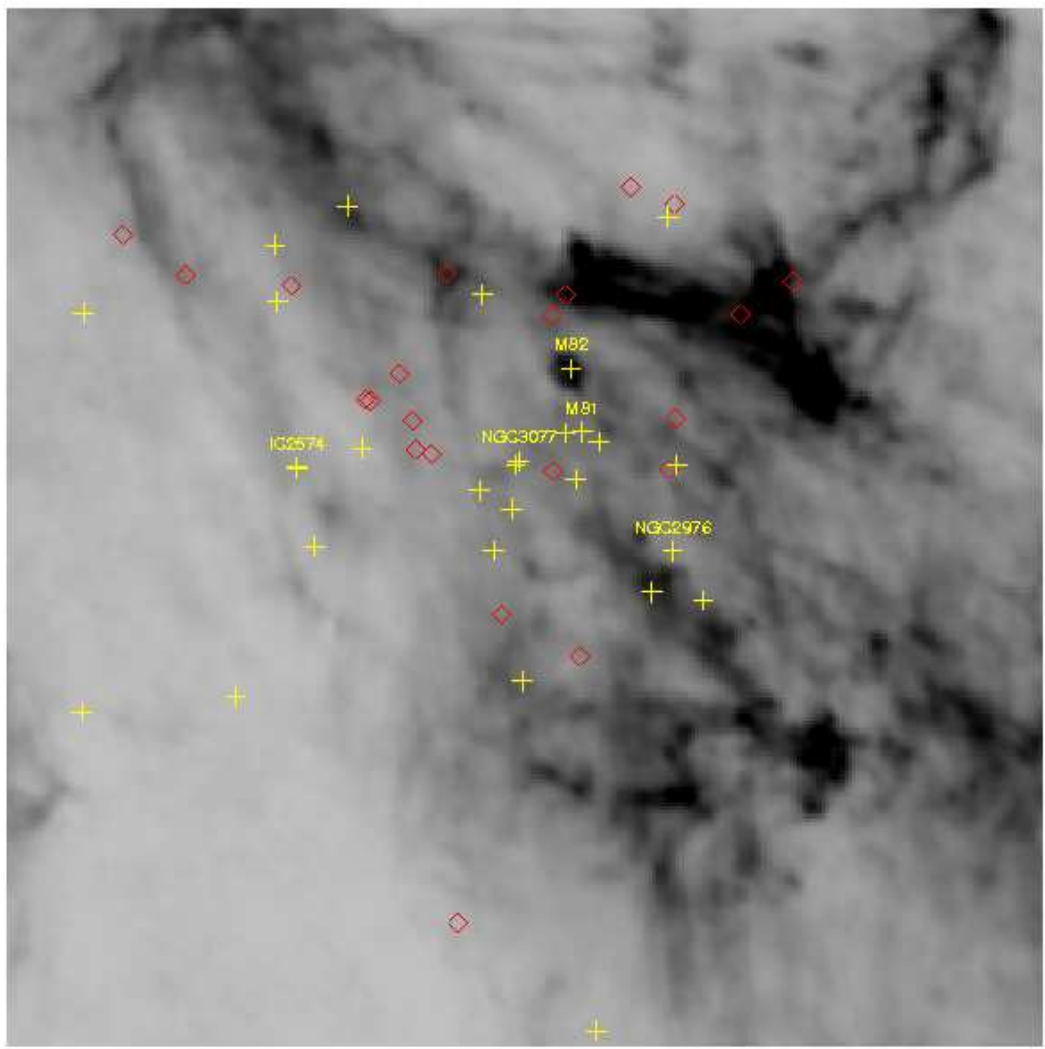}
\caption[M81 galaxy detections superimposed on Schlegel dust map]{We
display the M81 galaxy detections superimposed on the \citet{sfd98} dust map.
Red diamonds represent newly discovered candidate group members while
yellow crosses correspond to previously known group members.
\label{schlegmap}}
\end{centering}
\end{figure}

The asymmetric distribution of dSphs noted by \citet{KKp00} 
has become less pronounced with the addition of the new objects. As discussed
in that paper, a possible cause for the asymmetry is
contamination of the M81 Group area with many reflecting nebulae (galactic cirrus)
which made it difficult to find extreme LSB dwarfs that are unresolved into stars
or have only old stellar populations.
In Figure \ref{schlegmap} we show the M81 region of the sky in the \citet{sfd98}
galactic dust map. 
Superimposed are the locations of galaxies in
the group.  A number of new candidates (diamonds) can be seen to lie in 
regions of higher extinction and, given the dearth of detections in
regions with the highest extinction, it is possible we are still
missing faint M81 dwarfs concealed by these clouds.  However, it seems unlikely
we would be missing dwarfs of comparable brightness to the ones displaying the
noted asymmetry, given that the new candidates from this survey which are detected
in regions of high extinction are all 1-2 mags fainter.

\begin{figure}[t]
\begin{centering}
\includegraphics[angle=0,totalheight=8.0in]{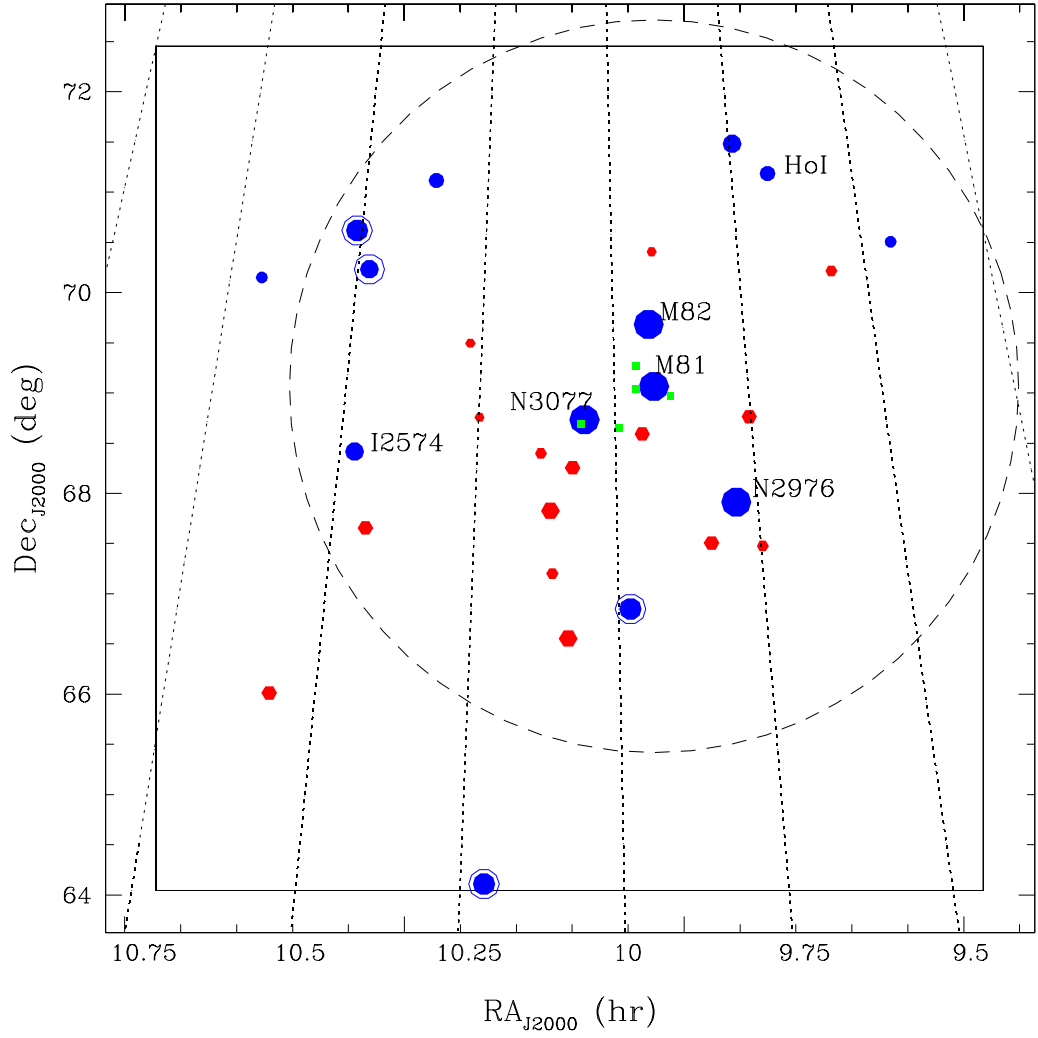}
\caption[Distribution of morphological types]{Projected distribution of early
type (red hexagons) and late type (blue circles) galaxies. Green squares indicate
candidate tidal dwarfs. The location of BCD candidates, including DDO 82,
are marked by blue rings.  We include in this plot only
previously known and best candidates.  Size indicates surface brightness
with larger points denoting brighter effective surface brightness.
\label{morloc}}
\end{centering}
\end{figure}

We compare the projected distribution of morphological types in Figure \ref{morloc}.
Late types, early types, and tidal candidates are distinguished by symbol types
while point sizes indicate surface brightness with
brighter surface brightnesses denoted by larger points.
All the bright giant galaxies are late types (circles) and all candidate
tidal dwarfs (squares) lie near the center of the cluster right around M81, M82,
and NGC 3077.  Early type dwarf spheroidals (hexagons) lie primarily towards the 
group core while all late type dwarfs lie towards the periphery 
of the group.  A clear morphology-density relation is seen.  We find the median
distance from M81 for late type dwarfs is 1.7 times that for dwarf spheroidals.

\subsection{Faint-end of the Group Environment Luminosity Function}\label{lfconc}

$\Lambda$CDM models predict a faint-end slope for the mass function of $-1.8$ \citep{tt02}.  
We measure a luminosity function faint-end slope, $\alpha \sim -1.28\pm 0.06$ for M81
group galaxies.  The steepest faint-end slopes consistent with our 
data are found to be $-1.39\pm0.05$ for our binned differential LF, constructed with all 
candidates and having presumably overestimated magnitude corrections under the
assumption of exponential profiles, and $-1.45\pm0.06$ when fitting a possible faint-end 
upturn observed in the full sample.  Even when we apply completeness corrections
at faint magnitudes, we find faint-end slopes
flatter than $-1.30$.  When we include only most likely member candidates in our
LF, we find a slightly shallower slope of $-1.27\pm0.06$.  We therefore
expect $-1.30$ to be an upper limit.

\begin{figure}[t]
\begin{centering}
\includegraphics[angle=270,totalheight=5.5in]{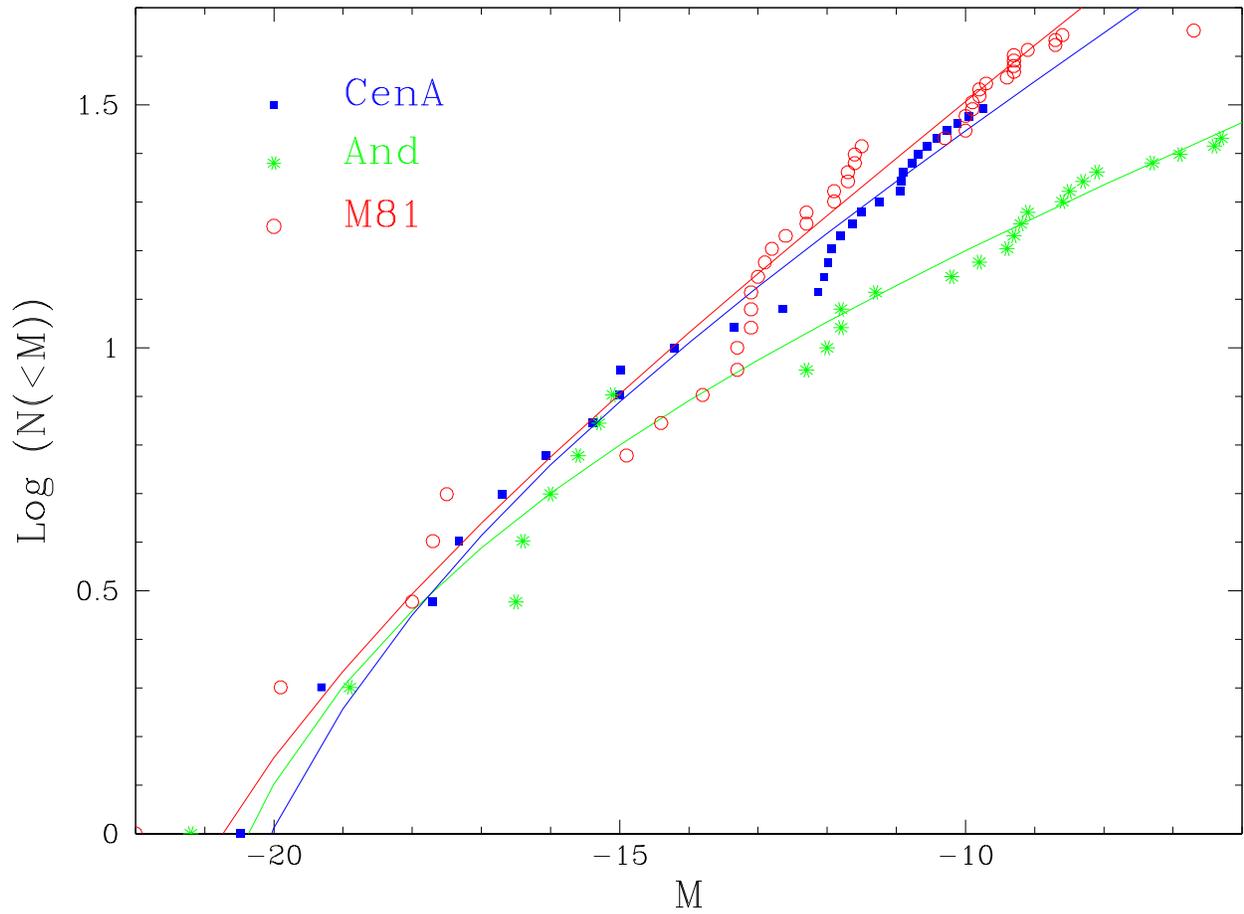}
\caption[Integrated LF for Andromeda, CenA satellites]{Cumulative
LFs for Cen A ($B$ band, squares), Andromeda ($V$ band, stars),  
and M81 ($r^{\prime}$ band, open circles) satellites.  For clarity, we
do not include error bars. Applying a cumulative
Schechter function to these data, we find faint-end slopes of $-1.13^{+0.06}_{-0.06}$
for Andromeda systems, $-1.23\pm^{+0.04}_{-0.10}$ for Cen A, and $-1.28^{+0.06}_{-0.06}$ 
for M81.
\label{andLF}}
\end{centering}
\end{figure}

We compare the M81 Group LF with that of other nearby groups.  Detections
of dwarf galaxies in the Local Group go much fainter than we recover in M81. 
However, because of the full sky coverage necessary to 
obtain a complete census, completeness is still quite low. 
We therefore look only at the LF for Andromeda and its satellites. 
Large areal coverage is still required, including probing into the
zone of obscuration, but progress is being made.  We construct a
cumulative LF for Andromeda satellites including all previously
known galaxies \citep{vdb06} and recent family additions \citep{imi07,maj07,
martin06,ifh08,And08} (Figure \ref{andLF}). A fit to the cumulative $V$ band counts finds
a shallow faint-end slope of $-1.13^{+0.06}_{-0.06}$.  This is very similar to the best Schechter 
function fit for Local Group galaxies found by \citet{pv99} with $\alpha = -1.1\pm0.1$.  The
recent discoveries of Local Group dwarfs have not increased the slope significantly.

The nearby Cen A group at 3.6 Mpc, in contrast to the spiral dominant Local Group and
M81 Groups, contains a dominant giant elliptical galaxy with radio loud AGN 
and a late type fraction of only $\sim38$\%.
Using $B$ band magnitudes for the Cen A group from \citet{ksd02}, we find
a faint-end slope $\alpha = -1.23^{+0.04}_{-0.10}$, nearly as steep as that of M81 
(Figure \ref{andLF}).

\citet{bpj01} find that steep LFs are generated by the combined number counts
of dI and dE galaxies, and when dIs are excluded from the construction of
the LF, the faint-end slope flattens.  This might imply that more evolved regions
with fewer late types should have shallower slopes.  Alternatively, from a comparison
of poor groups, clusters, and field environments, \citet{zm00} found that
the dwarf-to-giant ratio (and therefore the faint end of the LF)
increased in regions of higher density, and that this increase was specifically
due to an increase in the fraction of quiescent dwarfs.
In our work on M81 we find a steeper slope in the dynamically unevolved,
dI-rich M81 Group than these other two groups and in particular the dSph-rich Local Group.
However, recent work on the luminosity function for two nearby dynamically evolved
groups with large early type galaxy populations finds 
steep faint-end slopes with values of $-1.3\pm0.1$ for
the NGC 5846 group \citep{mtt05} and $-1.35$ for the NGC 1407 group 
\citep{ttm06}, slightly steeper than what we find for M81.
These works determine group membership on the basis
of morphology and surface brightness and both reach down to $M_{R} = -12$.

It is possible that the shallower Local Group LF is due 
in part to incompleteness in the surveyed region.  
In recent work, \citet{kop07} construct the LF for the Local Group 
including the recent discoveries of ultra-faint dwarfs with SDSS.  They
also include a volume correction factor, assuming different number density laws
for the distribution of halo satellites, and find faint-end slopes ranging
between $-1.29 < \alpha < -1.25$ for the magnitude range $-19 < M_{V} < -3$,
very similar to the slope we find for M81.  In this case, with all
groups discussed here having faint end slopes of $\alpha \sim -1.3$,
we find no clear trend in faint end slope as a function of environment within
groups of galaxies.  

A few recent studies have found steep faint-end slopes in group environments.
Using a statistical subtraction of the background for nearby groups and poor
clusters identified in SDSS data, \citet{gllv06} find
a sharp upturn in composite LFs at $M_{r} \sim -17$ with slopes 
of $-1.9 < \alpha < -1.6$.
They find this upturn regardless of group mass, number of members, or environment.
Using colors and morphology to establish membership, \citet{k06} also find a steep 
upturn with $\alpha \sim -1.7$ starting around $M_B > -15$ in Hickson compact groups. 
While we do see a possible increase in slope around $M_{r^{\prime}} > -15$, this 
appears to be followed by a flattening at fainter magnitudes.  We do not find any evidence
for a slope as steep as $-1.7$.  
If the slope were this steep in our survey region, we would expect to detect the 
large numbers of bright dwarf galaxies implied by this slope. 
It is hardly conceivable that our
faint-end slope measurement suffers from selection effects compatible with a
factor of 10 inconsistency in counts at $M_{r^{\prime}} \sim -10$.
Down to at least $M_{r^{\prime}} < -10$ we do not find such a steep upturn 
in the poor M81 Group when constructing the LF from our nearly complete survey 
of galaxies whose membership is established through more direct means.  
Alternatively, the M81 Group may represent a different
environment than probed in these two studies.
Our results are more consistent with the large field surveys such as
SDSS and COSMOS which find slopes of $-1.3$ (or $-1.5$ with corrections for surface
brightness selection effects) and $-1.2$ respectively \citep{bls05,liu08}.  
With our deeper survey limit we probe a fainter dwarf galaxy population
yet find a similar slope to that found in these redshift studies.

\subsection{Implication for Cosmological Models}\label{cosmo}

Assuming the LF follows a constant slope of $-1.3$, rather than turning
over at $r^{\prime} > 17.0$ where we start to become incomplete, we would expect to be
missing at least $\sim70$ M81 Group members in our survey area in the 
range $17 < r^{\prime} < 22$ ($-11 < M_{r^{\prime}} < -6$).
If the slope increased to $-1.8$ at faint magnitudes, beginning at 
$M_r^{\prime} = -12$, we would expect to be missing over 1700 galaxies
in our survey area. This is over an order of magnitude greater than we 
would expect from our data.  For the intrinsic faint-end slope
to be this steep, we would have to be missing a very large population of 
very faint and low surface brightness, low stellar concentration
galaxies, and based on our simulations, we do not find any evidence
that we could be missing such large numbers of galaxies brighter than
$M_{r^{\prime}} = -7.$

Recently there have been a number of successful attempts at bringing
model predictions closer into line with the lower number counts of observations. 
One idea is that the dark matter halo circular velocity
is larger than extrapolated from measured stellar velocity dispersions.  This could be
caused by dark matter halos having much larger extents than the luminous component of the
galaxy such that velocity dispersions only provide an estimate of the mass
in the interior of the halo \citep{pen08}.  Another explanation posits that
dark matter halos had more mass at higher redshift but have since lost mass through 
dynamical friction \citep{kgk04}.  In both of these scenarios,
the observed galaxies today correspond to larger dark matter halos in the simulations,
thus alleviating the problem of excess counts predicted at the measured luminosities.  
Another widely used solution to the discrepancy incorporates the effects of reionization on
the smallest mass halos in galaxy formation models.  Reionization suppresses the 
accumulation of gas and thus star formation, leaving lower mass halos dark.  
A number of recent models find
good agreement with observations \citep{simon07,kang08,bov08}.

\citet{kop07} compare the Milky Way LF to semi-analytical predictions and
find a reasonable match with \citet{bfbcl02} who include effects
of tidal disruption and photoionization with CDM theory predictions.
Counts to $M_{V} < -3$ agree well within the uncertainties.  This
provides some support for the hypothesis that at least part of the cause 
of the disparity between $\Lambda$CDM predictions and observations
is due to the suppression of gas accumulation in small galaxies collapsing
post reionization 
although other predictions from their model, such as central surface brightness
values, do not agree well with observations.  
In further support of this idea, \citet{simon07} compare the observed Local Group 
circular velocity distribution 
function (a proxy for the mass function) to that predicted from the Via Lactea N-body
simulation \citep{vialac07}. They find good agreement when assuming that the Local 
Group halos observed
at z=0 were the ones which collapsed prior to reionization and which
correspond to the model objects with the largest values of $v_{circ}$ at
various high redshifts.  

We find a faint-end slope of the M81 Group LF to a limiting magnitude
of $M_{r^{\prime}} = -10$ that
is much shallower than $\Lambda$CDM cosmology would predict but similar to that 
found for Milky Way halo satellites.   We do not reach to the same
ultra faint population as is currently being uncovered in the Local Group, and we have not
obtained kinematical measurements to directly compare the mass function with theoretical
predictions.  The M81 halo has a similar mass as the Milky Way, thus
solutions to the missing halo problem which bring model number counts 
into agreement with observations
by associating more massive halos with fainter galaxies while lower mass
halos remain dark could equally well explain the shallow slope we find.
Therefore, part of the
explanation for the 'missing galaxy' problem may be in the physics of how 
mass is converted to light.
Incorporating the effects of feedback and star formation efficiency
may help reconcile theory and observations.  Accounting for the suppression
of gas infall into the low mass halos of the
forming galaxies by reionization in the early universe may explain
the bulk of the discrepancy and naturally explains the noted
environmental dependence of the LF, with clusters and more dynamically
evolved regions exhibiting slopes with $\alpha$ steeper than $-1.4$ 
(see e.g. \citet{tsb05, con02,KCMM2,dp98,bntuw95}) as compared to the shallower
slopes typically measured in group and field environments, and which we 
find in the M81 Group.

\section{Summary and Conclusions}\label{conc}

We have discovered 22 new candidate dwarf galaxies in the M81 Group from 
our 65 square degree CFHT/MegaCam survey, designed to extend beyond
the second turnaround radius for the group.  Of these candidates, we 
believe 12 are likely members. The remainder may prove to be background
galaxies, foreground associations of stars, or artifacts in our data.
The 12 likely members consist of 3 candidate BCDs lying near the periphery 
of the group, 1 tidal dwarf candidate lying within the HI tidal bridge between
M81 and NGC 3077, and a number of dI and dSph candidates.
No new large diffuse or any ultra compact dwarfs are detected.
The large fraction of late to early types previously noted in the M81 Group
is reconfirmed with
our new objects.  Assuming all objects lie at the distance of M81,
we find $r^{\prime}$ absolute magnitudes for these new objects range 
from -6.7 to -13.3.  The faintest object, if real, has a measured
size of 90 pc that would cause it to encroach on the 40 - 100 pc size gap 
region between globular clusters and dwarf galaxies.
From false star/galaxy tests we expect to
detect nearly 100\% of the group member dwarfs down to $M_{r^{\prime}} = -10$
and over 50\% to $M_{r^{\prime}} = -9$,
not quite into the regime of the recently discovered ultra faint Local Group galaxies.
Including all 22 previously known M81 Group members in our survey
region, we construct the group differential and cumulative
luminosity functions and find modestly steep faint-end slopes of $\sim -1.30\pm 0.06$.
Including only the most likely members, we find a similar slope of $-1.27\pm0.06$.
This slope is steeper than what has been found in the case of the Andromeda satellites,
but may be consistent with the Local Group after taking into account all
new discoveries and including corrections for completeness.  Even with the 
addition of the 22 new candidate dwarf galaxies in the M81 Group, 
number counts remain an order of magnitude below cosmological 
predictions of halo counts.
 
\acknowledgments
This research was funded in part by NSF award AST03-07706, RFBR grant 07-02-00005, and
NASA/STScI grant HST-GO-11126.01-A.
Based on observations obtained with MegaPrime/MegaCam, a joint project of CFHT and 
CEA/DAPNIA, at the Canada-France-Hawaii Telescope (CFHT) which is operated by the 
National Research Council (NRC) of Canada, the Institut National des Sciences 
de l'Univers of the Centre National de la Recherche Scientifique (CNRS) of France, 
and the University of Hawaii. This work is based in part on data products produced 
at TERAPIX.
Thanks to Canada-France-Hawaii Telescope queue observers, TERAPIX at the
Institut d'Astrophysique de Paris, and Subaru Telescope support staff,
H\'{e}l\`{e}ne Courtois and Luca Rizzi for UH88 observations, Viktor Afanasiev
and Alex Moiseev for work with the BTA 6m data, and Elias Brinks for providing information
on an HI cloudlet near one of our candidates. We thank the referee for constructive
comments and suggestions.

\bibliographystyle{apj}
\bibliography{kc}

\begin{thebibliography}{103}
\expandafter\ifx\csname natexlab\endcsname\relax\def\natexlab#1{#1}\fi

\bibitem[{{Afanasiev} \& {Moiseev}(2005)}]{AM05}
{Afanasiev}, V.~L. \& {Moiseev}, A.~V. 2005, Astronomy Letters, 31, 194

\bibitem[{{Allen} \& {Shu}(1979)}]{as79}
{Allen}, R.~J. \& {Shu}, F.~H. 1979, \apj, 227, 67

\bibitem[{{Babul} \& {Rees}(1992)}]{br92}
{Babul}, A. \& {Rees}, M.~J. 1992, \mnras, 255, 346

\bibitem[{{Barkhouse} {et~al.}(2007){Barkhouse}, {Yee}, \&
  {L{\'o}pez-Cruz}}]{bark07}
{Barkhouse}, W.~A., {Yee}, H.~K.~C., \& {L{\'o}pez-Cruz}, O. 2007, \apj, 671,
  1471

\bibitem[{{Bekki} {et~al.}(2001{\natexlab{a}}){Bekki}, {Couch}, \&
  {Drinkwater}}]{bcd01}
{Bekki}, K., {Couch}, W.~J., \& {Drinkwater}, M.~J. 2001{\natexlab{a}}, \apjl,
  552, L105

\bibitem[{{Bekki} {et~al.}(2001{\natexlab{b}}){Bekki}, {Couch}, \&
  {Shioya}}]{bcs01}
{Bekki}, K., {Couch}, W.~J., \& {Shioya}, Y. 2001{\natexlab{b}}, \pasj, 53, 395

\bibitem[{{Belokurov} {et~al.}(2007{\natexlab{a}}){Belokurov}, {Zucker},
  {Evans}, {Kleyna}, {Koposov}, {Hodgkin}, {Irwin}, {Gilmore}, {Wilkinson},
  {Fellhauer}, {Bramich}, {Hewett}, {Vidrih}, {De Jong}, {Smith}, {Rix},
  {Bell}, {Wyse}, {Newberg}, {Mayeur}, {Yanny}, {Rockosi}, {Gnedin},
  {Schneider}, {Beers}, {Barentine}, {Brewington}, {Brinkmann}, {Harvanek},
  {Kleinman}, {Krzesinski}, {Long}, {Nitta}, \& {Snedden}}]{belo07}
{Belokurov}, V., {Zucker}, D.~B., {Evans}, N.~W., {Kleyna}, J.~T., {Koposov},
  S., {Hodgkin}, S.~T., {Irwin}, M.~J., {Gilmore}, G., {Wilkinson}, M.~I.,
  {Fellhauer}, M., {Bramich}, D.~M., {Hewett}, P.~C., {Vidrih}, S., {De Jong},
  J.~T.~A., {Smith}, J.~A., {Rix}, H.-W., {Bell}, E.~F., {Wyse}, R.~F.~G.,
  {Newberg}, H.~J., {Mayeur}, P.~A., {Yanny}, B., {Rockosi}, C.~M., {Gnedin},
  O.~Y., {Schneider}, D.~P., {Beers}, T.~C., {Barentine}, J.~C., {Brewington},
  H., {Brinkmann}, J., {Harvanek}, M., {Kleinman}, S.~J., {Krzesinski}, J.,
  {Long}, D., {Nitta}, A., \& {Snedden}, S.~A. 2007{\natexlab{a}}, \apj, 654,
  897

\bibitem[{{Belokurov} {et~al.}(2007{\natexlab{b}}){Belokurov}, {Zucker},
  {Evans}, {Kleyna}, {Koposov}, {Hodgkin}, {Irwin}, {Gilmore}, {Wilkinson},
  {Fellhauer}, {Bramich}, {Hewett}, {Vidrih}, {De Jong}, {Smith}, {Rix},
  {Bell}, {Wyse}, {Newberg}, {Mayeur}, {Yanny}, {Rockosi}, {Gnedin},
  {Schneider}, {Beers}, {Barentine}, {Brewington}, {Brinkmann}, {Harvanek},
  {Kleinman}, {Krzesinski}, {Long}, {Nitta}, \& {Snedden}}]{belo06}
---. 2007{\natexlab{b}}, \apj, 654, 897

\bibitem[{{Benson} {et~al.}(2003){Benson}, {Frenk}, {Baugh}, {Cole}, \&
  {Lacey}}]{bfbcl02}
{Benson}, A.~J., {Frenk}, C.~S., {Baugh}, C.~M., {Cole}, S., \& {Lacey}, C.~G.
  2003, \mnras, 343, 679

\bibitem[{{Bernstein} {et~al.}(1995){Bernstein}, {Nichol}, {Tyson}, {Ulmer}, \&
  {Wittman}}]{bntuw95}
{Bernstein}, G.~M., {Nichol}, R.~C., {Tyson}, J.~A., {Ulmer}, M.~P., \&
  {Wittman}, D. 1995, \aj, 110, 1507+

\bibitem[{{Binggeli} \& {Jerjen}(1998)}]{bj98}
{Binggeli}, B. \& {Jerjen}, H. 1998, \aap, 333, 17

\bibitem[{{Blanton} {et~al.}(2005){Blanton}, {Lupton}, {Schlegel}, {Strauss},
  {Brinkmann}, {Fukugita}, \& {Loveday}}]{bls05}
{Blanton}, M.~R., {Lupton}, R.~H., {Schlegel}, D.~J., {Strauss}, M.~A.,
  {Brinkmann}, J., {Fukugita}, M., \& {Loveday}, J. 2005, \apj, 631, 208

\bibitem[{{Boerngen} \& {Karachentseva}(1982)}]{BK82}
{Boerngen}, F. \& {Karachentseva}, V.~E. 1982, Astronomische Nachrichten, 303,
  189

\bibitem[{{Boerngen} \& {Karachentseva}(1985)}]{BK85}
---. 1985, Astronomische Nachrichten, 306, 301

\bibitem[{{Bovill} \& {Ricotti}(2008)}]{bov08}
{Bovill}, M.~S. \& {Ricotti}, M. 2008, ArXiv e-prints, 806

\bibitem[{{Boyce} {et~al.}(2001){Boyce}, {Phillipps}, {Jones}, {Driver},
  {Smith}, \& {Couch}}]{bpj01}
{Boyce}, P.~J., {Phillipps}, S., {Jones}, J.~B., {Driver}, S.~P., {Smith},
  R.~M., \& {Couch}, W.~J. 2001, \mnras, 328, 277

\bibitem[{{Brinks} {et~al.}(2007){Brinks}, {Walter}, \& {Skillman}}]{bws07}
{Brinks}, E., {Walter}, F., \& {Skillman}, E.~D. 2007, ArXiv e-prints, 708

\bibitem[{{Caldwell} {et~al.}(1998){Caldwell}, {Armandroff}, {Da Costa}, \&
  {Seitzer}}]{cads98}
{Caldwell}, N., {Armandroff}, T.~E., {Da Costa}, G.~S., \& {Seitzer}, P. 1998,
  \aj, 115, 535

\bibitem[{{Caon} {et~al.}(1993){Caon}, {Capaccioli}, \& {D'Onofrio}}]{ccd93}
{Caon}, N., {Capaccioli}, M., \& {D'Onofrio}, M. 1993, \mnras, 265, 1013

\bibitem[{{Chiboucas} \& {Mateo}(2006)}]{KCMM2}
{Chiboucas}, K. \& {Mateo}, M. 2006, \aj, 132, 347

\bibitem[{{Chiboucas} \& {Mateo}(2007)}]{KCMM1}
---. 2007, \apjs, 170, 95

\bibitem[{{Col{\'{\i}}n} {et~al.}(2000){Col{\'{\i}}n}, {Avila-Reese}, \&
  {Valenzuela}}]{colin00}
{Col{\'{\i}}n}, P., {Avila-Reese}, V., \& {Valenzuela}, O. 2000, \apj, 542, 622

\bibitem[{{Conselice}(2002)}]{con02}
{Conselice}, C.~J. 2002, \apjl, 573, L5

\bibitem[{{Davidge}(2008)}]{dav08}
{Davidge}, T.~J. 2008, ArXiv e-prints, 803

\bibitem[{{de Mello} {et~al.}(2008){de Mello}, {Smith}, {Sabbi}, {Gallagher},
  {Mountain}, \& {Harbeck}}]{dem08}
{de Mello}, D.~F., {Smith}, L.~J., {Sabbi}, E., {Gallagher}, J.~S., {Mountain},
  M., \& {Harbeck}, D.~R. 2008, \aj, 135, 548

\bibitem[{{de Propris} \& {Pritchet}(1998)}]{dp98}
{de Propris}, R. \& {Pritchet}, C.~J. 1998, \aj, 116, 1118

\bibitem[{{Diemand} {et~al.}(2007){Diemand}, {Kuhlen}, \& {Madau}}]{vialac07}
{Diemand}, J., {Kuhlen}, M., \& {Madau}, P. 2007, \apj, 657, 262

\bibitem[{{Drinkwater} {et~al.}(2003){Drinkwater}, {Gregg}, {Hilker}, {Couch},
  {Ferguson}, {Jones}, \& {Phillipps}}]{drink2}
{Drinkwater}, M.~J., {Gregg}, M.~D., {Hilker}, M., {Couch}, W.~J., {Ferguson},
  H.~C., {Jones}, B., \& {Phillipps}, S. 2003, The Cosmic Cauldron, 25th
  meeting of the IAU, Joint Discussion 10, 18 July 2003, Sydney, Australia, 10

\bibitem[{{Drinkwater} {et~al.}(2000){Drinkwater}, {Jones}, {Gregg}, \&
  {Phillipps}}]{djgp00}
{Drinkwater}, M.~J., {Jones}, J.~B., {Gregg}, M.~D., \& {Phillipps}, S. 2000,
  Publications of the Astronomical Society of Australia, 17, 227

\bibitem[{{Durrell} {et~al.}(2004){Durrell}, {Decesar}, {Ciardullo},
  {Hurley-Keller}, \& {Feldmeier}}]{durrell}
{Durrell}, P.~R., {Decesar}, M.~E., {Ciardullo}, R., {Hurley-Keller}, D., \&
  {Feldmeier}, J.~J. 2004, in IAU Symposium, Vol. 217, Recycling Intergalactic
  and Interstellar Matter, ed. P.-A. {Duc}, J.~{Braine}, \& E.~{Brinks}, 90--+

\bibitem[{{Fellhauer} \& {Kroupa}(2002)}]{fk02}
{Fellhauer}, M. \& {Kroupa}, P. 2002, \mnras, 330, 642

\bibitem[{{Froebrich} \& {Meusinger}(2000)}]{FM00}
{Froebrich}, D. \& {Meusinger}, H. 2000, \aaps, 145, 229

\bibitem[{{Fukugita} {et~al.}(1996{\natexlab{a}}){Fukugita}, {Ichikawa},
  {Gunn}, {Doi}, {Shimasaku}, \& {Schneider}}]{fuku}
{Fukugita}, M., {Ichikawa}, T., {Gunn}, J.~E., {Doi}, M., {Shimasaku}, K., \&
  {Schneider}, D.~P. 1996{\natexlab{a}}, \aj, 111, 1748

\bibitem[{{Fukugita} {et~al.}(1996{\natexlab{b}}){Fukugita}, {Ichikawa},
  {Gunn}, {Doi}, {Shimasaku}, \& {Schneider}}]{figdss96}
---. 1996{\natexlab{b}}, \aj, 111, 1748

\bibitem[{{Gonz{\'a}lez} {et~al.}(2006){Gonz{\'a}lez}, {Lares}, {Lambas}, \&
  {Valotto}}]{gllv06}
{Gonz{\'a}lez}, R.~E., {Lares}, M., {Lambas}, D.~G., \& {Valotto}, C. 2006,
  \aap, 445, 51

\bibitem[{{Haiman} {et~al.}(1996){Haiman}, {Thoul}, \& {Loeb}}]{haiman96}
{Haiman}, Z., {Thoul}, A.~A., \& {Loeb}, A. 1996, \apj, 464, 523

\bibitem[{{Ibata} {et~al.}(2007){Ibata}, {Martin}, {Irwin}, {Chapman},
  {Ferguson}, {Lewis}, \& {McConnachie}}]{imi07}
{Ibata}, R., {Martin}, N.~F., {Irwin}, M., {Chapman}, S., {Ferguson}, A.~M.~N.,
  {Lewis}, G.~F., \& {McConnachie}, A.~W. 2007, \apj, 671, 1591

\bibitem[{{Irwin} \& {Hatzidimitriou}(1995)}]{IH95}
{Irwin}, M. \& {Hatzidimitriou}, D. 1995, \mnras, 277, 1354

\bibitem[{{Irwin} {et~al.}(2008){Irwin}, {Ferguson}, {Huxor}, {Tanvir},
  {Ibata}, \& {Lewis}}]{ifh08}
{Irwin}, M.~J., {Ferguson}, A.~M.~N., {Huxor}, A.~P., {Tanvir}, N.~R., {Ibata},
  R.~A., \& {Lewis}, G.~F. 2008, \apjl, 676, L17

\bibitem[{{Jordi} {et~al.}(2006){Jordi}, {Grebel}, \& {Ammon}}]{jordicol06}
{Jordi}, K., {Grebel}, E.~K., \& {Ammon}, K. 2006, \aap, 460, 339

\bibitem[{{Kamionkowski} \& {Liddle}(2000)}]{kl00}
{Kamionkowski}, M. \& {Liddle}, A.~R. 2000, Physical Review Letters, 84, 4525

\bibitem[{{Kang}(2008)}]{kang08}
{Kang}, X. 2008, ArXiv e-prints, 806

\bibitem[{{Karachentsev} {et~al.}(2002{\natexlab{a}}){Karachentsev}, {Dolphin},
  {Geisler}, {Grebel}, {Guhathakurta}, {Hodge}, {Karachentseva}, {Sarajedini},
  {Seitzer}, \& {Sharina}}]{K2002}
{Karachentsev}, I.~D., {Dolphin}, A.~E., {Geisler}, D., {Grebel}, E.~K.,
  {Guhathakurta}, P., {Hodge}, P.~W., {Karachentseva}, V.~E., {Sarajedini}, A.,
  {Seitzer}, P., \& {Sharina}, M.~E. 2002{\natexlab{a}}, \aap, 383, 125

\bibitem[{{Karachentsev} {et~al.}(2000){Karachentsev}, {Karachentseva},
  {Dolphin}, {Geisler}, {Grebel}, {Guhathakurta}, {Hodge}, {Sarajedini},
  {Seitzer}, \& {Sharina}}]{KKp00}
{Karachentsev}, I.~D., {Karachentseva}, V.~E., {Dolphin}, A.~E., {Geisler}, D.,
  {Grebel}, E.~K., {Guhathakurta}, P., {Hodge}, P.~W., {Sarajedini}, A.,
  {Seitzer}, P., \& {Sharina}, M.~E. 2000, \aap, 363, 117

\bibitem[{{Karachentsev} {et~al.}(2004){Karachentsev}, {Karachentseva},
  {Huchtmeier}, \& {Makarov}}]{KKHM04}
{Karachentsev}, I.~D., {Karachentseva}, V.~E., {Huchtmeier}, W.~K., \&
  {Makarov}, D.~I. 2004, \aj, 127, 2031

\bibitem[{{Karachentsev} {et~al.}(2002{\natexlab{b}}){Karachentsev}, {Sharina},
  {Dolphin}, {Grebel}, {Geisler}, {Guhathakurta}, {Hodge}, {Karachentseva},
  {Sarajedini}, \& {Seitzer}}]{ksd02}
{Karachentsev}, I.~D., {Sharina}, M.~E., {Dolphin}, A.~E., {Grebel}, E.~K.,
  {Geisler}, D., {Guhathakurta}, P., {Hodge}, P.~W., {Karachentseva}, V.~E.,
  {Sarajedini}, A., \& {Seitzer}, P. 2002{\natexlab{b}}, \aap, 385, 21

\bibitem[{{Karachentseva}(1968)}]{Ka68}
{Karachentseva}, V.~E. 1968, Soobshcheniya Byurakanskoj Observatorii Akademiya
  Nauk Armyanskoj SSR Erevan, 39, 62

\bibitem[{{Kawata} \& {Mulchaey}(2008)}]{km08}
{Kawata}, D. \& {Mulchaey}, J.~S. 2008, \apjl, 672, L103

\bibitem[{{Kewley} \& {Dopita}(2002)}]{kd02}
{Kewley}, L.~J. \& {Dopita}, M.~A. 2002, \apjs, 142, 35

\bibitem[{{Klypin} {et~al.}(1999){Klypin}, {Kravtsov}, {Valenzuela}, \&
  {Prada}}]{klypin99}
{Klypin}, A., {Kravtsov}, A.~V., {Valenzuela}, O., \& {Prada}, F. 1999, \apj,
  522, 82

\bibitem[{{Koposov} {et~al.}(2007){Koposov}, {Belokurov}, {Evans}, {Hewett},
  {Irwin}, {Gilmore}, {Zucker}, {Rix}, {Fellhauer}, {Bell}, \&
  {Glushkova}}]{kop07}
{Koposov}, S., {Belokurov}, V., {Evans}, N.~W., {Hewett}, P.~C., {Irwin},
  M.~J., {Gilmore}, G., {Zucker}, D.~B., {Rix}, H.~., {Fellhauer}, M., {Bell},
  E.~F., \& {Glushkova}, E.~V. 2007, ArXiv e-prints, 706

\bibitem[{{Kravtsov} {et~al.}(2004){Kravtsov}, {Gnedin}, \& {Klypin}}]{kgk04}
{Kravtsov}, A.~V., {Gnedin}, O.~Y., \& {Klypin}, A.~A. 2004, \apj, 609, 482

\bibitem[{{Krusch} {et~al.}(2006){Krusch}, {Rosenbaum}, {Dettmar}, {Bomans},
  {Taylor}, {Aronica}, \& {Elwert}}]{k06}
{Krusch}, E., {Rosenbaum}, D., {Dettmar}, R.-J., {Bomans}, D.~J., {Taylor},
  C.~L., {Aronica}, G., \& {Elwert}, T. 2006, \aap, 459, 759

\bibitem[{{Kurtz} \& {Mink}(1998)}]{KM98}
{Kurtz}, M.~J. \& {Mink}, D.~J. 1998, \pasp, 110, 934

\bibitem[{{Liu} {et~al.}(2008){Liu}, {Capak}, {Mobasher}, {Paglione}, {Rich},
  {Scoville}, {Tribiano}, \& {Tyson}}]{liu08}
{Liu}, C.~T., {Capak}, P., {Mobasher}, B., {Paglione}, T.~A.~D., {Rich}, R.~M.,
  {Scoville}, N.~Z., {Tribiano}, S.~M., \& {Tyson}, N.~D. 2008, \apj, 672, 198

\bibitem[{{Lynds} {et~al.}(1998){Lynds}, {Tolstoy}, {O'Neil}, \&
  {Hunter}}]{lynd98}
{Lynds}, R., {Tolstoy}, E., {O'Neil}, Jr., E.~J., \& {Hunter}, D.~A. 1998, \aj,
  116, 146

\bibitem[{{Mahdavi} {et~al.}(2005){Mahdavi}, {Trentham}, \& {Tully}}]{mtt05}
{Mahdavi}, A., {Trentham}, N., \& {Tully}, R.~B. 2005, \aj, 130, 1502

\bibitem[{{Majewski} {et~al.}(2007){Majewski}, {Beaton}, {Patterson},
  {Kalirai}, {Geha}, {Mu{\~n}oz}, {Seigar}, {Guhathakurta}, {Gilbert}, {Rich},
  {Bullock}, \& {Reitzel}}]{maj07}
{Majewski}, S.~R., {Beaton}, R.~L., {Patterson}, R.~J., {Kalirai}, J.~S.,
  {Geha}, M.~C., {Mu{\~n}oz}, R.~R., {Seigar}, M.~S., {Guhathakurta}, P.,
  {Gilbert}, K.~M., {Rich}, R.~M., {Bullock}, J.~S., \& {Reitzel}, D.~B. 2007,
  \apjl, 670, L9

\bibitem[{{Makarov} {et~al.}(2006){Makarov}, {Makarova}, {Rizzi}, {Tully},
  {Dolphin}, {Sakai}, \& {Shaya}}]{mmrt06}
{Makarov}, D., {Makarova}, L., {Rizzi}, L., {Tully}, R.~B., {Dolphin}, A.~E.,
  {Sakai}, S., \& {Shaya}, E.~J. 2006, \aj, 132, 2729

\bibitem[{{Makarova} {et~al.}(2002){Makarova}, {Grebel}, {Karachentsev},
  {Dolphin}, {Karachentseva}, {Sharina}, {Geisler}, {Guhathakurta}, {Hodge},
  {Sarajedini}, \& {Seitzer}}]{mak02}
{Makarova}, L.~N., {Grebel}, E.~K., {Karachentsev}, I.~D., {Dolphin}, A.~E.,
  {Karachentseva}, V.~E., {Sharina}, M.~E., {Geisler}, D., {Guhathakurta}, P.,
  {Hodge}, P.~W., {Sarajedini}, A., \& {Seitzer}, P. 2002, \aap, 396, 473

\bibitem[{{Martin} {et~al.}(2006){Martin}, {Ibata}, {Irwin}, {Chapman},
  {Lewis}, {Ferguson}, {Tanvir}, \& {McConnachie}}]{martin06}
{Martin}, N.~F., {Ibata}, R.~A., {Irwin}, M.~J., {Chapman}, S., {Lewis}, G.~F.,
  {Ferguson}, A.~M.~N., {Tanvir}, N., \& {McConnachie}, A.~W. 2006, \mnras,
  371, 1983

\bibitem[{{Mateo}(1998)}]{mm98}
{Mateo}, M.~L. 1998, \araa, 36, 435

\bibitem[{{McConnachie} {et~al.}(2008){McConnachie}, {Huxor}, {Martin},
  {Irwin}, {Chapman}, {Fahlman}, {Ferguson}, {Ibata}, {Lewis}, {Richer}, \&
  {Tanvir}}]{And08}
{McConnachie}, A., {Huxor}, A., {Martin}, N., {Irwin}, M., {Chapman}, S.,
  {Fahlman}, G., {Ferguson}, A., {Ibata}, R., {Lewis}, G., {Richer}, H., \&
  {Tanvir}, N. 2008, ArXiv e-prints, 806

\bibitem[{{McConnachie} \& {Irwin}(2006)}]{mci06}
{McConnachie}, A.~W. \& {Irwin}, M.~J. 2006, \mnras, 365, 1263

\bibitem[{{Monet} {et~al.}(2003){Monet}, {Levine}, {Canzian}, {Ables}, {Bird},
  {Dahn}, {Guetter}, {Harris}, {Henden}, {Leggett}, {Levison}, {Luginbuhl},
  {Martini}, {Monet}, {Munn}, {Pier}, {Rhodes}, {Riepe}, {Sell}, {Stone},
  {Vrba}, {Walker}, {Westerhout}, {Brucato}, {Reid}, {Schoening}, {Hartley},
  {Read}, \& {Tritton}}]{Monet03}
{Monet}, D.~G., {Levine}, S.~E., {Canzian}, B., {Ables}, H.~D., {Bird}, A.~R.,
  {Dahn}, C.~C., {Guetter}, H.~H., {Harris}, H.~C., {Henden}, A.~A., {Leggett},
  S.~K., {Levison}, H.~F., {Luginbuhl}, C.~B., {Martini}, J., {Monet},
  A.~K.~B., {Munn}, J.~A., {Pier}, J.~R., {Rhodes}, A.~R., {Riepe}, B., {Sell},
  S., {Stone}, R.~C., {Vrba}, F.~J., {Walker}, R.~L., {Westerhout}, G.,
  {Brucato}, R.~J., {Reid}, I.~N., {Schoening}, W., {Hartley}, M., {Read},
  M.~A., \& {Tritton}, S.~B. 2003, \aj, 125, 984

\bibitem[{{Moore} {et~al.}(1999){Moore}, {Ghigna}, {Governato}, {Lake},
  {Quinn}, {Stadel}, \& {Tozzi}}]{moore99}
{Moore}, B., {Ghigna}, S., {Governato}, F., {Lake}, G., {Quinn}, T., {Stadel},
  J., \& {Tozzi}, P. 1999, \apjl, 524, L19

\bibitem[{{Moore} {et~al.}(1996){Moore}, {Katz}, {Lake}, {Dressler}, \&
  {Oemler}}]{mkldo96}
{Moore}, B., {Katz}, N., {Lake}, G., {Dressler}, A., \& {Oemler}, A. 1996,
  \nat, 379, 613

\bibitem[{{Moore} {et~al.}(1998){Moore}, {Lake}, \& {Katz}}]{mlk98}
{Moore}, B., {Lake}, G., \& {Katz}, N. 1998, \apj, 495, 139

\bibitem[{{Mori} \& {Burkert}(2000)}]{mb00}
{Mori}, M. \& {Burkert}, A. 2000, \apj, 538, 559

\bibitem[{{Pe{\~n}arrubia} {et~al.}(2008){Pe{\~n}arrubia}, {McConnachie}, \&
  {Navarro}}]{pen08}
{Pe{\~n}arrubia}, J., {McConnachie}, A.~W., \& {Navarro}, J.~F. 2008, \apj,
  672, 904

\bibitem[{{Perelmuter} \& {Racine}(1995)}]{perel}
{Perelmuter}, J.-M. \& {Racine}, R. 1995, \aj, 109, 1055

\bibitem[{{Pettini} \& {Pagel}(2004)}]{pp04}
{Pettini}, M. \& {Pagel}, B.~E.~J. 2004, \mnras, 348, L59

\bibitem[{{Press} {et~al.}(1992){Press}, {Teukolsky}, {Vetterling}, \&
  {Flannery}}]{press92}
{Press}, W.~H., {Teukolsky}, S.~A., {Vetterling}, W.~T., \& {Flannery}, B.~P.
  1992, {Numerical recipes in FORTRAN. The art of scientific computing}
  (Cambridge: University Press, |c1992, 2nd ed.)

\bibitem[{{Pritchet} \& {van den Bergh}(1999)}]{pv99}
{Pritchet}, C.~J. \& {van den Bergh}, S. 1999, \aj, 118, 883

\bibitem[{{Roberts} {et~al.}(2007){Roberts}, {Davies}, {Sabatini}, {Auld}, \&
  {Smith}}]{rds07}
{Roberts}, S., {Davies}, J., {Sabatini}, S., {Auld}, R., \& {Smith}, R. 2007,
  \mnras, 379, 1053

\bibitem[{{Roberts} {et~al.}(2004){Roberts}, {Davies}, {Sabatini}, {van Driel},
  {O'Neil}, {Baes}, {Linder}, {Smith}, \& {Evans}}]{rds04}
{Roberts}, S., {Davies}, J., {Sabatini}, S., {van Driel}, W., {O'Neil}, K.,
  {Baes}, M., {Linder}, S., {Smith}, R., \& {Evans}, R. 2004, \mnras, 352, 478

\bibitem[{{Sabbi} {et~al.}(2008){Sabbi}, {Gallagher}, {Smith}, {de Mello}, \&
  {Mountain}}]{sbsdm08}
{Sabbi}, E., {Gallagher}, J.~S., {Smith}, L.~J., {de Mello}, D.~F., \&
  {Mountain}, M. 2008, ArXiv e-prints, 802

\bibitem[{{Schechter}(1976)}]{s76}
{Schechter}, P. 1976, \apj, 203, 297

\bibitem[{{Schlegel} {et~al.}(1998){Schlegel}, {Finkbeiner}, \&
  {Davis}}]{sfd98}
{Schlegel}, D.~J., {Finkbeiner}, D.~P., \& {Davis}, M. 1998, \apj, 500, 525+

\bibitem[{{Secker} \& {Harris}(1997)}]{sh97}
{Secker}, J. \& {Harris}, W.~E. 1997, \pasp, 109, 1364

\bibitem[{{Secker} {et~al.}(1997){Secker}, {Harris}, \& {Plummer}}]{shp97}
{Secker}, J., {Harris}, W.~E., \& {Plummer}, J.~D. 1997, \pasp, 109, 1377

\bibitem[{{Shaya} \& {Tully}(1984)}]{st84}
{Shaya}, E.~J. \& {Tully}, R.~B. 1984, \apj, 281, 56

\bibitem[{{Simon} \& {Geha}(2007)}]{simon07}
{Simon}, J.~D. \& {Geha}, M. 2007, ArXiv e-prints, 706

\bibitem[{{Stoehr} {et~al.}(2002){Stoehr}, {White}, {Tormen}, \&
  {Springel}}]{stoehr02}
{Stoehr}, F., {White}, S.~D.~M., {Tormen}, G., \& {Springel}, V. 2002, \mnras,
  335, L84

\bibitem[{{Strigari} {et~al.}(2007){Strigari}, {Bullock}, {Kaplinghat},
  {Diemand}, {Kuhlen}, \& {Madau}}]{sbkd07}
{Strigari}, L.~E., {Bullock}, J.~S., {Kaplinghat}, M., {Diemand}, J., {Kuhlen},
  M., \& {Madau}, P. 2007, \apj, 669, 676

\bibitem[{{Thoul} \& {Weinberg}(1996)}]{tw96}
{Thoul}, A.~A. \& {Weinberg}, D.~H. 1996, \apj, 465, 608

\bibitem[{{Trentham}(1997)}]{t97}
{Trentham}, N. 1997, \mnras, 286, 133

\bibitem[{{Trentham} {et~al.}(2005){Trentham}, {Sampson}, \& {Banerji}}]{tsb05}
{Trentham}, N., {Sampson}, L., \& {Banerji}, M. 2005, \mnras, 357, 783

\bibitem[{{Trentham} \& {Tully}(2002)}]{tt02}
{Trentham}, N. \& {Tully}, R.~B. 2002, \mnras, 335, 712

\bibitem[{{Trentham} {et~al.}(2006){Trentham}, {Tully}, \& {Mahdavi}}]{ttm06}
{Trentham}, N., {Tully}, R.~B., \& {Mahdavi}, A. 2006, \mnras, 369, 1375

\bibitem[{{Tully} {et~al.}(1981){Tully}, {Boesgaard}, {Dyck}, \&
  {Schempp}}]{tul81}
{Tully}, R.~B., {Boesgaard}, A.~M., {Dyck}, H.~M., \& {Schempp}, W.~V. 1981,
  \apj, 246, 38

\bibitem[{{Tully} \& {Pierce}(2000)}]{tp00}
{Tully}, R.~B. \& {Pierce}, M.~J. 2000, \apj, 533, 744

\bibitem[{{Tully} {et~al.}(2002){Tully}, {Somerville}, {Trentham}, \&
  {Verheijen}}]{tstv02}
{Tully}, R.~B., {Somerville}, R.~S., {Trentham}, N., \& {Verheijen}, M.~A.~W.
  2002, \apj, 569, 573

\bibitem[{{Tully} {et~al.}(1996){Tully}, {Verheijen}, {Pierce}, {Huang}, \&
  {Wainscoat}}]{TullyUM}
{Tully}, R.~B., {Verheijen}, M.~A.~W., {Pierce}, M.~J., {Huang}, J.-S., \&
  {Wainscoat}, R.~J. 1996, \aj, 112, 2471

\bibitem[{{van den Bergh}(1966)}]{vdB71}
{van den Bergh}, S. 1966, \aj, 71, 922

\bibitem[{{van den Bergh}(2006)}]{vdb06}
---. 2006, \aj, 132, 1571

\bibitem[{{van Zee} {et~al.}(2001){van Zee}, {Salzer}, \& {Skillman}}]{vz01}
{van Zee}, L., {Salzer}, J.~J., \& {Skillman}, E.~D. 2001, \aj, 122, 121

\bibitem[{{Walsh} {et~al.}(2007){Walsh}, {Jerjen}, \& {Willman}}]{bootes07}
{Walsh}, S.~M., {Jerjen}, H., \& {Willman}, B. 2007, \apjl, 662, L83

\bibitem[{{Willman} {et~al.}(2005){Willman}, {Blanton}, {West}, {Dalcanton},
  {Hogg}, {Schneider}, {Wherry}, {Yanny}, \& {Brinkmann}}]{will1}
{Willman}, B., {Blanton}, M.~R., {West}, A.~A., {Dalcanton}, J.~J., {Hogg},
  D.~W., {Schneider}, D.~P., {Wherry}, N., {Yanny}, B., \& {Brinkmann}, J.
  2005, \aj, 129, 2692

\bibitem[{{Young} {et~al.}(2007){Young}, {Skillman}, {Weisz}, \&
  {Dolphin}}]{yswd07}
{Young}, L.~M., {Skillman}, E.~D., {Weisz}, D.~R., \& {Dolphin}, A.~E. 2007,
  \apj, 659, 331

\bibitem[{{Yun} {et~al.}(1994){Yun}, {Ho}, \& {Lo}}]{yun94}
{Yun}, M.~S., {Ho}, P.~T.~P., \& {Lo}, K.~Y. 1994, \nat, 372, 530

\bibitem[{{Zabludoff} \& {Mulchaey}(2000)}]{zm00}
{Zabludoff}, A.~I. \& {Mulchaey}, J.~S. 2000, \apj, 539, 136

\bibitem[{{Zucker} {et~al.}(2006){Zucker}, {Belokurov}, {Evans}, {Kleyna},
  {Irwin}, {Wilkinson}, {Fellhauer}, {Bramich}, {Gilmore}, {Newberg}, {Yanny},
  {Smith}, {Hewett}, {Bell}, {Rix}, {Gnedin}, {Vidrih}, {Wyse}, {Willman},
  {Grebel}, {Schneider}, {Beers}, {Kniazev}, {Barentine}, {Brewington},
  {Brinkmann}, {Harvanek}, {Kleinman}, {Krzesinski}, {Long}, {Nitta}, \&
  {Snedden}}]{zucker06}
{Zucker}, D.~B., {Belokurov}, V., {Evans}, N.~W., {Kleyna}, J.~T., {Irwin},
  M.~J., {Wilkinson}, M.~I., {Fellhauer}, M., {Bramich}, D.~M., {Gilmore}, G.,
  {Newberg}, H.~J., {Yanny}, B., {Smith}, J.~A., {Hewett}, P.~C., {Bell},
  E.~F., {Rix}, H.-W., {Gnedin}, O.~Y., {Vidrih}, S., {Wyse}, R.~F.~G.,
  {Willman}, B., {Grebel}, E.~K., {Schneider}, D.~P., {Beers}, T.~C.,
  {Kniazev}, A.~Y., {Barentine}, J.~C., {Brewington}, H., {Brinkmann}, J.,
  {Harvanek}, M., {Kleinman}, S.~J., {Krzesinski}, J., {Long}, D., {Nitta}, A.,
  \& {Snedden}, S.~A. 2006, \apjl, 650, L41

\end{thebibliography}

\clearpage
\begin{deluxetable}{lllllll}
\tabletypesize{\scriptsize}
\tablewidth{0pt}
\tablecaption{New M81 Dwarf Candidates\label{dwarftab1}}
\tablehead{
\colhead{Name} &
\colhead{$\alpha$} &
\colhead{$\delta$ J2000.0} &
\colhead{$B_{t}$\tablenotemark{a}} &
\colhead{$a^{\prime} \times b^{\prime}$\tablenotemark{b}} &
\colhead{Type} &
\colhead{Comments} \\
}
\startdata
  d0926+70 & 09 26 27.9 & +70 30 24 & 19.0 & $0.8\times0.6$ & dI &  \\
  d0934+70 & 09 34 03.7 & +70 12 57 & 19.3 & $0.7\times0.6$ & dSph & behind cirrus \\
  d0939+71 & 09 39 15.9 & +71 18 42 & 20.0 & $0.7\times0.6$ & dSph? & 9.4 arcmin from HoI \\
  d0944+69 & 09 44 22.5 & +69 12 40 & 21.0& $0.2\times0.2$ & dSph & \\
  d0944+71 & 09 44 34.4 & +71 28 57 & 17.5 & $1.3\times0.8$ & dI & behind brt foreground star \\
  d0946+68 & 09 46 13.0 & +68 42 55 & 20.5 & $0.2\times0.2$ & dI  & background? \\
  d0955+70 & 09 55 13.6 & +70 24 29 & 20.5 & $0.8\times0.7$ & dSph &  \\
  d0957+70 & 09 57 12.4 & +70 12 35 & 21.0 & $0.3\times0.3$ & dSph? &  \\
  d0958+66 & 09 58 48.5 & +66 50 59 & 16.0 & $1.6\times0.8$ & BCD &  KUG 0954+670 \\
  d0959+68 & 09 59 33.1 & +68 39 25 & 18.0 & $0.8\times0.6$ & dI/tdl &  tidal? \citet{durrell}\\
  d1006+67 & 10 06 46.2 & +67 12 04 & 21.0 & $0.7\times0.7$ & dSph & \\
  d1009+70 & 10 09 34.9 & +70 32 55 & 18.5 & $0.6\times0.5$ & dI & semi-resolved, background? \\
  d1012+64 & 10 12 48.4 & +64 06 27 & 15.6 & $1.4\times0.9$ & BCD  & UGC5497 \\
  d1013+68 & 10 13 11.7 & +68 43 45 & 20.5 &  $1.5\times0.9$  & dI? & artifact?  \\
  d1014+68 & 10 14 55.8 & +68 45 27 & 21.0& $0.4\times0.4$ & dSph & \\
  d1015+69 & 10 15 06.9 & +69 02 15 & 20.5 &  $0.4\times0.3$  & dI  &  backgrnd? \\
  d1016+69 & 10 16 18.3 & +69 29 45 & 21.0 & $0.4\times0.4$ & dSph  &  \\
  d1019+69 & 10 19 52.9 & +69 11 19 & 18.5 &  $0.6\times0.5$  & dI &  backgrnd? \\
  d1020+69 & 10 20 25.0 & +69 11 50 & 20.5 &  $1.5\times0.7$  & dSph? & artifact? \\
  d1028+70 & 10 28 39.7 & +70 14 01 & 16.2 & $1.4\times0.9$ & BCD  & \\
  d1041+70 & 10 41 16.8 & +70 09 03 & 20.5 & $0.8\times0.4$ & dI   &  pear-shaped  \\
  d1048+70 & 10 48 57.0 & +70 25 38 & 21.0 & $1.0\times0.9$    & dSph  & background? \\
                                                                                          
\enddata
                                                                                          
\tablenotetext{a}{$B_{t}$ are estimated by eye on POSS-II}
\tablenotetext{b}{angular dimensions correspond to an isophote ~28m arcsec$^{-2}$}
                                                                                          
\end{deluxetable}

\begin{deluxetable}{lrrrrrrrrr}
\tabletypesize{\scriptsize}
\tablewidth{0pt}
\tablecaption{Spectroscopy of 5 dwarfs in M81\label{tabspec}}
\tablehead{
\colhead{Name} &
\colhead{$\alpha$} &
\colhead{$\delta$} &
\colhead{telescope} &
\colhead{date} &
\colhead{exposure} &
\colhead{grating} &
\colhead{objects} &
\colhead{v$_{rad(Sub)}$} &
\colhead{v$_{rad(BTA)}$} \\
\colhead{} &
\colhead{} &
\colhead{(J2000.0)} &
\colhead{} &
\colhead{} &
\colhead{(s)} &
\colhead{} &
\colhead{} &
\colhead{(km/s)} &
\colhead{(km/s)} \\
}

\startdata
  Previously known: & & & & & & & & & \\
  F8D1 & 9 44 39.3 & +67 26 06  & Subaru & 24 Nov 06 & 1800 & VPH & globular cluster & $-125 \pm 130$ & \\
  IKN & 10 08 05.2  & +68 24 33 & Subaru & 24 Nov 06 & 1800 & VPH & globular cluster & $-140 \pm 64$ & \\
  New Candidates: & & & & & & & & & \\
  d0958+66 & 09 58 48.5 & +66 50 59 & Subaru & 22 Nov 06 & 600 & R300 & galaxy & $+33 \pm 94$ & \\
   &  &  & BTA & 11 Nov 06 & 600 & VPHG400 & galaxy &  & $+90 \pm 50$   \\
  d1028+70 & 10 28 39.7 & +70 14 01 & Subaru & 23 Nov 06 & 600 & VPH & galaxy & $-90 \pm 79$ & \\
   &  &  & BTA & 11 Nov 06 & 600 & VPHG400 & galaxy & & $-114 \pm 50$  \\
  d1012+64 & 10 12 48.4  & +64 06 27 & BTA & 20 Oct 07 & 900 & VPHG400 & galaxy & & $+150 \pm 50$ \\
                                                                                          
\enddata
                                                                                          
\end{deluxetable}

\begin{deluxetable}{rrrrrrrrrrrrrrrr}
\rotate
\tabletypesize{\footnotesize}
\tablewidth{0pt}
\tablecaption{Structural and Photometric Properties of M81 Group Candidates\label{tabX}}
\tablehead{
\colhead{\small galaxy} &
\colhead{\small $r^{\prime}_{ap}$\tablenotemark{a}} &
\colhead{\small $r^{\prime}_{ap_2}$\tablenotemark{b}} &
\colhead{\small $r^{\prime}_{fit}$\tablenotemark{c}} &
\colhead{\small ($V-I$)\tablenotemark{d}} &
\colhead{\small $\epsilon$} &
\colhead{\small PA} &
\colhead{\small $\mu_{\circ}$} &
\colhead{\small $\langle \mu_{e} \rangle$} &
\colhead{\small R$_{e}$} &
\colhead{\small R$_{e}$} &
\colhead{\small n} &
\colhead{\small A$_{r^{\prime}}$} &
\colhead{\small $r^{\prime}_{cor_{e}}$\tablenotemark{e}} &
\colhead{\small $r^{\prime}_{cor_{s}}$\tablenotemark{f}} &
\colhead{\small M$_{r^{\prime}_{cor_{s}}}$\tablenotemark{g}} \\
\colhead{} &
\colhead{} &
\colhead{mag} &
\colhead{} &
\colhead{} &
\colhead{} &
\colhead{} &
\colhead{$r^{\prime} arcsec^{-2}$} &
\colhead{$r^{\prime} arcsec^{-2}$} &
\colhead{arcsec} &
\colhead{kpc} &
\colhead{} &
\colhead{mag} &
\colhead{} &
\colhead{mag} &
\colhead{} \\
 }

\startdata

d0926+70\tablenotemark{\dag} & 18.3 &  18.9 &  18.4 & 0.87 & 0.18 & -23.9 & 26.0 & 26.2 & 14.3 & 0.25 & 0.42 & 0.48 & 17.4 & 17.9 & -9.9 \\
d0934+70\tablenotemark{\dag} & 18.9 & 18.9 & 18.5 & 1.07 & 0.07 & 63.5 & 26.2 & 26.4 & 15.3 & 0.27 & 0.42 & 0.66 & 17.5 & 18.0 & -9.8 \\
d0939+71 & 18.9 & 19.1 & 18.2 &  & 0.05 & -80.9 & 25.5 & 26.1 & 15.3 & 0.27 & 0.66 & 0.09 & 18.3 &  18.5 & -9.3 \\
d0944+69 & 21.6 & 21.9 &  21.2 &  & 0.09 & -9.7 & 25.9 & 26.7 & 4.9 & 0.09 & 0.78 & 0.22 & 20.7 & 21.1 & -6.7 \\
d0944+71 & 15.6 & 15.9 & 15.6 &  & 0.20 & -27.4 &  23.4 & 23.9 & 18.3 & 0.32 & 0.61 & 0.09 & 15.0 & 15.2 & -12.6 \\
d0946+68 & 19.4 & 19.5 & 19.6 &  & 0.04 & 48.6 & 24.8 & 25.6 & 7.5 & 0.13 & 0.76 & 0.21 & 19.0 & 19.2 & -8.6 \\
d0955+70 & 18.5 & 18.6 & 18.8 & 1.03 & 0.04 & 86.8 & 26.6 & 27.1 & 18.6 & 0.32 & 0.61 & 0.43 & 17.5 & 17.9 & -9.9 \\
d0957+70 &  19.5 & 19.3 &  19.9 &  & 0.04 & -27.6 & 27.1 & 27.3 & 11.3 & 0.20 & 0.36 & 0.24 & 18.2 & 19.1 & -8.7 \\
d0958+66 & 14.8 & 14.8 & 15.0 & 0.84 & 0.36 & -6.9 & 21.6 & 22.3 & 11.6 & 0.20 & 0.71 & 0.17 & 14.7 & 14.7 & -13.1 \\
d0959+68\tablenotemark{\dag} & 16.4 & 16.5 &  16.2 & 0.74 & 0.10 & 35.1 & 26.0 & 25.5 & 26.6 & 0.46 & 0.21 & 0.20 & 16.2 & 16.2 & -11.6 \\
d1006+67 & 18.3 & 18.5 & 18.1 &  & 0.05 & 86.1 & 26.2 & 26.5 & 18.1 & 0.32 & 0.44 & 0.17 & 17.6 & 18.0 & -9.8 \\
d1009+70 & 17.9 & 17.8 & 17.8 &  0.68 & 0.19 & -17.6 & 24.9 & 25.2 & 11.6 & 0.20 & 0.50 & 0.38 & 17.4 & 17.5 & -10.3 \\
d1012+64 & 14.6 & 14.6 &  14.6 &  & 0.15 & 17.0 & 21.5 & 22.3 & 13.9 & 0.24 & 0.88 & 0.05 & 14.5 & 14.5 & -13.3 \\
d1013+68\tablenotemark{\dag} & 19.0 & 19.6 &  18.6 &  & 0.10 & -34.5 & 26.9 & 27.9 & 29.5 & 0.51 & 1.00 & 0.11 & 18.5 & 18.5 & -9.3 \\
d1014+68 & 18.7 & 18.8 &  18.8 &  & 0.01 & -8.9 & 27.2 & 27.3 & 19.7 & 0.34 & 0.29 & 0.13 & 17.5 & 18.5 & -9.3 \\
d1015+69\tablenotemark{\dag} & 19.4 & 19.5 &  19.4 &  & 0.07 & -49.7 & 25.5 & 26.0 & 8.5 & 0.15 & 0.68 & 0.13 & 18.9 & 19.1 & -8.7 \\
d1016+69 & 18.9 & 19.0 &  19.1 &  & 0.01 & 87.6 & 27.1 & 27.6 & 19.3 & 0.34 &  0.60 & 0.13 & 17.8 & 18.4 & -9.4 \\
d1019+69 & 18.1 & 17.5 &  18.1 &  & 0.09 & 38.1 & 24.5 & 25.0 & 9.8 & 0.17 & 0.63 & 0.11 & 17.7 & 17.8 & -10.0 \\
d1020+69 & 18.6 & 18.3 &  18.1 &  & 0.13 & -31.5 & 26.7 & 26.9 & 22.3 & 0.39 & 0.41 & 0.10 & 17.5 & 18.1 & -9.7 \\
d1028+70 & 15.7 & 15.7 &  15.7 & 0.77 & 0.20 & -54.8 & 22.4 & 23.3 & 13.4 & 0.23 & 0.88 & 0.11 & 15.5 & 15.5 & -12.3 \\
d1041+70 & 18.5 & 18.6 &  19.0 & 1.00 & 0.18 & 26.0 & 26.1 & 26.2 & 10.9 & 0.19 & 0.30 & 0.15 & 18.1 & 18.5 & -9.3 \\
d1048+70\tablenotemark{\dag} & 18.3 & 19.5 &  19.4 &  & 0.02 & -10.6 & 27.3 & 27.6 & 17.5 & 0.30 & 0.42 & 0.10  & 18.2 &  18.7 & -9.1 \\

\enddata
                                                                                
\tablecomments {Measured apparent magnitudes and surface brightnesses are uncorrected for extinction.}
\tablenotetext{a}{Magnitude measured in circular aperture, mean sky measured in circular annulus $>2r_{galaxy}$}
\tablenotetext{b}{Aperture magnitude with sky+foreground contamination estimated from counts in nearby circular apertures}
\tablenotetext{c}{Magnitude from curve of growth fitting with cumulative Sersic function}
\tablenotetext{d}{Integrated color within R$_{e}$}
\tablenotetext{e}{Average measured $r^{\prime}$ corrected to total magnitudes assuming 
an exponential profile, extinction corrected. See text.}
\tablenotetext{f}{Average measured $r^{\prime}$ corrected to total magnitudes assuming a Sersic profile, extinction corrected. See text.}
\tablenotetext{g}{Extinction corrected absolute magnitude assuming galaxy is at the distance of M81, 3.6Mpc, and taking the Sersic profile corrected $r^{\prime}$ measurement} 
\tablenotetext{\dag}{Measurements affected by cirrus or foreground/background objects, 
or not well fit by a Sersic function}

\end{deluxetable}

\begin{deluxetable}{lrrrrrrrrrrrr}
\tabletypesize{\scriptsize}
\tablewidth{0pt}
\tablecaption{Photometry of previously known M81 group members\label{knownpar}}
\tablehead{
\colhead{Name} &
\colhead{$\alpha$} &
\colhead{$\delta$} &
\colhead{type} &
\colhead{$\epsilon$} &
\colhead{$\mu_{o}$} &
\colhead{$\mu_{e}$} &
\colhead{R$_{e}$} &
\colhead{$r^{\prime}_{meas}$} &
\colhead{A$_{r^{\prime}}$} &
\colhead{$r^{\prime}_{cor_{e}}$\tablenotemark{a}} &
\colhead{$r^{\prime}_{cor_{s}}$\tablenotemark{b}} &
\colhead{$M_{r^{\prime}_{cor_{s}}}$\tablenotemark{c}} \\
\colhead{} &
\colhead{(J2000.0)} &
\colhead{} &
\colhead{} &
\colhead{} &
\colhead{$r^{\prime}$ arcsec$^{-2}$} &
\colhead{$r^{\prime}$ arcsec$^{-2}$} &
\colhead{arcsec} &
\colhead{} &
\colhead{mag} &
\colhead{} &
\colhead{mag} &
\colhead{} \\
}
\startdata
  HoI & 9 40 28.2 & +71 11 11 & dI &   & 25.0  & 25.2 & 66.7  & 13.9  & 0.1  & 13.4 & 13.4 & -14.4 \\
  F8D1 & 9 44 50.0 & +67 28 32 & dSph &   & 26.5 & 26.6 & 67.3  & 15.3  & 0.2 & 14.0 & 14.7 & -13.0  \\
  FM1 & 9 45 10.0 & +68 45 54 & dSph & 0.14 & 24.8 & 25.2 & 21.1 & 16.5 & 0.2 & 15.9 & 16.2 & -11.5  \\
  N2976 & 9 47 15.6 & +67 54 49 & Sc pec & 0.60 & 20.4 & 20.7 & 43.6 & 10.5  & 0.2  & 9.7 & 9.8 & -18.0  \\
  KK77 & 9 50 10.0 & +67 30 24  & dSph & 0.29 & 24.5 & 25.2 & 36.7 & 15.4 & 0.4 & 14.7 & 14.9 & -12.8  \\
  BK3N & 9 53 48.5 & +68 58 09 & dI & 0.01 & 25.9 & 24.8 & 6.7 & 18.6  & 0.2  & 17.8 & 17.8 & -10.2  \\
  M81 & 9 55 33.5 & +69 04 00 & Sb &  & 17.1 & 19.3 & 81.1 & 7.8  & 0.2  & 7.2 & 7.2 & -20.6  \\
  M82 & 9 55 53.9 & +69 40 57  & Irr &  & 17.4 & 19.1 & 60.0  & 8.4  & 0.4   & 7.9 & 7.9 & -19.9  \\
  K61 & 9 57 02.7 & +68 35 30 & dSph & 0.25 & 23.9 & 24.7 & 38.4 & 14.8  & 0.2 & 14.5 & 14.5 & -13.3 \\
  A0952+69 & 9 57 29.0 & +69 16 20 & pec  & &  & &   & 16.3  & 0.2   &   &  \\
  HoIX & 9 57 32.4 & +69 02 35 & dI & 0.24 & 24.1 & 24.7 & 46.8 & 14.3  & 0.2  & 13.9 & 14.0 & -13.8 \\
  N3077 & 10 03 21.0 & +68 44 02 & Irr & 0.24 & 19.6 & 20.2 & 37.5 & 10.3  & 0.2  & 10.0 & 10.1 & -17.8  \\
  Garland & 10 03 42.0 & +68 41 36 & pec &  &  & &   &  16.5  & 0.2  &   &  \\
  BK5N & 10 04 40.3 & +68 15 20 & dSph & 0.20  & 24.5 & 25.1 & 22.1  & 16.4  & 0.1  & 16.1 & 16.1 & -11.8  \\
  K63 & 10 05 07.3 & +66 33 18 & dSph & 0.18 & 23.9 & 24.1 & 24.9 & 15.1  & 0.2  & 14.4 & 14.8 & -13.0  \\
  K64 & 10 07 01.9 & +67 49 39 & dSph & 0.33 & 23.2 & 23.8 & 22.1  & 15.1  & 0.2  & 14.6 & 14.7 & -13.1  \\
  IKN & 10 08 05.9 & +68 23 57 & dSph  &  & & &  & (15.7)  & 0.2  & (15.5) & (15.5) & (-12.3)  \\
  HS117 & 10 21 25.2 & +71 06 58 & dI & 0.39 & 24.3 & 24.8 & 19.9 & 16.3  & 0.3  & 15.5 & 15.9 & -11.9  \\
  DDO78 & 10 26 27.9 & +67 39 24 & dSph & (0.10) & (24.8) & (25.0) & (25.0) & (15.9)  & 0.1  & (15.0) & (15.0) & (-12.8) \\
  IC2574 & 10 28 22.4 & +68 24 58 & SABm &  & 22.6 & 24.0 & 142.0 & 11.5  & 0.3 & 10.2 & 10.3 & -17.7  \\
  DDO82 & 10 30 35.0 & +70 37 10 & Im & 0.40 & 21.8 & 22.7 & 33.9 & 13.1  & 0.1  & 12.9 & 12.9 & -15.1  \\
  BK6N & 10 34 31.9 & +66 00 42 & dSph & 0.32 & 24.6 & 24.6 & 16.6 & 16.2   & 0.0  & 15.9 & 15.9 & -12.0  \\

\enddata

\tablecomments{Values in parentheses highly uncertain due to presence of
bright foreground stars.  Values for brightest galaxies may be underestimates
due to sky subtraction errors in the TERAPIX reduction pipeline.}
\tablenotetext{a}{Corrected for extinction and flux lost in sky assuming an exponential profile.}
\tablenotetext{b}{Corrected for extinction and flux lost in sky assuming a Sersic profile.}
\tablenotetext{c}{Absolute Sersic profile corrected magnitude, using distance moduli from 
Table 2 of \citet{K2002} or assuming 27.8 where not provided.}
                                                                                          
\end{deluxetable}

\begin{deluxetable}{lrrr}
\tabletypesize{\scriptsize}
\tablewidth{0pt}
\tablecaption{Faint-end slope from differential luminosity functions\label{slopediff}}
\tablehead{
\colhead{Sample} &
\colhead{$\alpha$ ($r^{\prime}_{meas}$)} &
\colhead{$\alpha$ ($r^{\prime}_{cor(exp)}$)} &
\colhead{$\alpha$ ($r^{\prime}_{cor(sersic)}$)} \\
}
\startdata
All (completeness corr.) & $-1.29^{+0.04}_{-0.04}$ & $-1.28^{+0.04}_{-0.05}$ & $-1.32^{+0.04}_{-0.05}$ \\
All  & $-1.26^{+0.09}_{-0.08}$ & $-1.39^{+0.05}_{-0.05}$ & $-1.36^{+0.09}_{-0.08}$ \\
Best & $-1.20^{+0.10}_{-0.10}$ & $-1.34^{+0.07}_{-0.07}$ & $-1.30^{+0.10}_{-0.10}$ \\
Original & $-1.19^{+0.07}_{-0.07}$ & $-1.24^{+0.09}_{-0.08}$ & $-1.26^{+0.08}_{-0.08}$ \\
All ($-15 \leq r^{\prime} < -10$) & $-1.43^{+0.09}_{-0.09}$ & $-1.44^{+0.07}_{-0.07}$  & $-1.46^{+0.10}_{-0.10}$ \\
                                                                                          
\enddata
                                                                                          
\end{deluxetable}

\begin{deluxetable}{lrrr}
\tabletypesize{\scriptsize}
\tablewidth{0pt}
\tablecaption{Faint-end slope from integrated luminosity functions\label{slopeint}}
\tablehead{
\colhead{Sample} &
\colhead{$\alpha$ ($r^{\prime}_{meas}$)} &
\colhead{$\alpha$ ($r^{\prime}_{cor(exp)}$)} &
\colhead{$\alpha$ ($r^{\prime}_{cor(sersic)}$)} \\
}
\startdata
All (completeness corr.) & $-1.29^{+0.02}_{-0.05}$ & $-1.28^{+0.03}_{-0.05}$ & $-1.28^{+0.02}_{-0.05}$ \\
All  & $-1.29^{+0.04}_{-0.06}$ & $-1.29^{+0.05}_{-0.08}$ & $-1.28^{+0.04}_{-0.04}$ \\
Best & $-1.27^{+0.05}_{-0.06}$ & $-1.27^{+0.04}_{-0.07}$ & $-1.26^{+0.05}_{-0.04}$ \\
Original & $-1.26^{+0.07}_{-0.10}$ & $-1.24^{+0.06}_{-0.07}$ & $-1.25^{+0.06}_{-0.06}$ \\
                                                                                          
\enddata
                                                                                          
\end{deluxetable}

\clearpage


\clearpage

\clearpage
                                                                                 
\clearpage

\end{document}